\newcommand{\uselipics}[1]{}
\newcommand{\useieee}[1]{#1}
\uselipics{\documentclass[letterpaper,USenglish,cleveref, autoref,thm-restate]{lipics-v2021}
\hideLIPIcs  %uncomment to remove references to LIPIcs series (logo, DOI, ...), e.g. when preparing a pre-final version to be uploaded to arXiv or another public repository
}
\useieee{
\documentclass[conference]{IEEEtran}
\IEEEoverridecommandlockouts
\usepackage{amsthm,amsmath,amssymb,amsfonts}
\usepackage{thm-restate}
\usepackage{enumitem}

\author{\IEEEauthorblockN{Max Bannach}
\IEEEauthorblockA{\textit{AI and Data Science Section} \\
\textit{European Space Agency}\\
Noordwijk, The Netherlands \\
{max.bannach@esa.int}}

\and

\IEEEauthorblockN{Erik D. Demaine, Timothy Gomez}
\IEEEauthorblockA{\textit{Computer Science and AI Lab} \\
\textit{Massachusetts Institute of Technology}\\
Cambridge, United States \\
demaine@mit.edu, tagomez7@mit.edu}

\and

\IEEEauthorblockN{Markus Hecher}
\IEEEauthorblockA{\textit{CNRS, CRIL, University of Artois, Lens}, France \\
\textit{Massachusetts Institute of Technology}\\
Cambridge, United States \\
hecher@mit.edu, hecher@cril.fr}
}

}
\title{\#P is Sandwiched by One and Two \#2DNF Calls: Is Subtraction Stronger Than We Thought?*\thanks{* This is a self-archived version of a paper that will appear at LICS 2025. Author names are given in alphabetical order. This research was carried out while Hecher was a PostDoc at MIT. It was funded by the Austrian Science Fund (FWF), grants J4656 and P32830, the Society for Research Funding in Lower Austria (GFF NOE) grant ExzF-0004, as well as the Vienna Science and Technology Fund (WWTF) grant ICT19-065.}}

\uselipics{
\titlerunning{\#P is Sandwiched by One and Two \#2DNF Calls}

\author{Max Bannach}{European Space Agency, Advanced Concepts Team, Noordwijk, The Netherlands}{max.bannach@esa.int}{https://orcid.org/0000-0002-6475-5512}{}
\author{Erik D. Demaine}{Massachusetts Institute of Technology, Computer Science and Artificial Intelligence Lab, USA}{edemaine@mit.edu}{https://orcid.org/0000-0003-3803-5703}{}{}
\author{Timothy Gomez}{Massachusetts Institute of Technology, Computer Science and Artificial Intelligence Lab, USA}{tagomez7@mit.edu}{}{}{}
\author{Markus Hecher}{Massachusetts Institute of Technology, Computer Science and Artificial Intelligence Lab, USA}{hecher@mit.edu}{https://orcid.org/0000-0003-0131-6771}{}

\authorrunning{Bannach, Demaine, Gomez, and Hecher}
\Copyright{Max Bannach, Erik D. Demaine, Timothy Gomez, and Markus Hecher} 


\category{} %optional, e.g. invited paper
\relatedversion{} %optional, e.g. full version hosted on arXiv, HAL, or other respository/website
\EventEditors{John Q. Open and Joan R. Access}
\EventNoEds{2}
\EventLongTitle{42nd Conference on Very Important Topics (CCC 2024)}
\EventShortTitle{CCC 2024}
\EventAcronym{CCC}
\EventYear{2024}
\EventDate{July 22--25, 2024}
\EventLocation{Ann Arbor, Michigan, USA}
\EventLogo{}
\SeriesVolume{??}
\ArticleNo{??}
}

\useieee{
\usepackage{hyperref}

\newtheorem{theorem}{Theorem}
\newtheorem{lemma}[theorem]{Lemma}
\newtheorem{observation}[theorem]{Observation}
\newtheorem{corollary}[theorem]{Corollary}
\newtheorem{proposition}[theorem]{Proposition}

\newtheorem*{conjecture*}{Conjecture}
\newtheorem{example}[theorem]{Example}
\newtheorem{definition}[theorem]{Definition}

}
\usepackage{booktabs}
\usepackage{pifont}% http://ctan.org/pkg/pifont
\usepackage{mathtools}
\usepackage{xspace}
\usepackage{tikz}
\usetikzlibrary{graphs, arrows.meta}
\usetikzlibrary{decorations.pathreplacing, backgrounds}

\usetikzlibrary{fit,positioning,arrows,shapes,decorations,automata,%
                petri,%
                topaths,calc,patterns} %backgrounds}

\usepackage{booktabs}

\usepackage{array}
\newcolumntype{H}{>{\setbox0=\hbox\bgroup}c<{\egroup}@{}}

\tikzstyle{tdnode} = [draw,rounded corners,top color=vertexTopColor,bottom color=vertexBottomColor,minimum size=1.5em]
\tikzstyle{stdnode} = [tdnode, font=\scriptsize]
\tikzstyle{stdnodecompact} = [stdnode, inner sep = 1.5pt, outer sep = 0.1pt]
\tikzstyle{stdnodetable} = [stdnode, inner sep = 1.5pt, outer sep = 0]
\tikzstyle{stdnodenum} = [minimum size=1.5em, font=\scriptsize]
\tikzstyle{tdedge} = [-,draw,thick]
\tikzstyle{tdlabel} = [draw=none, rectangle, fill=none, inner sep=0pt, font=\scriptsize]

\tikzstyle{squigarrow} = [->,line join=round,decorate, decoration={
    zigzag,segment length=4,amplitude=.9,post=lineto,post length=2pt}]

\tikzstyle{dashedarrow} = [->,dashed]
\colorlet{vertexTopColor}{white}
\colorlet{vertexBottomColor}{black!10}
\definecolor{jade}{rgb}{0.0, 0.66, 0.42}
\definecolor{cerise}{HTML}{CE4760}
\colorlet{fg}{jade!75!black}
\colorlet{bg}{cerise!75!black}
\definecolor{deepspace}{RGB}{0,50,71}
\definecolor{excitered}{RGB}{237,27,47}
\definecolor{trustyazure}{RGB}{0,155,219}
\definecolor{enlightyellow}{RGB}{251,171,24}
\definecolor{pureteal}{RGB}{0,174,157}

\newcommand\Gap{\mathrm{gap}}
\newcommand\Span{\mathrm{span}}
\newcommand\Tot{\mathrm{tot}}
\newcommand\Class[1]{%
  \mathchoice%
  {\text{\normalfont\small$\mathrm{#1}$}}%
  {\text{\normalfont\small$\mathrm{#1}$}}%
  {\text{\normalfont$\mathrm{#1}$}}%
  {\text{\normalfont$\mathrm{#1}$}}%
}
\newcommand{\Lang}[1]{\text{\normalfont\textsc{#1}}}

\DeclarePairedDelimiter\floor{\lfloor}{\rfloor}

\newcommand{\cnf}

\def\phi{\varphi}
\DeclareMathOperator{\itw}{itw}
\DeclareMathOperator{\ibw}{ibw}
\DeclareMathOperator{\tw}{tw}
\DeclareMathOperator{\bw}{bw}
\DeclareMathOperator{\vars}{vars}
\DeclareMathOperator{\clauses}{clauses}

\DeclareMathOperator{\bag}{\chi}

\DeclareMathOperator{\rootOf}{root}
\DeclareMathOperator{\children}{children}

\newcommand\coloneq{\mathrel{\raise.4pt\hbox{:}{=}}}
\newcommand\eqcolon{\mathrel{{=}\raise.4pt\hbox{:}}}
\DeclareMathOperator{\assignment}{\subseteq}

\allowdisplaybreaks

\begin{document}
\fontsize{11}{13}\selectfont %% 11pt

\maketitle

\begin{abstract}
The canonical class in the realm of counting complexity is
$\Class{\#P}$. It is well known that the problem of counting the
models of a propositional formula in disjunctive normal form
(\Lang{\#dnf}) is complete for $\Class{\#P}$ under Turing
reductions. On the other hand, $\Lang{\#dnf}\in\Span\Class{L}$ and
$\Span\Class{L}\subsetneq\Class{\#P}$ unless
$\Class{NL}=\Class{NP}$. Hence, the class of functions
logspace-reducible to \Lang{\#dnf} is a strict subset of $\Class{\#P}$ under
plausible complexity-theoretic assumptions. By contrast, we show that
\emph{two} calls to a (restricted) \Lang{\#2dnf} oracle suffice to capture
$\Gap\Class{P}$, namely, that the logspace many-one closure of the
subtraction between the results of two $\Lang{\#2dnf}$ calls %.
%\frqq$\Lang{\#2dnf}-\Lang{\#2dnf}$\flqq\ 
\emph{is} $\Gap\Class{P}$. Because $\Class{\#P}\subsetneq\Gap\Class{P}$,
$\Class{\#P}$ is strictly contained between one and
two \Lang{\#2dnf} oracle calls. 

Surprisingly, the propositional formulas needed in 
both calls are \emph{linear-time} computable,
and the reduction preserves interesting structural as well as symmetry properties, leading to algorithmic applications.
We show that a \emph{single subtraction} suffices to compensate for the absence of negation while still capturing $\Gap\Class{P}$, i.e., our results carry over to the monotone fragments of \Lang{\#2sat} and \Lang{\#2dnf}.
Since our reduction is linear-time, it preserves sparsity and, as a consequence we obtain a \emph{sparsification lemma} for both \Lang{\#2sat} and \Lang{\#2dnf}. This has only been known for \Lang{$k$sat} with $k\geq 3$ and respective counting versions.

%% A consequence of our result is a new characterization of $\Gap\Class{P}$, which
%% turns out to be equal to the subtraction of the output of two $\Span\Lang{L}$ calls. 
We further show that both \Lang{\#2dnf} calls can be combined into a single call if we allow a little
postprocessing (computable by $\Class{AC}^0$- or $\Class{TC}^0$-circuits). Consequently,
we derive refined versions of Toda's Theorem: $\Class{PH}\subseteq
[\Lang{\#mon2sat}]^{\mathrm{log}}_{\Class{TC}^0}=[\Lang{\#mon2dnf}]^{\mathrm{log}}_{\Class{TC}^0}$ and
$\Class{PH}\subseteq [\Lang{\#impl2sat}]^{\mathrm{log}}_{\Class{AC}^0}$.
Our route to these results is via structure-aware reductions that preserve parameters like treewidth up to an additive overhead. The absence of multiplicative overhead indeed yields parameterized $\Class{SETH}$-tight lower bounds.
\end{abstract}

\useieee{
\begin{IEEEkeywords}
fine-grained counting complexity, sparsification, sharp-p, span-l, satisfiability, sharp-sat, SETH, fixed-parameter tractability, treewidth, linear-time log-space reduction, lower bound, arithmetic postprocessing, \#2cnf, \#2dnf, monotone 
\end{IEEEkeywords}}
\section{Introduction}

%\textbf{todo 1: contribution \#eth, as we showed equivalence between \#eth for \#3sat and \#eth for \#mon2dnf; todo 2: $C_=^P$-completeness for $\#mono2dnf$, which is tight (hard for only 1 dnf term / 1 variable difference).
%todo 3: under eth we can't improve to a planar version without quadratic blow-up! todo 4: https://arxiv.org/pdf/1511.07480 todo 5: replace $t_v \vee f_v$ by $t_v \vee x_v$, $f_v \vee y_v$ for fresh $x_v, y_v$, enabling bipartite, degree 4 for mon2dnf$_{tc0}$}

The function problem \Lang{\#sat} asks, given a propositional formula
$\phi$ in conjunctive normal form (a \Lang{cnf}), how
many of the $2^n$ possible assignments%
\footnote{An \emph{assignment} 
$\beta\assignment\vars(\phi)$ is interpreted as the subset of \emph{true variables}.}
$\beta\assignment\vars(\phi)$
satisfy $\phi$, that
is, the task is to determine the number $\#(\phi)$ of \emph{models} of $\phi$. It is well known that
\Lang{\#sat} is complete for~$\Class{\#P}$, the class of functions
definable as the number of accepting paths of a polynomial-time
nondeterministic Turing machine. In fact, \Lang{\#sat} is
$\Class{\#P}$-complete under logspace many-one reduction, because the Cook-Levin construction
is solution preserving~\cite[Lemma 3.2]{Valiant79}.
Denoting the closure under logspace many-one reduction\footnote{\label{sec:cmonious}Many-one reductions (no postprocessing) imply parsimony. A reduction is \emph{$c$-monious} if it preserves the solution count up to a factor $c$.}
by $[\cdot]^{\log}$, we can \emph{characterize}~$\Class{\#P}$~as:

\begin{observation}\label{observation:3sat}
  $\Class{\#P}=[\Lang{\#sat}]^{\log}=[\Lang{\#3sat}]^{\log}$.
\end{observation}

In stark contrast to the decision version of the problem,
$\Lang{\#sat}$ remains hard even for heavily restricted fragments of
propositional logic. For instance, \Lang{\#sat} trivially reduces to~\Lang{\#dnf}, which asks for the number of models of a formula in
\emph{disjunctive} normal form: observe that $\neg\phi$ is a
\Lang{dnf} and that $\#(\phi)=2^n-\#(\neg\phi)$.
This reduction has two additional features: \emph{(i)}
we require only \emph{one call} to a \Lang{\#dnf} oracle, and \emph{(ii)}
we need to perform \emph{one subtraction} in a postprocessing step
after querying the oracle. That is, we did \emph{not} show
a logspace many-one reduction from $\Lang{\#sat}$ to
$\Lang{\#dnf}$, but from $\Lang{\#sat}$ to the problem of computing
$2^n-\#(\psi)$ for a formula $\psi \in \#\Lang{dnf}$, which we denote as ``$2^n-\#\Lang{dnf}$'':
%We will denote this fact as:

\begin{observation}\label{observation:sharpdnf}
  $\Class{\#P}=[2^n-\Lang{\#dnf}]^{\log}$.
\end{observation}

One may be tempted to think
that this is just a slight technicality, but in fact this \emph{subtraction is crucial:}
$\Lang{\#dnf}$ lies in $\Span\Class{L}$, the class of functions
expressible as the \emph{span of a nondeterministic logspace} Turing machine, i.e., the number of distinct outputs that an
$\Class{NL}$-transducer can produce~\cite{AlvarezJenner93}. We know
$\Span\Class{L}\subsetneq\Class{\#P}$ unless
$\Class{NL}=\Class{NP}$~\cite[Proposition 4.10]{AlvarezJenner93}~\cite{DyerEtAl03}, %see also \cite{DyerEtAl03}, 
hence, $\Class{\#P}=[\Lang{\#dnf}]^{\log}$ is unlikely under plausible
complexity-theoretic assumptions.

\smallskip
\noindent\textbf{The quest for understanding the complexity of~\#2SAT and~\#2DNF.}
The examples illustrate that counting remains hard on syntactically restricted formulas,
and that they do so by surprisingly simple reductions.
One usual suspect that seems to be missing is \Lang{\#2sat},
for which one would expect a similar reduction. 
The seminal work by Valiant~\cite{Valiant79,Valiant79b} proved that
\Lang{\#2sat} is $\Class{\#P}$-hard by a sophisticated chain of reductions from
\Lang{\#sat}, via several variations of the problem of computing a
permanent, to the task of counting matchings in graphs, and then finally
to \Lang{\#2sat}. This chain of reductions results in a time effort of at least $\Omega(n^3 \log n)$ because of formulas
of size $\Omega(n^2)$ (the reduction from computing perfect
matchings to imperfect matchings~\cite[Step 6 in
  Theorem~1]{Valiant79b}) and a \emph{polynomial number} of
oracle calls as well as the ability to postprocess the results
\emph{modulo} a polynomially bounded
number~\cite[Proposition~3.4]{Valiant79}. The insight that even a
simple ``minus'' in postprocessing can have dramatic impacts on the
complexity of counting problems raises the question of how much of the
complexity of \Lang{\#2sat} is ``hidden'' by this seemingly involved reduction.
This question leads to the quest for
a direct reduction from \Lang{\#sat} to
\Lang{\#2sat} and, in the light of Observation~\ref{observation:sharpdnf},
to \Lang{\#2dnf}%, and Observation~\ref{observation:mon-mon} motivates the study of \Lang{\#mondnf}
. \emph{What} makes \Lang{\#2sat} and \Lang{\#2dnf} hard? 

There are multiple complexity classes to characterize these
so-called ``easy to decide and hard to count problems''
\cite{bakali2022guest}. Almost all of these classes collapse under
Turing reductions, making it necessary to study parsimony. We mention
two classes here: $\Tot\Class{P}$ and $\Span\Class{L}$. The
first is the class of counting problems corresponding to the number of
\emph{all} paths of a polynomial time NDTM. It was proven in
\cite{pagourtzis2006complexity} that $\Tot\Class{P}$ is exactly the
set of problems which (1) have an easy decision version\footnote{This class
is called \#PE. The subset of \#P with easy decision versions. } and
(2) are self-reducible. Two subclasses of $\Tot\Class{P}$ were studied in~\cite{bakali2020characterizations} with second order logic showing connections to generalizations of \Lang{\#2sat}. A number of $\Tot\Class{P}$-complete problems
were found in \cite{antonopoulos2022completeness}. The class
$\Span\Class{L}$ is contained in $\Tot\Class{P}$. It is known
that $\Span\Class{L}$ admits a \emph{fully
polynomial randomized approximation scheme} ($\Class{FPRAS}$) via %the %for the 
%$\Span\Class{L}$-complete
%problem~\cite{} of 
counting the number of strings of
length $n$ accepted by an NFA~\cite{AlvarezJenner93,ArenasEtAl21,meel2024faster}. Since this reduction is parsimonious, it preserves the approximation. On the other hand, 
$\Tot\Class{P}$ contains many important problems such as \Lang{\#2sat}
and \Lang{\#perfect-matching}, and the former is sufficient to show that $\Tot\Class{P}$ is \emph{not} a subset of $\Class{FPRAS}$ (class of functions with an $\Class{FPRAS}$). Neither \Lang{\#2sat} nor \Lang{\#dnf} are known to be complete for either class. 

%~\\[-3em]
\subsection{Contribution I: \useieee{Reducing }\uselipics{A Reduction from }\#SAT to Two Calls of \#2DNF}
We provide a new reduction from $\Lang{\#sat}$ to \emph{two} calls to a
\Lang{\#2sat} or \Lang{\#2dnf} oracle. Crucially, we only need a \emph{single subtraction} to combine the
results (no involved postprocessing or modulo computations). All our reductions are logspace computable and,
thus, we can phrase our $\Lang{\#2sat}$ reduction as follows:

\begin{restatable}{theorem}{maintheorem}\label{theorem:main}%
\begin{enumerate}[label=(\arabic*)]
    \item $\Lang{\#sat}\in[\Lang{\#2sat}-\Lang{\#2sat}]^{\log}$
    \item $[\Lang{\#2sat}-\Lang{\#2sat}]^{\log}=[\Lang{\#2dnf}-\Lang{\#2dnf}]^{\log}$
\end{enumerate}  
\end{restatable}

Since \Lang{\#2sat} does not admit an $\Class{FPRAS}$ under common assumptions~\cite{sly2010computational}, we have
$\Lang{\#2sat}\not\in[\Lang{\#2dnf}]^{\log}$ and, hence, an
Immerman-Szelepcsényi-type theorem does \emph{not hold} in the counting
world. Theorem~\ref{theorem:main}, however, establishes a theorem of this type
if \emph{two} calls are permitted. 
In fact, we prove a stronger form of the first part of the theorem:
we reduce to restricted versions of \Lang{\#2sat}; the reduction
can be implemented either in logspace or in linear time; and it %the reduction
preserves important structural and symmetry properties of the input.

\begin{restatable}[Main Lemma]{lemma}{mainlemma}\label{lemma:main}%  
  There is a $\{$ linear-time, logspace $\}$ algorithm mapping a \Lang{cnf} $\phi$ and a corresponding tree decomposition to \Lang{cnf}s $\psi_1$ and $\psi_2$ with at most two variables per clause such that, for $\rho\in\{\tw, \itw, \bw, \ibw\}$,%
  \footnote{Here $\tw(\phi), \itw(\phi), \bw(\phi), \ibw(\phi)$ denote treewidth and bandwidth of two different graphs associated with formula $\phi$; see Section~\ref{sec:contrib-iv}.}
  ~\\[-1em]\[ %~\\[-2.5em]
  \#(\phi) = \#(\psi_1) - \#(\psi_2)%
  \quad\text{and}\quad\useieee{\]\vspace{-1.7em}\[}%
  \max\{\rho(\psi_1),\rho(\psi_2)\}\leq \alpha\cdot\rho(\phi) +
  14\text{ with } %\rho\in\{\bw, \itw\}, 
  \alpha = 1.%
  \]
  \noindent For $\rho\in\{\tw,\itw\}$, the resulting formulas can be restricted to the following fragments:
  \begin{enumerate}[label=(\Alph*)]
  \item $\alpha=1$ and $\psi_1, \psi_2$ are \emph{monotone}, i.e., do not contain negations; or
  \item $\alpha=3$ and $\psi_1, \psi_2$ comprise binary \emph{implications} and are \emph{cubic} and \emph{bipartite}, i.e., every variable occurs at most three times and the primal graph does not contain an odd~cycle.
  \end{enumerate}  
\end{restatable}
%\medskip

The second part of Theorem~\ref{theorem:main} follows because
$[\Lang{\#2sat}-\Lang{\#2sat}]^{\log}$ and
$[\Lang{\#2dnf}-\Lang{\#2dnf}]^{\log}$ both turn out to be precisely
$\Gap\Class{P}$. We discuss this further in the next subsection.
While Lemma~\ref{lemma:main} is the key to all our contributions, a direct consequence is the following:

\begin{corollary}\label{cor:sharpmon}
  $\Class{\#P}{\,\subsetneq\,}[\Lang{\#mon2sat}{-}\Lang{\#mon2sat}]^{\log}{\,\subseteq\,}\Gap\Class{P}$.
\end{corollary}
\begin{proof}
  Containment follows from Lemma~\ref{lemma:main}; it must be
  strict as functions in $\Class{\#P}$ cannot map to
  negative numbers but functions in $[\Lang{\#mon2sat}-\Lang{\#mon2sat}]^{\log}$ can. 
\end{proof}

\noindent Interestingly, we do not expect to improve the reduction of Lemma~\ref{lemma:main} for planarity while keeping linear time.

\begin{proposition}\label{proposition:planar}
	There is no $c$-monious$^{\ref{sec:cmonious}}$ linear-time Turing reduction from $\Lang{3sat}$ to $\Lang{\#planar3sat}$ (under $\Class{ETH}$).
	%\textcolor{red}{todo 3: under eth we can't improve to a planar version without quadratic blow-up!}
\end{proposition}

\subsection{Contribution II: New Characterization of GapP}

%\textcolor{red}{Focus on \#2DNF is the easiest problem with known gapP-C. FP - TotP is also GapP-C, but we have an open problem about one call to spanL. (maybe progress toward negative result)}

Because subtracting the model counts of two monotone formulas is
enough to capture~$\Class{\#P}$, the natural next question is
what is needed to capture $\Gap\Class{P}$. While $\Gap\Class{P}$ is still defined on non-deterministic Turing machines, in contrast to $\#\Class{P}$ it amounts to the number of \emph{accepting paths minus} the number of \emph{rejecting paths}. We show that even this class can
be characterized by two calls to oracles of restricted fragments of
$\Lang{\#2sat}$ or $\Lang{\#2dnf}$. In the following theorem,
$\Class{\#P} - \Class{\#P}$ (respectively $\Span\Class{L} - \Span\Class{L}$)
refers to the subtraction of the results of two~$\Class{\#P}$
(respectively $\Span\Class{L}$) oracle calls. Since it is open whether
$\Lang{\#dnf}$ is $\Span\Class{L}$-hard, and since it is \emph{not expected}~\cite[Theorem 2]{sly2010computational} for $\Lang{\#2sat}$ to be in $\Span\Class{L}$ (unless we have $\Class{RP}=\Class{NP}$), this makes our characterization of $\Class{\#P}-\Class{\#P}$ 
via $[\Lang{\#2sat}-\Lang{\#2sat}]^{\log}$ \emph{and its dual}
$[\Lang{\#2dnf}-\Lang{\#2dnf}]^{\log}$ significant; the result holds even in the absence of negation. %, even without negation.
%problems and their duals. % possibly stronger than spanL-spanL. 
%
Below, $\Lang{\#impl2sat}$ is strictly in $\Lang{\#Horn2sat}$ and $\Lang{\#0,1-2dnf}$ is its dual over~$\Lang{dnf}$.

\begin{restatable}[Characterization of GapP]{theorem}{gaptheorem}\label{theorem:gapp}\useieee{\quad\quad$\Gap\Class{P}=$}%
  ~\\[-1.75em]\begin{align*}
    %~\\[-2em]&\Gap\Class{P} \\ %\Class{\#P} - \Class{\#P} = [\#3\Lang{sat} - \#3\Lang{sat}]^{\log} \\
    %%&= [\#3\Lang{dnf} - O(1)]^{\log}
   %% = [O(1)-\#3\Lang{dnf}]^{\log}\\
    \uselipics{\Gap\Class{P}}&= [\Lang{\#2sat}{-}\Lang{\#2sat}]^{\log} = [\Lang{\#impl2sat}{-}\Lang{\#impl2sat}]^{\log}\\[-0.25em]
    &= [\Lang{\#0,1-2dnf}{-}\Lang{\#0,1-2dnf}]^{\log} = [\Lang{\#2dnf}{-}\Lang{\#2dnf}]^{\log} \useieee{\\[-0.25em]&}= [\Lang{\#dnf}{-}\Lang{\#dnf}]^{\log}\uselipics{\\
    &}= [\#\Lang{mon2sat} {-} \#\Lang{mon2sat}]^{\log} \useieee{\\[-0.25em]&}= [\#\Lang{mon2dnf} {-} \#\Lang{mon2dnf}]^{\log}
    = \Span\Class{L} {-} \Span\Class{L}.
    \end{align*}~\\[-1.5em]
The characterization extends to \emph{cubic} and \emph{bipartite} restrictions of $\Lang{\#impl2sat}$ and $\Lang{\#0,1-2dnf}$;
	even if both formulas use the same variables and differ by only one literal/variable occurrence. %if restricted to ~formulas.
\end{restatable}

This result illustrates the
\emph{power of subtraction}, which by the theorem compensates for both the
absence of negation \emph{and} clauses of size at least three. Since
it is known that
$\Lang{\#2dnf}\in\Span\Class{L}$ with
$\Span\Class{L}\subsetneq\Class{\#P}$ (unless
$\Class{NL}=\Class{NP}$~\cite[Proposition 4.10]{AlvarezJenner93}) and $\Class{\#P}\subsetneq\Gap\Class{P}$ by
Corollary~\ref{cor:sharpmon}, Theorem~\ref{theorem:gapp} implies
that $\Class{\#P}$ is \emph{strictly sandwiched} between one and two 
\Lang{\#2dnf} oracle calls: 

\begin{corollary}
  $[\Lang{\#2dnf}]^{\log}\subsetneq \Class{\#P}$ unless $\Class{NL}=\Class{NP}$; and $\Class{\#P} \subsetneq [\Lang{\#2dnf}-\Lang{\#2dnf}]^{\log}$ unless $\Class{UP}=\Class{SPP}$.
\end{corollary}

%While Theorem~\ref{theorem:gapp} shows that $[\Lang{\#3dnf}-O(1)]^{\log}$ coincides with $\Gap\Class{P}$ under many-one logspace reductions, we do not expect that the difference between a logspace-computable number and the result of a \Lang{\#2dnf} or \Lang{\#2cnf} computation is sufficient to capture $\#\Class{P}$:
%
%\begin{conjecture}\label{conj:insufficient}
%    $[O(1)-\Lang{\#2sat}]^{\log}\subsetneq \#\Class{P}$ and $[O(1)-\Lang{\#2dnf}]^{\log}\subsetneq \#\Class{P}$.
%\end{conjecture}

%Using the tools for proving 
%Theorem~\ref{theorem:gapp}, we also prove
%$\Class{\#P} - \Class{\#P} = \Span\Class{L} - \Span\Class{L}$, where
%$\Class{\#P} - \Class{\#P}$
%(respectively $\Span\Class{L} - \Span\Class{L}$)
%refers to the difference between the results of two~$\Class{\#P}$
%(respectively $\Span\Class{L}$) calls.
%We obtain a second characterization of $\Gap\Class{P}$:

%\begin{restatable}[Characterization of GapP II]{theorem}{gapspanltheorem}\label{theorem:gapspanl}%
%    \begin{align*}
%    \Gap\Class{P} &= \Class{\#P} - \Class{\#P} = \Span\Class{L} - \Span\Class{L}
%    = [\#3\Lang{dnf} - \#3\Lang{dnf}]^{\log}\\
%    &= [\#3\Lang{dnf} - O(1)]^{\log}
%    = [O(1)-\#3\Lang{dnf}]^{\log}\\
%    &= [\#\Lang{mon2dnf} - \#\Lang{mon2dnf}]^{\log}
%    = [\#\Lang{mon2sat} - \#\Lang{mon2sat}]^{\log}.
%    \end{align*}
%\end{restatable}

\useieee{\begin{figure*}[tb]}\uselipics{\begin{figure}[tb]}
  \centering
~\\[-1em]
  \resizebox{0.9\textwidth}{!}{\begin{tikzpicture}[scale=0.95]
    %
    % the cut
    %
    \draw[color=fg, thick, rounded corners]  (1.5,-4.5) -- (1.5,-4)
    .. controls (1.5,2) and (10,-5) .. (10,3) -- (10,4.45);

    %
    % base classes
    %
    \node[baseline] (spanl)  at (0,0)   {$\Span\Class{L}$};
    \node[baseline] (sharpp) at (5,0)   {$\Class{\#P}$};
    \node[baseline] (gapp+)  at (8.5,0) {$\Gap\Class{P}^+$};
    \node[baseline] (gapp)   at (12,0)  {$\Gap\Class{P}$};
    \node[baseline] (spanL2) at (15,0)  {$\Span\Class{L}-\Span\Class{L}$};

    %
    % logic based classes for gapP
    %
    \node[baseline] (2impl2ac) at (12,4)  {$[\Lang{\#impl2sat}]^{\log}_{\text{AC}^0}$};
    \node[baseline, fill=white, inner sep=2pt] (2impl2)   at (12,2)  {$[\Lang{\#impl2sat}-\Lang{\#impl2sat}]^{\log}$};
    \node[baseline] (01dnf2)   at (12,-4) {$[\Lang{\#mon2dnf}]^{\log}_{\text{TC}^0}$ %{$[\Lang{\#0,1-2dnf}-\Lang{\#0,1-2dnf}]^{\log}$
    };
    %\node[baseline] (2sat2)    at (12,2)  {$[\Lang{\#2sat}-\Lang{\#2sat}]^{\log}$};
    \node[baseline] (2dnf2)    at (12,-2) {$[\#\Lang{mon2dnf}-\#\Lang{mon2dnf}]^{\log}$};

    %
    % logic based classes for sharpP
    %
    \node[baseline] (3sat)     at (5,2)    {$[\Lang{\#3sat}]^{\log}$};
    \node[baseline] (2mon2)    at (5,-2)   {$[\Lang{\#mon2sat}-\Lang{\#mon2sat}]^{\log}$};
    %\node[baseline] (2mondnf2) at (9,-1.5){$[\Lang{\#mon2dnf}-\Lang{\#mon2dnf}]^{\log}$};
    \node[baseline] (2monTC)   at (5,-4)   {$[\Lang{\#mon2sat}]^{\log}_{\text{TC}^0}$};
    \node[baseline] (3dnf)     at (5,4)    {$[2^n-\Lang{\#3dnf}]^{\log}$};

    %
    % logic based classes in spanL
    %
    \node[baseline] (dnf)  at (0,-2)   {$[\Lang{\#dnf}]^{\log}$};
    \node[baseline] (2dnf) at (0,-4)   {$[\Lang{\#2dnf}]^{\log}$};

    %
    % #2sat
    %
    \node[baseline] (2sat)  at (0,2)   {$[\Lang{\#2sat}]^{\log}$};
    \node[baseline] (n2dnf) at (0,4)   {$[2^n-\Lang{\#2dnf}]^{\log}$};
    
    %
    % class containment
    %
    \graph[use existing nodes, edges = {semithick, >={[round,sep]Stealth}}]{
      2dnf -> dnf -> spanl;
      %sharpp -> 
      2impl2 -> 2impl2ac;
      2mon2 -> 2monTC;
      2dnf2 -> 01dnf2;
      %2mondnf2 -> 2dnf2;
    };

    \draw[semithick, ->, >={[round,sep]Stealth}] 
    (2sat) to node[midway, fill=white, inner sep=0pt] {$\scriptstyle\Class{RP}\neq\Class{NP}$} (spanl);
    \draw[semithick] ($(spanl)+(-0.25,0.25)$) to ($(spanl)+(0.25,0.75)$);
    
%\coordinate (southmon2) at ($(2mondnf2.south)+(-.15,+.145)$);
    
 %   \graph[use existing nodes, edges = {ultra thick, double,double distance=2pt, >={[round,sep]Stealth}}]{
 %     southmon2 -- 2dnf2;
  %  };
    %
    \graph[use existing nodes, edges = {semithick, double,double distance=2pt, >={[round,sep]Stealth}}]{
      3dnf -- 3sat -- sharpp;
      2impl2 -- %2sat2 -- 
      gapp -- 2dnf2; % -- 01dnf2;
      n2dnf -- 2sat;
        2monTC -- 01dnf2;
      %gapp -- spanL2;
    };

    \graph[use existing nodes, edges = {ultra thick, >={[round,sep]Stealth}}]{
      2dnf -> dnf -> spanl;
      %sharpp -> 
    };
    
    \graph[use existing nodes, edges = {ultra thick, double,double distance=2pt, >={[round,sep]Stealth}}]{
        2dnf2 -- gapp -- spanL2;
      2mon2 -- 2dnf2;

    };

    \draw[semithick, densely dashed, ->, >={[round,sep]Stealth}] (sharpp)  to node[midway,fill=white,inner sep=1pt,baseline] {%$\scriptstyle\Class{NL}\neq\Class{NP}$
    } 
    (2mon2);
    %

    % highlight main results
    \begin{scope}[on background layer]
    \draw[color=bg, fill=bg!25, rounded corners] (2mon2.south west)
    -- (2dnf2.south east)
    -- (2dnf2.north east)
    -- (gapp.north east)
    -- (gapp.north west)
    -- (gapp.south west)
    -- (2dnf2.north west)
    -- (2mon2.north west)
    -- cycle;
    \end{scope}

%\coordinate (northmon2) at ($(2mondnf2.north)+(-0,-.2)$);
    
    %\draw[ultra thick, densely dashed, ->, >={[round,sep]Stealth}] (sharpp)  to[] (northmon2);

    \draw[ultra thick, densely dashed, ->, >={[round,sep]Stealth}] (spanl)  to node[midway,fill=white,inner sep=1pt,baseline] {$\scriptstyle\Class{NL}\neq\Class{NP}$} (sharpp);
    \draw[ultra thick, densely dashed, ->, >={[round,sep]Stealth}] (sharpp) to node[midway,fill=white,inner sep=1pt,baseline] {$\scriptstyle\Class{UP}\neq\Class{SPP}$} (gapp+);
    \draw[ultra thick, densely dashed, ->, >={[round,sep]Stealth}] (gapp+)  to (gapp);
    \draw[semithick, densely dashed, ->, >={[round,sep]Stealth}] (2sat)   to node[midway,fill=white,inner sep=1pt,baseline] {$\scriptstyle\Class{NL}\neq\Class{NP}$} (3sat);

    %
    % References
    %
    \node[baseline, anchor=west, color=gray]  at (5,-1)      {\scriptsize Cor.~\ref{cor:sharpmon}}; % / Obs.~\ref{lem:sharp-no-neg}};
    \node[baseline, anchor=west, color=gray]  at (12,-1)     {\scriptsize Thm.~\ref{theorem:gapp}};
    %\node[baseline, anchor=west, color=gray]  at (12,3)      {\scriptsize Thm.~\ref{theorem:gapp}};
    \node[baseline, anchor=west, color=gray]  at (12,1)      {\scriptsize Thm.~\ref{theorem:gapp}};
    \node[baseline, anchor=west, color=gray]  at (12,-3)     {\scriptsize Thm.~\ref{theorem:singlecalla}, Lem.~\ref{lemma:main}};
    \node[baseline, anchor=south, color=gray] at (13,0.1)    {\scriptsize Thm.~\ref{theorem:gapp}};
    \node[baseline, anchor=south, color=gray] at (8.5,-2)    {\scriptsize Thm.~\ref{theorem:gapp}};

    \node[baseline, anchor=south, color=gray] at (6.75,0.25)   {\scriptsize Obs.~\ref{observation:gapp+}};
    \node[baseline, anchor=south, color=gray] at (10.25,0.1)   {\scriptsize Obs.~\ref{observation:gapp+}};
    
    \node[baseline, anchor=west, color=gray]  at (12,3)      {\scriptsize Thm.~\ref{theorem:singlecallb}, Thm.~\ref{theorem:gapp}};
    \node[baseline, anchor=east,  color=gray] at (5,-3)      {\scriptsize Thm.~\ref{theorem:singlecalla}, Lem.~\ref{lemma:main}};
    
    \node[baseline, anchor=east,  color=gray] at (9.25,-3.6)      {\scriptsize Lem.~\ref{cor:singlecall}};
    
    \node[baseline, anchor=south, color=gray] at (2.5,2.25)  {\scriptsize Lem.~\ref{lemma:2satvs3ssat}};
  \end{tikzpicture}}
~\\[-2.25em]
  \caption{%
    Overview of complexity classes considered in this paper. An arrow
    $A\,\raisebox{0.2ex}{\protect\tikz{\protect\draw[semithick, ->, >={[round,sep]Stealth}] (0,0) to (0.5,0);}}\,B$
    indicates $A\subseteq B$, a dashed arrow means that 
    the implication is strict under the assumption shown on the line
    (e.g., $\text{NL}\neq\text{NP}$ implies
    $\text{span}\text{L}\subsetneq\text{\#P}$), and 
    $A\,\raisebox{0.2ex}{\protect\tikz{\protect\draw[semithick, double, double distance=2pt] (0,0) to (0.5,0);}}\,B$
    stands for $A=B$. If the arrow tip is crossed out, $A$ is not contained in $B$. We use $[\cdot]^{\log}$ to indicate the logspace
    closure of the problem in brackets (e.g., $[\Lang{\#3sat}]^{\log}$
    is the logspace (many-one, parsimonious) closure of
    $\Lang{\#3sat}$). The shorthand $\Lang{A}-\Lang{B}$ indicates
    that two oracle calls are allowed (with potentially different
    instances), and the result is the difference between both calls. A
    class in the subscript such as in
    $[\text{\#Mon2SAT}]^{\log}_{\text{TC}^0}$ indicates that
    $\text{TC}^0$ postprocessing is allowed \emph{after} the oracle
    call. Sets to the left of the green line contain only
    positive functions, while sets on the right contain
    functions that can map to negative numbers. Hence, the left side
    is strictly contained in the right. 
    \textbf{Thick edges} are fundamental, highlighting differences between one and two $\Lang{\#2dnf}$~calls. The \textcolor{bg}{{red area}} marks central insights of our contribution.%
    }
  \label{figure:overview}
\useieee{\end{figure*}}\uselipics{\end{figure}}

Figure~\ref{figure:overview} depicts an overview of counting
classes and their logical description under logspace many-one
reductions. We added the \emph{positive} part of
$\Gap\Class{P}$ defined as
$\Gap\Class{P}^+=\{\,f\in\Gap\Class{P}\mid\forall x:
f(x)\geq0\,\}$, which, unless $\Class{UP}=\Class{SPP}$, contains functions that are not in $\Class{\#P}$~\cite{OgiwaraHemachandra93}.

\begin{observation}\label{observation:gapp+}
  $\Class{\#P}\subseteq\Gap\Class{P}^+\subsetneq\Gap\Class{P}$
  and $\Class{\#P}\subsetneq\Gap\Class{P}^+$ unless $\Class{UP}=\Class{SPP}$.
\end{observation}

Figure~\ref{figure:overview} also incorporates the following lemma, which observes
that logspace-computable $c$-monious
reductions between
$\Lang{\#2sat}$ and $\Lang{\#3sat}$ are not possible unless
$\Class{NL}$ and $\Class{NP}$ collapse.

\begin{lemma}\label{lemma:2satvs3ssat}
    $[\Lang{\#2sat}]^{\log}\subsetneq [\Lang{\#3sat}]^{\log}$
    unless $\Class{NL}=\Class{NP}$.
\end{lemma}
\begin{proof}
We show a stronger result.
Suppose we have a \emph{$c$-monious}
%\footnote{A many-one reduction is \emph{$c$-monious} if it increases the
%  number of solutions by precisely a factor of $c$.}
logspace reduction $R$ from
$\Lang{\#3sat}$ to $\Lang{\#2sat}$ for a positive integer
constant~$c$, i.e., $R$ changes the number of solutions
by precisely a multiplicative factor of~$c$.
Then the following algorithm decides $\Lang{3sat}$ in
$\Class{NL}$: On input $\phi$, first compute $R(\phi)$ with
$\#(\phi)=c\cdot\#(R(\phi))$, which is possible by
assumption and since logspace is closed under composition. If $\phi$ is unsatisfiable, we
have $\#(\phi)=c\cdot\#(R(\phi))=0$, and otherwise we have
$\#(R(\phi))>0$ since $c>0$. We decide whether $\#(R(\phi))>0$ by solving $\Lang{2sat}$, which is
possible since $\Class{NL}$ is closed under complement~\cite{Immerman88}.
\end{proof}

\subsection{Contribution III: Characterization of Polynom.\ Hierarchy}

%\textbf{TODO: tone up a bit, actually a single \#2dnf (in spanL) count is enough. However, this is quite exciting, given that spanL is in   FPRAS~\cite{ArenasEtAl21}!}

Toda's theorem~\cite{Toda91} states that the whole polynomial-time hierarchy
can be solved by a polynomial-time Turing machine equipped with a single call
to a $\Class{\#P}$ oracle.
In fact, a $\Span\Class{L}$ oracle suffices~\cite[Corollary~4.11]{AlvarezJenner93}:

\begin{observation}
  $\Class{PH}\subseteq \Class{P}^{\Class{\#P[1]}}\subseteq \Class{P}^{\Class{\#P}}=\Class{P}^{\Span\Class{L}}$.
\end{observation}

In the framework of logspace many-one reductions, the emerging
question is how much computation needs to be carried out by the
polynomial-time Turing machine. Is it sufficient to ``just'' prepare
the oracle call, or is significant postprocessing necessary? 
%Despite Conjecture~\ref{conj:insufficient}
Below we prove that we can combine the two calls to \Lang{\#2sat} oracles
needed in Theorem~\ref{theorem:main} into a \emph{single
call} if we allow \emph{divisions} afterwards. More precisely,
we show that $\Class{\#P}$ can be simulated by a logspace reduction to $\Lang{\#mon2sat}$ followed by $\Class{TC}^0$ postprocessing:

\begin{restatable}{theorem}{singlecalltheorema}\label{theorem:singlecalla}%
$\Gap\Class{P}{\subseteq} [\Lang{\#mon2sat}]^{\log}_{\Class{TC}^0}{=} [\Lang{\#mon2dnf}]^{\log}_{\Class{TC}^0}$.
\end{restatable}

A crucial part of the postprocessing in Theorem~\ref{theorem:singlecalla} is
\emph{division} and, thus, we do not expect to be able to lower the
postprocessing power since division is $\Class{TC}^0$-complete~\cite{HesseEtAl02}
and $\Class{AC}^0\subsetneq\Class{TC}^0$~\cite{Hastad86}. 
However, if
we allow a slightly more powerful fragment of propositional logic, we
can prepare a count that we just need to divide by a power of 2, which
\emph{is} possible in $\Class{AC}^0$:

\begin{restatable}{theorem}{singlecalltheoremb}\label{theorem:singlecallb}%
  $\Gap\Class{P}{\subseteq} [\Lang{\#impl2sat}]^{\log}_{\Class{AC}^0}{=} [\Lang{\#0,1-2dnf}]^{\log}_{\Class{AC}^0}$.
  This statement holds even if $\Lang{\#impl2sat}$ is restricted to cubic and bipartite formulas.
\end{restatable}

Finally, we can use these two results to obtain a stronger
variant of Toda's celebrated result~\cite{Toda91} using logspace
reductions to counting problems of restricted fragments of
propositional logic (even contained in $\Lang{\#2dnf}$) with only little postprocessing:

%\textcolor{red}{highlight new memberships (the left side)}

\begin{restatable}[Characterization of PH]{theorem}{todaimproved}\label{theorem:todaimproved}\hfill

  \begin{tikzpicture}
    \node[baseline] (a) at (0,0)  {\Lang{ph}};
    \node[baseline] (b) at (2.15,0.75)  {$[\Lang{\#mon2sat}]^{\mathrm{log}}_{\Class{TC}^0}\hspace{.15ex}=$};
    \node[baseline] (bb) at (5,0.75)  {$[\Lang{\#mon2dnf}]^{\mathrm{log}}_{\Class{TC}^0}$};
    \node[baseline] (c) at (2.15,-0.75) {$[\Lang{\#impl2sat}]^{\mathrm{log}}_{\Class{AC}^0}\hspace{.15ex}=$};
    \node[baseline] (cc) at (5,-0.75) {$[\Lang{\#0,1-2dnf}]^{\mathrm{log}}_{\Class{AC}^0}$};
    \node[baseline] (d) at (7.4,0)  {$\Class{P}^{\Lang{\#2dnf}[1]}$};
    %\node[baseline] (e) at (9.4,0)  {$\Class{P}^{\Span\Class{L}[1]}$};
    %\node[baseline] (f) at (11.2,0)  {$\Class{P}^{\#\Class{P}[1]}$.};
    %
    \draw[opacity=0] (a) -- node[midway,opacity=1,sloped] {$\subseteq$} (b.south west);
    \draw[opacity=0] (a) -- node[midway,opacity=1,sloped] {$\subseteq$} (c.north west);
    \draw[opacity=0] (bb.south east) -- node[midway,opacity=1,sloped] {$\subseteq$} (d.north west);
    \draw[opacity=0] (cc.north east) -- node[midway,opacity=1,sloped] {$\subseteq$} (d.south west);
    %\draw[opacity=0] (d) -- node[midway,opacity=1]        {$\subseteq$} (e);
    %\draw[opacity=0] (e) -- node[midway,opacity=1]        {$\subseteq$} (f);
  \end{tikzpicture}
 \end{restatable}

\subsection{Contribution IV: New Upper and Lower Bounds for \#SAT}
\label{sec:contrib-iv}

Finally, we observe that the reductions used to prove  Theorem~\ref{theorem:main} can be implemented in linear time and
that they preserve important structural parameters of the input such
as its treewidth (the details of Lemma~\ref{lemma:main}).
The lemma has some immediate algorithmic consequences.
Interestingly, we obtain tight ($\Class{SETH}$-based) lower bounds, %cf.\ Corollary~\ref{sec:sethnlb}, 
\emph{via parameterized complexity}, as we establish strong parameterized guarantees with \emph{only additive overhead} (already a multiplicative factor larger than~$1$ is problematic). Indeed, without this parameterized route, obtaining such tight bounds is challenging, as there exists an $\mathcal{O}(1.3^n)$ algorithm~\cite{DBLP:journals/tcs/DahllofJW05}.

First note that fine distinctions can be made when defining
structural properties of propositional formulas. Usually, parameters
such as the treewidth $\tw(\phi)$ are defined over the \emph{primal
graph,} which is the graph that contains a vertex for every variable
and that connects two variables if they appear together in a
clause. Another graphical representation of a formula is the
\emph{incidence graph,} which contains a vertex for every variable
\emph{and every clause} and that connects two vertices if the variable
appears in the clause. The latter representation gives rise to the
definition of \emph{incidence treewidth} $\itw(\phi)$ for
which it is known that
$\itw(\phi)\leq\tw(\phi)+1$~\cite[Chapter~17]{HandbookSAT}.

It is relatively easy to show that $\#(\phi)$ can be computed with
$O\big(2^{\tw(\phi)}|\phi|\big)$ or
$O\big(4^{\itw(\phi)}|\phi|\big)$ arithmetic operations. It
was a long-standing open problem whether the exponential dependency on
$\itw(\phi)$ can be improved to $O\big(2^{\itw(\phi)}|\phi|\big)$, which
Slivovsky and Szeider~\cite{SlivovskyS20} answered affirmatively
with an involved algorithm utilizing zeta and Möbius transforms to
compute covering products. We obtain the result as a corollary from
Lemma~\ref{lemma:main} because our reduction to \Lang{\#2sat} increases the
incidence treewidth \emph{only by a concrete additive constant}:

\begin{corollary}\label{corollary:sharp-itw}
  There is an algorithm computing $\#(\phi)$ in
  $O(2^{\itw(\phi)}|\phi|)$ arithmetic operations.
\end{corollary}
\begin{proof}
  By Lemma~\ref{lemma:main}, we can reduce a \Lang{cnf} $\phi$ to
  \Lang{2cnf}s $\psi_1$ and $\psi_2$ with $\itw(\psi_i)\leq \itw(\phi)+14$.
  In the incidence graph of a \Lang{2cnf}, all vertices corresponding
  to clauses have maximum degree~2. By the \emph{almost simplicial
  rule}, contracting such a vertex to one of its neighbors cannot
  increase the treewidth past~2~\cite{BodlaenderKEG01}. However,
  contracting all vertices corresponding to a clause to one of their
  neighbors yields exactly the primal graph, hence, we have
  $\tw(\psi_i)\leq\itw(\psi_i)+1\leq\itw(\phi)+15$.
  Finally, we compute $\#(\psi_i)$ by dynamic
  programming over a tree decomposition of the primal graph, requiring
  $O\big(2^{\tw(\psi_i)}|\psi_i|\big)$ arithmetic operations~\cite{BodlaenderBL13,SamerS10}.
\end{proof}

\noindent\uselipics{\subparagraph*}\useieee{\paragraph*}{SETH-Tight Lower Bounds} On the other hand, since the reduction preserves structural properties
up to an \emph{additive} constant factor, 
we can complement the upper bound with a tight
lower bound under the (strong) exponential time hypothesis
($\Class{SETH}$)~\cite{ImpagliazzoPaturi01}.

\begin{restatable}[SETH LB]{theorem}{sethlb}
%\begin{theorem}
\label{sec:sethlb}
  Under $\Class{SETH}$, % fails,
  $\#(\phi)$ cannot be computed with $o\big(2^{\rho}\big)\cdot|\phi|^{O(1)}$ arithmetic
  operations on formulas with at most two variables per clause for any
  $\rho\in\{\bw(\phi)$, $\ibw(\phi)$, $\tw(\phi)$, $\itw(\phi)\}$.
  The results extend to bipartite monotone formulas for~$\rho\in\{\tw(\phi), \itw(\phi)\}$. % without negation. %  of constant degree  or (B) bipartite formulas of degree~$3$.
\end{restatable}

% \begin{corollary}\label{sec:sethnlb}
%   Unless $\Class{SETH}$ fails,
%   $\#(\phi)$ cannot be computed in time  $o\big(2^{n}\big)\cdot|\phi|^{O(1)}$ on bipartite monotone formulas~$\phi$ with at most two variables per clause and
%   $n$ variables.
%   %
%   %The result holds on %(A)  bipartite %constant-degree   formulas without negation. % and (B) bipartite degree-$3$ implications.
% \end{corollary}
% \begin{proof}
%     Result follows from guarantees of Lemma~\ref{lemma:main} and Theorem~\ref{sec:sethlb}, as $\rho{=}n$ (worst case). 
% \end{proof}
% % \begin{proof}
% % 	\textcolor{red}{TODO: replace $t_v \vee f_v$ by $t_v \vee x_v$, $f_v \vee y_v$ for fresh $x_v, y_v$, enabling bipartite, degree 4 for mon2dnf$_{tc0}$}
% % \end{proof}

Note that for $\Class{SETH}$-based bounds, already a linear factor as in case (B) in Lemma~\ref{lemma:main} is problematic. However, under $\Class{ETH}$ we obtain these constant-degree results.

\begin{restatable}[ETH LB]{theorem}{ethlb}
%\begin{theorem}[ETH LB]
  Unless $\Class{ETH}$ fails,
  $\#(\phi)$ cannot be computed with $\big(2^{o(\rho)}\big)\cdot|\phi|^{O(1)}$ arithmetic
  operations on formulas with at most two variables per clause for $\rho\in\{\bw(\phi)$, $\ibw(\phi)$, $\tw(\phi)$, $\itw(\phi)\}$.
  The result still holds for~$\rho\in\{\tw(\phi), \itw(\phi)\}$ and (A) bipartite constant-degree formulas without negation or (B) bipartite implication formulas of degree~$3$.
\end{restatable}

We also obtain the following non-parameterized bound, which significantly improves the lower bound of~\cite[Corollary 4.4]{Curticapean18}, as our reduction \emph{preserves parameters}. Without our parameterized detour, we would \emph{not directly obtain} such a strong bound. %, but instead derived a significantly weaker bound, e.g., $o\big(2^{\frac{|\vars(\varphi)|}{11}}\big)\cdot|\phi|^{O(1)}$.
%We derive stronger results than~\cite[Corollary 4.4]{Curticapean18} by sparsifying first~\cite{ImpagliazzoPaturi01}. %$\hspace{-1em}$

\begin{corollary}\label{corr:sethnlb}
  Under $\Class{ETH}$,
  $\#(\phi)$ cannot be computed in time  $2^{o(n)}\cdot|\phi|^{O(1)}$ on formulas with at most two variables per clause,
  $n$ variables, and $\mathcal{O}(n)$ clauses.
  The result holds on (A) bipartite constant-degree formulas without negation and (B) bipartite degree-$3$ implications.
\end{corollary}
% \begin{proof}
% First, we apply sparsification lemma~\cite{ImpagliazzoPaturi01}, which allows us to restrict to cases that $m$ is in $\mathcal{O}(n)$.

All bounds carry over to stronger results under weaker assumptions than $\Class{ETH}$ and $\Class{SETH}$, namely counting-based versions $\Class{\#ETH}$~\cite{DellEtAl14} and
$\Class{\#SETH}$~\cite{FockeMarxRzazewski24}, respectively.

%
%
%
%
%
%~\\[-2em]
%
\subsection{Structure of the Paper}
In Section~\ref{sec:tech} we provide an overview of our techniques, which is followed by concluding remarks and discussions in Section~\ref{sec:concl}.
Section~\ref{sec:prelim} recalls preliminaries and defines common notation.
Then, Section~\ref{sec:main} focuses on Contribution I and establishes our main reduction
from \Lang{\#sat} to (two calls of)~$\Lang{\#2sat}$, as well as Theorem~\ref{theorem:main}. In Section~\ref{sec:sharpsub}, %\ref{sec:structure}, 
we show claimed structural properties leading to Contribution IV, followed by extensions of our reduction to
restricted variants in Sections~\ref{section:monotone}. %and~\ref{section:cubic-bipartite}. %which finally establishes Theorem~\ref{theorem:main}.   
Then, Section~\ref{sec:gapp} covers Contribution II, thereby showing
consequences of our reductions (also for~$\Lang{\#2dnf}$) and its relationship to $\Gap\Class{P}$.
In Section~\ref{sec:toda} we show Contribution III, where we
demonstrate how  to reduce \Lang{\#sat} to a single~$\Lang{\#2sat}$ (or $\Lang{\#2dnf}$)
oracle call, followed by $\Class{AC}^0$ or $\Class{TC}^0$~postprocessing.
Finally, Section~\ref{sec:rel} briefly discusses related work and Section~\ref{sec:concl} contains  concluding remarks and discussions.

\section{Overview of Used Techniques}\label{sec:tech}

The backbone of our reduction is the  fact that \Lang{\#sat}
can be reduced to \emph{weighted} \Lang{\#2sat} by encoding the
\emph{inclusion-exclusion principle.} In \emph{weighted \Lang{\#sat}},
also known as \Lang{\#wsat} or \Lang{wmc}, the input is a weighted
\Lang{cnf} (a \Lang{wcnf}), i.e., a \Lang{cnf} $\psi$ together with 
weights
$w\colon\vars(\psi)\rightarrow\mathbb{R}$. The goal is to compute the
\emph{weighted} (or \emph{scaled}) count: \vspace{-1.5em}
\[
\quad\qquad\qquad\qquad\#_w(\psi) \coloneq
\sum_{\substack{\beta\assignment\vars(\psi)\\ \beta\models\psi}}
\prod_{x\in\beta}
w(x).
\]

\vspace{-.25em}\noindent The reduction sets $w(x)=1$ for all $x\in\vars(\phi)$ and introduces for every clause
$c=\ell_1\vee\dots\vee\ell_k$ a fresh variable \d{$c$} with $w(\text{\d{$c$}})=-1$ as
well as the new set of clauses $\bigwedge_{i=1}^k(\neg \text{\d{$c$}}\vee \neg \ell_i)$. Intuitively, the variable $\text{\d{$c$}}$ indicates that the clause $c$ is
\emph{not} satisfied by the assignment, i.e., if we set $\text{\d{$c$}}$ to
true we have to falsify all literals in $c$. 
Let the resulting \Lang{wcnf} be $\psi$, then
there are $2^n$ assignments for $\psi$ that
contribute $1$ to the weighted count (those setting \d{$c$} variables to false).

On the other hand, every assignment that sets exactly one $\text{\d{$c$}}$ to
true (i.e., that falsifies at least one clause in $\phi$) will contribute $-1$ to
the weighted count (and, crucially, is \emph{not} a model of
$\phi$). Hence, from $2^n$ we will automatically
\emph{subtract} all assignments that do not satisfy \emph{one}
clause. All assignments that falsify \emph{two} clauses will
again contribute one (because the $-1$ cancel out in the product), all
assignments that falsify three clauses will subtract one, for four
clauses they add one, and so on. By the inclusion-exclusion principle
we conclude: \vspace{-.5em}
\[
\#(\phi)=\#_w(\psi)=2^{n}
-
|\{\,
\beta\mid\text{$\beta\assignment\vars(\phi)$ with $\beta\not\models\phi$}
\,\}|.%\vspace{-2.25em} %\\[-4em]
\]
%~\\[-3em]
\vspace{-1.83em}
\begin{example}\label{ex:running}
Consider~$\phi = c_1 \wedge c_2\wedge c_3$ with clauses $c_1=\neg a \vee b \vee c$,
$c_2=a \vee \neg b\vee c$, and $c_3=\neg c$. 
% The following table shows
% which of the eight possible assignments to $\{a,b,c\}$ satisfies
% (\check) or falsifies (\uncheck) one of the clauses $c_1$, $c_2$, or
% $c_3$; or many simultaneously:
% \smallskip

% {\scriptsize
% \begin{tabular}{ccc|ccc|ccc|c}
%   $a$ & $b$ & $c$ & $c_1$    & $c_2$    & $c_3$    & $c_1\vee c_2$ & $c_1\vee c_3$ & $c_2\vee c_3$ & $c_1\vee c_2\vee c_3$ \\
%   \cmidrule(lr){1-10}
%   $0$ & $0$ & $0$ & \check   & \check   & \check   & \check        & \check        & \check        & \check               \\
%   $0$ & $0$ & $1$ & \check   & \check   & \uncheck & \check        & \check        & \check        & \check               \\
%   $0$ & $1$ & $0$ & \check   & \uncheck & \check   & \check        & \check        & \check        & \check               \\
%   $0$ & $1$ & $1$ & \check   & \check & \uncheck & \check        & \check        & \check      & \check               \\
%   $1$ & $0$ & $0$ & \uncheck & \check   & \check   & \check        & \check        & \check        & \check               \\
%   $1$ & $0$ & $1$ & \check   & \check   & \uncheck & \check        & \check        & \check        & \check               \\
%   $1$ & $1$ & $0$ & \check   & \check   & \check   & \check        & \check        & \check        & \check               \\
%   $1$ & $1$ & $1$ & \check   & \check   & \uncheck & \check        & \check        & \check        & \check               \\
% \end{tabular}
% }
% \smallskip

\noindent
The inclusion–exclusion reduction produces the formula:
\[
\psi =
       (\neg\text{\d{$c_1$}}\vee a)
\wedge (\neg\text{\d{$c_1$}}\vee \neg b)
\wedge (\neg\text{\d{$c_1$}}\vee \neg c)
\wedge (\neg\text{\d{$c_2$}}\vee \neg a)\useieee{\]\vspace{-1.75em}\[}
\wedge (\neg\text{\d{$c_2$}}\vee b)
\wedge  (\neg\text{\d{$c_2$}}\vee \neg c)
\wedge (\neg\text{\d{$c_3$}}\vee c).
\]
with weight $w(\text{\d{$c_1$}})=w(\text{\d{$c_2$}})=w(\text{\d{$c_3$}})=-1$ and
$w(a)=w(b)=w(c)=1$. The number of
assignments that falsify $c_1$ is $1$, for $c_2$ is $1$, and $4$ for
$c_3$. There are no assignments that falsify $c_1$ and $c_2$ or $c_1$
and $c_3$ simultaneously, respectively; there is also no
assignment that falsifies $c_2$ and $c_3$. Since, finally, there is no
assignment that falsifies all three clauses, we obtain
$\#_w(\psi)=2^3-1-1-4+0+0+0-0=2=\#(\phi)$. 
\end{example}

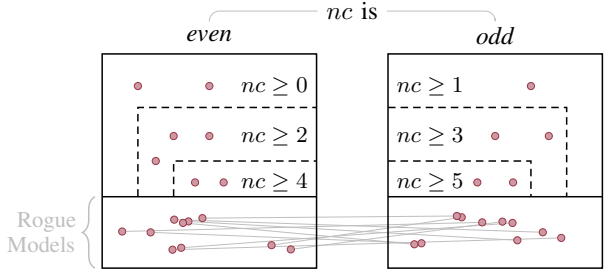
\begin{figure}[tbp]
    \centering
    \scalebox{\uselipics{0.85}\useieee{0.95}}{\begin{tikzpicture}[
        model/.style = {
          draw = bg,
          fill = bg!50,
          circle,
          inner sep = 0pt,
          minimum width = 0.1cm
        }
      ]

      \node (nc)   at (3.5,3.6)    {$nc$ is};
      \node (even) at (1.5,3.25)   {\emph{even}};
      \node (odd)  at (1.5+4,3.25) {\emph{odd}};
      \foreach \x in {even, odd}{
        \draw[thin, color=lightgray, rounded corners] (nc) -| (\x);
      }
            
      \draw[semithick] (0,0) rectangle (3,3);
      \draw[semithick] (0+4,0) rectangle (3+4,3);

      \draw[semithick] (0,1) -- (3,1);
      \draw[semithick] (0+4,1) -- (3+4,1);
      \draw [semithick, color=lightgray, decorate, decoration={brace, amplitude=5pt, mirror}]
      (-0.1,1) -- node[midway, left, text width=1.1cm] {\small\centering Rogue\\[-2pt] Models} (-0.1,0);

      \node[model] (v1) at (2.370,0.321)  {};
      \node[model] (v2) at (1.106,0.285)  {};
      \node[model] (v3) at (2.640,0.262)  {};
      \node[model] (v4) at (0.678,0.501)  {};
      \node[model] (v5) at (0.984,0.258)  {};
      \node[model] (v6) at (1.207,0.654)  {};
      \node[model] (v7) at (1.131,0.633)  {};
      \node[model] (v8) at (1.008,0.679)  {};
      \node[model] (v9) at (0.280,0.513)  {};
      \node[model] (v10) at (1.403,0.700) {};

      \node[model] (w1) at (1+4.031,0.712)  {};
      \node[model] (w2) at (1+4.742,0.633)  {};
      \node[model] (w3) at (1+4.601,0.651)  {};
      \node[model] (w4) at (1+3.468,0.348)  {};
      \node[model] (w5) at (1+5.424,0.423)  {};
      \node[model] (w6) at (1+4.810,0.390)  {};
      \node[model] (w7) at (1+3.369,0.334)  {};
      \node[model] (w8) at (1+5.166,0.502)  {};
      \node[model] (w9) at (1+4.318,0.641)  {};
      \node[model] (w10) at (1+3.963,0.730) {};

      \begin{scope}[on background layer]
        \foreach \i in {1,...,10}{
          \draw[thin, color=lightgray] (v\i) -- (w\i);
        }       
      \end{scope}

      \draw[semithick, densely dashed] (1,1)   |-  (3,1.5);
      \draw[semithick, densely dashed] (0.5,1) |-  (3,2.25);
      \draw[semithick, densely dashed] (4,1.5)  -| (2+4,1);
      \draw[semithick, densely dashed] (4,2.25) -| (2.5+4,1);

      \node[anchor=east] at (3,2.55) {\small$nc\geq0$};
      \node[anchor=east] at (3,1.85) {\small$nc\geq2$};
      \node[anchor=east] at (3,1.2)  {\small$nc\geq4$};
      \node[anchor=west] at (1+3,2.55) {\small$nc\geq1$};
      \node[anchor=west] at (1+3,1.85) {\small$nc\geq3$};
      \node[anchor=west] at (1+3,1.2)  {\small$nc\geq5$};

      \node[model] at (3-1.5, 2.55)  {};
      \node[model] at (3-2.5, 2.55)  {};
      \node[model] at (3-1.5, 1.85)  {};
      \node[model] at (3-2,   1.85)  {};
      \node[model] at (3-2.25, 1.5)  {};
      \node[model] at (3-1.3, 1.2)   {};
      \node[model] at (3-1.7, 1.2)   {};

      \node[model] at (3+3, 2.55)  {};
      \node[model] at (3+2.5, 1.85)  {};
      \node[model] at (3+3.25, 1.85)  {};
      \node[model] at (3+2.25, 1.2)  {};
      \node[model] at (3+2.75, 1.2)  {};      
    \end{tikzpicture}}\vspace{-.5em}    
    \caption{Simplified illustration of the relation of satisfying
      assignments (models) between our constructed formulas. Each box indicates a set of models of a
      formula. Models are represented by circles. The models can be
      separated into non-rogue models on the top and rogue models on
      the bottom. Rogue models have a line indicating they have a
      bijection with a model in the other formula. Non-rogue models
      are divided based on dissatisfying at least~$nc$ many
      clauses. Dotted lines indicate the set of models inside the line
      is a subset of the larger set.}%~\\[-1.2em]
    \label{fig:bijection}
\end{figure}

\begin{figure}[tbp]\vspace{-1.5em}
\hspace{-.9em}
    \begin{tikzpicture}[node distance=1mm, scale=0.6]%
\def\nodedist{0.7em}
%\tikzset{every node/.style=tdnode}
\tikzset{every path/.style=semithick}
\node (ableft) [tdnode,label={[yshift=-0.25em,xshift=0.25em] left:$ $}] {$\textcolor{fg}{\chi(t_3)}$};
\node (leaf1) [below=\nodedist of ableft,xshift=-2em, tdnode,label={[yshift=-0.25em,xshift=0.1em]left:$ $}] {$\textcolor{fg}{\chi(t_1)}$};
\node (leaf12) [below=\nodedist of ableft,xshift=1.5em, tdnode,label={[yshift=-0.25em,xshift=-0.1em]right:$ $}] {$\textcolor{fg}{\chi(t_2)}$};
\node (leaf2) [tdnode,label={[xshift=-1.0em, yshift=-0.15em]above right:$ $}, right = 0.5em of ableft]  {$\textcolor{fg}{\chi(t_4)}$};
\coordinate (middle) at ($ (ableft.north east)!.5!(leaf2.north west) $);
\node (root) [tdnode,thick,label={[]left:$ $}, above = \nodedist of middle] {$\textcolor{fg}{\chi(t_5)}$};
\node (llabel) [left=of root,xshift=-1.5em] {$\mathcal{T}$:};
\coordinate (top) at ($ (root.north east)+(3.5em,0) $);
\coordinate (bot) at ($ (top)+(0,-4em) $);
%
%\draw [dashed] (top) to (bot);
\draw [-] (leaf1) to (ableft);
\draw [-] ($(leaf12.north)+(0.2em,-0.1em)$) to ($(ableft.south)+(-0.25em,0.0em)$);
%\draw [-stealth'] (leaf2) to (leaf3);
\draw [-] (root) to (ableft);
\draw [-] (root) to (leaf2);
%
%
%SECOND PICTURE (right)
\node (rleaf1) [right=7em of ableft,tdnode,label={[yshift=-0.25em,xshift=0.25em] left:$ $}] {$\qquad\qquad\quad\quad$};
\node (rleaf1p) [right=7em of ableft,xshift=0.35em,inner sep=0.5] {$g(\textcolor{fg}{\chi(t_3)}, {\ldots})$};
\node (rem1) [below=1em of rleaf1,xshift=-2em, tdnode,label={[yshift=-0.25em,xshift=0.3em]left:$ $}] {$\qquad\qquad\qquad$};
%\node (rem1pointer) [draw,inner sep=2,rectangle,\statePredColor,below=3.5em of rleaf1,xshift=-0.5em,label={[yshift=-0.25em,xshift=-0.1em]}] {$\cdot$};
%\node (rem1a) [below=\nodedist of rleaf1,yshift=-0.5em,inner sep=0.5,xshift=-3.35em] {\textcolor{\outputPredColor}{$a_1$},};
\node (rem1p) [below=\nodedist of rleaf1,yshift=-0.5em,inner sep=0.5,xshift=-2.05em] {$g(\textcolor{fg}{\chi(t_1),{\ldots}})$};
\node (remab) [below=\nodedist of rleaf1,yshift=-0.3em,xshift=5.5em, tdnode,label={[yshift=-0.25em,xshift=-0.1em]right:$ $}] {$\quad\qquad\qquad\quad$};
\node (remabp) [below=\nodedist of rleaf1,yshift=-0.5em,inner sep=0.5,xshift=5.5em] {$g(\textcolor{fg}{\chi(t_2)},{\ldots})$};
\node (rleaf2) [tdnode,label={[xshift=-0.0em, yshift=-0.15em]above right:$ $}, right = 0.5em of rleaf1]  {$\qquad\qquad\qquad$};
\node (rleaf2p) [right=0.5em of rleaf1,yshift=-0.4em,inner sep=0.5,xshift=.5em,yshift=0.4em,] {$g(\textcolor{fg}{\chi(t_4)},{\ldots})$};
\coordinate (middle) at ($ (rleaf1.north east)!.5!(rleaf2.north west) $);
\node (join) [tdnode,thick,label={[xshift=-0.3em]right:$ $}, above  = \nodedist of middle] {\qquad\qquad\qquad\qquad\qquad\qquad\qquad};
\node (llabel) [left=of join,xshift= -1.25em] {$\mathcal{T}'$:};
\node (joinp) [above = \nodedist of middle,yshift=0.25em,xshift=-.05em,inner sep=0.5] {$g(\textcolor{fg}{\chi(t_5)}, {\ldots})$};
\coordinate (top) at ($ (join.north east)+(3.5em,0) $);
\coordinate (bot) at ($ (top)+(0,-4em) $);
\draw [-] (rem1) to (rleaf1);
\draw [-] ($(remab.north)+(0.2em,-0.1em)$) to ( $(rleaf1.south)+(-0.0em,0.0em)$);
\draw [-] (join) to (rleaf1);
\draw [-] ($(join.south)+(0em,0em)$) to (rleaf2);
\draw[dashedarrow,out=-170,in=-50,fg, >={[sep,round]Stealth}] (rem1) to (leaf1);
\draw[dashedarrow,out=-168,in=-40,fg, >={[sep,round]Stealth}] (remab) to (leaf12);
\draw[dashedarrow,out=-165,in=-15,fg, >={[sep,round]Stealth}] (rleaf1) to ($(ableft.south east)+(-2.0em,0.0em)$);
\draw[dashedarrow,fg,out=-191,in=12,fg, >={[sep,round]Stealth}] (rleaf2) to (leaf2) node[xshift=4.5em] {$g$};
\draw[dashedarrow,out=-180,in=30,fg, >={[sep,round]Stealth}] (join) to ($(root.north)$);
\end{tikzpicture}%}%%
~\\[-3em]
\caption{Illustration of our reduction that is guided along any given tree decomposition~$\mathcal{T}$ of the given formula.
The reduction uses structural dependencies of~$\mathcal{T}$ to ensure strong guarantees on the (tree)width of the resulting instance. 
$\mathcal{T}'$ is a tree decomposition of the resulting $\Lang{2dnf}$ formula(s).}\label{fig:tdguided}~\\[-1.6em]
\end{figure}
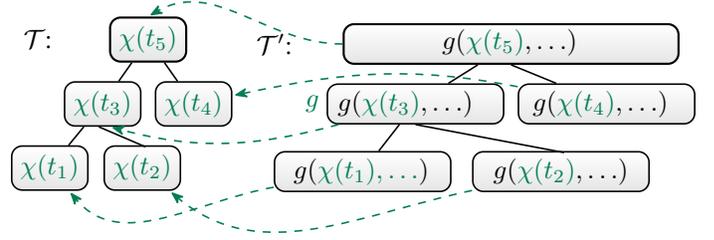

\noindent\textbf{A fault-tolerant version of inclusion-exclusion.} While inclusion-exclusion provides a reduction from $\Lang{\#sat}$ to weighted
$\Lang{\#2sat}$, we have the new problem of getting away with \emph{negative} weights. We indirectly realize inclusion-exclusion (and, thus, shave off weights) with a novel \emph{fault-tolerant version of
inclusion-exclusion}. The idea is that the first count may make errors (e.g.,
over- or under-count), but these errors can be carefully controlled to be well-behaved. We then count these errors using the second formula such that subtracting the results of both calls results in the correct model count. %counted (and later subtracted) using the second call. 
We call these errors \emph{rogue models} and outline the concept in Figure \ref{fig:bijection}. 
While we cannot properly quantify (and express) these rogue models via one counting operation, aligning rogue models in a symmetric way allows us to ``redo'' errors, which then indirectly paves the way for separating models from rogue models.
By construction, the formulas used in both calls are \emph{almost identical} (just a single fact differs). In fact, both formulas share the \emph{same} number of variables, which immediately gives \emph{closure under negation}: $\#(\psi_1)-\#(\psi_2) = 2^n - \#(\neg\psi_1) - (2^n - \#(\neg\psi_2)) = \#(\neg\psi_2) - \#(\neg\psi_1)$. This yields further results and insights even for fragments in which ``padding'' might not be expressible.
The more restrictive the $\Lang{\#2cnf}$ ($\Lang{\#2dnf}$) fragment gets, the easier it is to break this symmetry, potentially yielding incorrect results.
We guide the two
calls along a structural representation of the formula (say, a tree
decomposition), but do not directly utilize the width of the
decomposition (e.g., the reductions work for unstructured instances
as well).  See Figure~\ref{fig:tdguided} for an illustration, which highlights functional dependencies.

\smallskip
\noindent\textbf{How \useieee{to }\uselipics{can we }simulate PH with a single \#Mon2DNF call?} Theorems \ref{theorem:singlecalla} and \ref{theorem:singlecallb} can be proven using the idea of creating a new formula $\varphi$ by merging all the clauses of two formulas $\varphi_1$ and $\varphi_2$ with new variables. The key technique used is a way to \emph{switch} between the two formulas resulting in $\#(\varphi) = f(\#(\varphi_1),\#(\varphi_2))$. It is easy to create a reduction that results in $\#(\varphi_1)\cdot\#(\varphi_2)$; however, due to the commutative property of multiplication, we cannot tell which count is from which formula. Thus, we design specific \emph{switch} constructions that overcome this limitation by creating default assignments that fix the variables of one formula while allowing the others to be set freely. For restricted fragments, this is indeed challenging. %We rely on different variants of this switch. 
For $\Lang{\#impl2sat}$ to capture $\Gap\Class{P}$ we can encode both counts in the function $2^n\cdot\#(\varphi_1) + \#(\varphi_2)$ with the default assignments of all variables set to $1$ for one formula (and all $0$s for the other formula). This uses $n$ additional variables to scale by $2^n$. This function is then simple enough that its inverse is computable in~$\Class{AC}^0$.  

For $\Lang{\#mon2sat}$, $f$ hides $\#(\varphi_2)$ since it is multiplied by $\#(\varphi_1)$. The power we are lacking is the ability to enforce that variables are set to $0$ and, thus, we are limited to all $1$s for the default assignment. This makes it difficult to avoid multiplying both counts. Thus, we must use $\Class{TC}^0$ to extract both counts by performing integer division to simulate $[\Lang{\#mon2sat} - \Lang{\#mon2sat}]^{\log}$. 
The results for the corresponding $\Lang{\#2dnf}$ fragments follow by the fact that these classes are closed under negation, see Lemma~\ref{cor:singlecall}.
Nevertheless, in Theorem~\ref{theorem:todaimproved} we show that $\Lang{\#impl2dnf}$ enriched with~$\Class{AC}^0$ postprocessing and $\Lang{\#mon2dnf}$  with~$\Class{TC}^0$ postprocessing already contains~$\Class{PH}$.

\section{Preliminaries}\label{sec:prelim}
We consider proportional formulas in conjunctive normal form
(\Lang{cnf}s) like $\phi = (\neg a \vee b \vee c) \wedge (a \vee \neg b\vee c)\wedge
(\neg c)$ as set-of-sets $\{\{\neg a,b,c\},\{a,\neg b,c\},\{\neg
c\}\}$ and refer to its variables and clauses with
$\vars(\phi)$ and $\clauses(\phi)$; respectively. We use the notation
$\beta\assignment\vars(\phi)$ to refer to a subset $\beta$ of the \emph{variables} interpreted as those set to true.
%$|\{x,\neg x\}\cap\beta|\leq 1$. 
Such a set is called an
\emph{assignment}, and we say an assignment \emph{satisfies} (is a
\emph{model} of) a clause $c\in\clauses(\phi)$ if $\beta\cap
c\neq\emptyset$ or $(\{x \mid \neg x \in c\})\setminus\beta\neq\emptyset$. An assignment that satisfies \emph{all} clauses of
$\phi$ is a model of $\phi$, which we denote by
$\beta\models\phi$. The \emph{number of models} of
$\phi$ is defined as
\(
\#(\phi)\coloneq
|\{\,
\beta\mid\text{$\beta\assignment\vars(\phi)$ and $\beta\models\phi$}
\,\}|.
\)

\subsection{Fragments of Propositional Formulas}
Every clause in $\Lang{2cnf}$ %formula %every clause 
contains at most two literals, i.e., for
two variables $a$, $b$ the following clauses are allowed:

~\\[-2.25em]\[
(a\vee b), (\neg a\vee b), (a\vee\neg b), (\neg a\vee\neg b), (a), (\neg a), (b), (\neg b).
\]
A \Lang{horn2cnf} does not contain $(\neg a\vee\neg b)$, a
\Lang{mon2cnf} only contains $(a\vee b)$ (i.e., no negation and
no facts), and an \Lang{impl2cnf} does only contain $(\neg a\vee
b)\equiv(a\rightarrow b)$
and $(a\vee\neg b)\equiv(b\rightarrow a)$, i.e., only positive
implications. We make the same definitions for \Lang{\#dnf}s, but instead refer to
``\Lang{impl2dnf}'' by \Lang{0,1-2dnf}, as these are not implications. 
To be conform with the terminology used in
the literature, we call counting problems over \Lang{cnf}s
always \Lang{\#sat} (e.g., \Lang{\#impl2sat}) and over \Lang{dnf}s
just~\Lang{\#dnf}.

\subsection{Background in Structural Graph Theory}
A \emph{graph} $G$ consists of a set of \emph{vertices} $V(G)$
and a set of \emph{edges} $E(G)\subseteq\binom{V(G)}{2}$. The
\emph{neighbors} of a vertex $v\in V(G)$ are $N(v)=\{\,w\mid\{v,w\}\in
E(G)\,\}$ and its \emph{degree} is $|N(v)|$. %Further let $N[v]\coloneq N(v)\cup\{v\}$. 
%These
This definition extends to vertex sets. % of vertices. 

A \emph{tree decomposition}~$(T,\bag)$ of a graph~$G$ consists of a rooted tree~$T$ and a mapping~$\bag: V(T) \rightarrow 2^{V(G)}$ s.t.:

\uselipics{\vspace{-.75em}}
\begin{enumerate}%\small
\item for every $v\in V(G)$ the set $\{\, x\mid v\in\bag(x)\,\}$ is
  non-empty and connected in $T$;
\item for every $\{u,v\}\in E(G)$ there is at least one node $x\in V(T)$ with
  $\{u,v\}\subseteq \bag(x)$.
\end{enumerate}
\uselipics{\vspace{-.75em}}

The \emph{width} of a tree decomposition is the maximum size of its
bags minus one, i.e., $\max_{x\in
  V(T)}|\bag(x)|-1$. The \emph{treewidth} $\tw(G)$ of $G$ is
the minimum width among every decomposition of $G$. We let $\children(t)$ ne the set of child nodes of a node~$t$ in~$T$. 

\begin{example}
  The treewidth of the Ursa Major constellation (as graph shown on the
  left) is at most two, as proven by the tree decomposition on the right:
  \smallskip
  
  \scalebox{0.9}
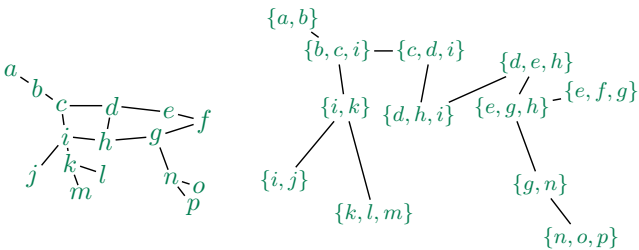
{\begin{tikzpicture}[scale=\uselipics{.92}\useieee{.65},
        star/.style = {
          color     = fg,
          inner sep = 1pt
        }
      ]
      \node[star] (a) at (-2.6,1.4)    {$a$};
      \node[star] (b) at (-2,1)        {$b$};
      \node[star] (c) at (-1.45,0.6)   {$c$};
      \node[star] (d) at (-0.3,0.6)    {$d$};
      \node[star] (e) at (1,0.45)      {$e$};
      \node[star] (f) at (1.8,0.25)    {$f$};
      \node[star] (g) at (0.7,-0.1)    {$g$};
      \node[star] (h) at (-0.45,-0.2)  {$h$};
      \node[star] (i) at (-1.35,-0.1)  {$i$};
      \node[star] (j) at (-2.1,-1)     {$j$};
      \node[star] (k) at (-1.25,-0.7)  {$k$};
      \node[star] (l) at (-0.5,-.95)   {$l$};
      \node[star] (m) at (-1,-1.4)     {$m$};
      \node[star] (n) at (1.05,-1.05)  {$n$};
      \node[star] (o) at (1.675,-1.25) {$o$};
      \node[star] (p) at (1.55,-1.7)   {$p$};      
      \graph[use existing nodes, edges = {semithick, color=black}]{
        a -- b -- c -- d -- e -- f -- g -- h -- i -- c;
        i -- {j, k -- {l,m}};
        g -- n -- {o,p};
        d -- h;
      };
      %
      % bb
      \node at (0,-2.5) {};
  \end{tikzpicture}}
  \hfill
  \scalebox{0.9}{\begin{tikzpicture}[
      scale=\uselipics{1.3}\useieee{1.2},
      bag/.style = {
        color     = fg,
        inner sep = 1pt,
        font      = \small
      }
    ]
    \node[bag] (b1)  at (-2,1)       {$\{a,b\}$};
    \node[bag] (b2)  at (-1.45,0.6)  {$\{b,c,i\}$};
    \node[bag] (b3)  at (-1.35,-0.1) {$\{i,k\}$};
    \node[bag] (b4)  at (-2.1,-1)    {$\{i,j\}$};
    \node[bag] (b5)  at (-1,-1.4)    {$\{k,l,m\}$};
    \node[bag] (b6)  at (-0.3,0.6)   {$\{c,d,i\}$};
    \node[bag] (b7)  at (-0.45,-0.2) {$\{d,h,i\}$};
    \node[bag] (b8)  at (1,0.45)     {$\{d,e,h\}$};
    \node[bag] (b9)  at (0.7,-0.1)   {$\{e,g,h\}$};
    \node[bag] (b10) at (1.8,0.1)   {$\{e,f,g\}$};
    \node[bag] (b11) at (1.05,-1.05) {$\{g,n\}$};
    \node[bag] (b12) at (1.55,-1.7)  {$\{n,o,p\}$};        
    \graph[use existing nodes, edges = {semithick, color=black}]{
      b1 -- b2 -- b3 -- {b4, b5};
      b2 -- b6 -- b7 -- b8 -- b9 -- {b10, b11 -- b12};
    };
  \end{tikzpicture}}
\end{example}

\vspace{-.75em}
Let~$f: V(G) \rightarrow \{1, \ldots, |V(G)|\}$
be a \emph{bijective mapping}.
The \emph{dilation} of~$G$ and~$f$ is the maximum (absolute) difference between integers assigned to
adjacent vertices, i.e., $\max_{\{u,v\}\in E(G)} |f(u)-f(v)|$. 
The \emph{bandwidth} of~$G$ is the minimum dilation of~$G$ among any such bijection.

\subsection{Structure of Propositional Formulas}
The \emph{primal graph} of a \Lang{cnf} $\phi$ is the graph $G_\phi$
with $V(G_\phi)=\vars(\phi)$ that contains an edge between two
vertices if the corresponding variables appear together in a
clause. Parameters for formulas can be defined via the primal graph, e.g., $\tw(\phi)\coloneq\tw(G_\phi)$.

Another representation \uselipics{of $\phi$} 
is the
\emph{incidence graph} $I_\phi$ with
$V(I_\phi)=\vars(\phi)\cup\clauses(\phi)$ and
$E(I_\phi)=\{\,\{x,c\}\mid\text{$x\in\vars(\phi)$,
  $c\in\clauses(\phi)$, and $\{x,\neg x\}\cap
  c\neq\emptyset$}\,\}$. This definition gives rise to \emph{incidence
parameters}, e.g., $\itw(\phi)\coloneq\tw(I_\phi)$

A \emph{labeled} tree decomposition~$(T,\bag,\delta)$ of~$\phi$ is
a tree decomposition~$(T,\bag)$ of~$G_\phi$, 
where every node gets assigned a set of labels using a
function~$\delta\colon T\rightarrow 2^{\clauses(\phi)}$.
A labeled tree decomposition requires \emph{(i)} for every node~$t$
of~$T$ and every $c\in \delta(t)$ that
$\vars(c)\subseteq\bag(t)$ and \emph{(ii)} $\clauses(\phi) = \bigcup_{t\in V(T)}\delta(t)$.
By introducing dummy nodes where necessary, we may assume without loss
of generality that $|\delta(t)|\leq 1$ for all $t\in V(T)$.

\section{Ingredients of the Main Lemma}\label{sec:main}

In this section we discuss a new reduction from \Lang{\#sat} to
\Lang{\#2sat} that increases the formula only linearly and preserves
the input's treewidth up to an additive constant. Thereby
we require two \Lang{\#2sat} oracle calls, followed by a
subtraction, which establishes the main lemma.

\mainlemma*

%\medskip
Let~$\phi$ be a propositional formula
and~$\mathcal{T}=(T,\bag,\delta)$ be a labeled tree decomposition of
$G_\varphi$.  For the ease of presentation, we first show the case in which $T$
is a path and every node (except the leaf) of $T$ gets assigned a clause label, i.e., $\delta(t)$ is not empty for non-leaf nodes. 
Note that if one is not interested in structural properties of a tree decomposition, one could construct a trivial decomposition~$\mathcal{T}$ in linear time, given \emph{any ordering} among the clauses of~$\phi$. Indeed, such a $\mathcal{T}$ could then use a node for every clause, put every variable in every bag, and the labeling assigns every clause to its node.

We use variables\footnote{By $\text{\d{$c$}}$ we just highlight the usage of $c$ as a variable (symbol) and not refer to the object $c$ itself.} $v, \text{\d{$c$}}, \overline{\text{\d{$c$}}}$ for every variable~$v\in \vars(\phi)$ and every clause $c\in \clauses(\phi)$.
Further, we use auxiliary variables $o_t$ and $e_t$ for every node~$t$ in $T$
to indicate that from the leaves of~$T$ up to node $t$, 
we assigned an even and odd number of clause variables to true, respectively.
Additionally, we require auxiliary variables $o1_t, o2_t$ for $o_t$
and $e1_t$, $e2_t$ for $e_t$. These auxiliary variables model auxiliary cases
when defining $o_t$ and $e_t$, which will be encoded symmetrically.

Our reduction is similar to the inclusion-exclusion reduction from
the introduction. For every given clause $c=l_1 \vee \cdots \vee l_k$,
we construct the following implications\footnote{\label{foot:pos}We would like to express $\text{\d{$c$}}\rightarrow \neg v$, which, unfortunately, is not an \Lang{impl2cnf}. However, recall that $v\rightarrow \overline{\text{\d{$c$}}}$ is equivalent to its contraposition~$\neg \overline{\text{\d{$c$}}} \rightarrow \neg v$. While $\text{\d{$c$}}$ and $\overline{\text{\d{$c$}}}$ are different symbols,  correctness is ensured by our definition of rogue models, see Definitions~\ref{def:rogue} and~\ref{def:roguesymm}.}.
\begin{flalign}
    \label{lab:wcnf}& \text{\d{$c$}} \rightarrow v && \text{for every negative }l_i\in c,\text{i.e., }l_i=\neg v,\\
    \label{lab:wcnf3}& v\rightarrow \overline{\text{\d{$c$}}} && \text{for every positive}^{\ref{foot:pos}}\text{ }l_i\in c,\text{i.e., }l_i=v.%\\
\end{flalign}

\begin{figure}[tbp]
    \centering
    \scalebox{0.75}{\begin{tikzpicture}
      \draw[thick, rounded corners, densely dashed, color=fg, fill=fg!25] (  0,0)   rectangle (    2,2);
      \draw[thick, rounded corners, densely dashed, color=fg, fill=fg!25] (  3,0)   rectangle (  3+2,2);
      \draw[thick, rounded corners, densely dashed, color=fg, fill=fg!25] (0+7,0)   rectangle (  7+2,2);
      \draw[thick, rounded corners, densely dashed, color=fg, fill=fg!25] (3+7,0)   rectangle (7+3+2,2);
      \draw[thick, rounded corners, densely dashed, color=fg, fill=fg!25] (  0,0-2.5) rectangle (    2,2-2.5);
      \draw[thick, rounded corners, densely dashed, color=fg, fill=fg!25] (  3,0-2.5) rectangle (  3+2,2-2.5);
      \draw[thick, rounded corners, densely dashed, color=fg, fill=fg!25] (0+7,0-2.5) rectangle (  7+2,2-2.5);
      \draw[thick, rounded corners, densely dashed, color=fg, fill=fg!25] (3+7,0-2.5) rectangle (7+3+2,2-2.5);

      \node[draw, semithick, inner sep=0pt, minimum width=0.5cm, minimum height=0.5cm] (a1) at (0.6,1.25)      {\small $o_t$};
      \node[draw, semithick, inner sep=0pt, minimum width=0.5cm, minimum height=0.5cm] (b1) at (0.5+0.9,1.25)  {\small $o1_t$};
      \node[draw, semithick, inner sep=0pt, minimum width=0.5cm, minimum height=0.5cm] (c1) at (0.6,1.25-0.75) {\small \d{$c$}};
      \node[draw, semithick, inner sep=0pt, minimum width=0.5cm, minimum height=0.5cm] (d1) at (0.6+3,1.25)    {\small $o_{t'}$};

      \node[draw, semithick, inner sep=0pt, minimum width=0.5cm, minimum height=0.5cm] (a2) at (0.6,1.25-2.5)      {\small $o_t$};
      \node[draw, semithick, inner sep=0pt, minimum width=0.5cm, minimum height=0.5cm] (b2) at (0.5+0.9,1.25-2.5)  {\small $o2_t$};
      \node[draw, semithick, inner sep=0pt, minimum width=0.5cm, minimum height=0.5cm] (c2) at (0.6,1.25-0.75-2.5) {\small \d{$\bar c$}};
      \node[draw, semithick, inner sep=0pt, minimum width=0.5cm, minimum height=0.5cm] (d2) at (0.6+3,1.25-2.5)    {\small $e_{t'}$};

      \node[draw, semithick, inner sep=0pt, minimum width=0.5cm, minimum height=0.5cm] (a3) at (7+0.6,1.25)      {\small $e_t$};
      \node[draw, semithick, inner sep=0pt, minimum width=0.5cm, minimum height=0.5cm] (b3) at (7+0.5+0.9,1.25)  {\small $e1_t$};
      \node[draw, semithick, inner sep=0pt, minimum width=0.5cm, minimum height=0.5cm] (c3) at (7+0.6,1.25-0.75) {\small \d{$c$}};
      \node[draw, semithick, inner sep=0pt, minimum width=0.5cm, minimum height=0.5cm] (d3) at (7+0.6+3,1.25)    {\small $e_{t'}$};

      \node[draw, semithick, inner sep=0pt, minimum width=0.5cm, minimum height=0.5cm] (a4) at (7+0.6,1.25-2.5)      {\small $e_t$};
      \node[draw, semithick, inner sep=0pt, minimum width=0.5cm, minimum height=0.5cm] (b4) at (7+0.5+0.9,1.25-2.5)  {\small $e2_t$};
      \node[draw, semithick, inner sep=0pt, minimum width=0.5cm, minimum height=0.5cm] (c4) at (7+0.6,1.25-0.75-2.5) {\small \d{$\bar c$}};
      \node[draw, semithick, inner sep=0pt, minimum width=0.5cm, minimum height=0.5cm] (d4) at (7+0.6+3,1.25-2.5)    {\small $o_{t'}$};

      \foreach \i in {1,2,3,4}{
        \foreach \v in {a,c,d}{
          \draw[semithick, ->, >={[round, sep]Stealth}] (b\i) to (\v\i);
        }
      }

      \draw[thin, lightgray] (0,-.25) -- (12,-.25);
      \draw[thin, lightgray] (6,2) -- (6,-2.5);      
    \end{tikzpicture}}    ~\\[-1.8em]
    \caption{
    An illustration of the four different \emph{non-rogue cases} per tree decomposition node~$t$ (with child node~$t'$), based on the single choice of variable in set $\{o1_t, o2_t, e1_t, e2_t\}$. Indeed, Equations~(\ref{eq:o1})--(\ref{eq:e2}) model all four potential cases, but also add many more \emph{rogue models} we eliminate by subtraction. This works due to symmetry (see Definitions~\ref{def:rogue},~\ref{def:roguesymm}).}
\label{fig:nonrogueenter-label}
\end{figure}
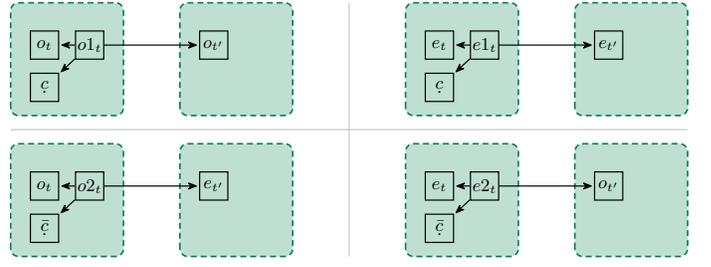

\noindent Intuitively, we guide the status of even (odd) along the tree decomposition. For a node~$t$, we define four possible cases of being even or odd due to invalidating a clause or not, see Figure~\ref{fig:nonrogueenter-label}.
So we construct clauses
for every node~$t$ in~$T$ with $t'\in\children(t)$ and $c\in\delta(t)$:
\begin{flalign}
    & o1_t \rightarrow e_{t'} & o1_t \rightarrow \text{\d{$c$}} && o1_t \rightarrow o_t &&& \text{Case 1: odd by}\notag\\[-.5em] &&&&&&&\text{choosing $c$}\label{eq:o1}\\
    & o2_t \rightarrow o_{t'} & o2_t \rightarrow \overline{\text{\d{$c$}}} && o2_t \rightarrow o_t &&& \text{Case 2: odd by}\notag\\[-.5em]&&&&&&&\text{not choosing $c$}\label{eq:o2}\\[.75em]
    & e1_t \rightarrow o_{t'} & e1_t \rightarrow \text{\d{$c$}} && e1_t \rightarrow e_t &&& \text{Case 1: even by}\notag\\[-.5em] &&&&&&&\text{choosing $c$}\label{eq:e1}\\
    & e2_t \rightarrow e_{t'} & e2_t \rightarrow \overline{\text{\d{$c$}}} && e2_t \rightarrow e_t &&& \text{Case 2: even by}\notag\\[-.5em] &&&&&&&\text{not choosing $c$}\label{eq:e2}%\\[1.5em]
\end{flalign}

\noindent We use an additional auxiliary variable\footnote{Note that~$\Lang{impl2cnf}$ is a strict subset of~$\Lang{horn2cnf}$, only allowing implications of the form~$a\rightarrow b$. Hence, we can't just add simple facts and need an additional auxiliary variable, which we refer to by~$x$.} $x$ and add for every leaf node~$t$ in $T$:
\begin{flalign}
    \label{red:leaf} & x \rightarrow e_t && \text{Initially, we choose $0$ clauses (even).}
\end{flalign}

%Let us denote with $R(\phi,\mathcal{T})$ the formulas obtained by applying the above process to a formula $\phi$ and a corresponding tree decomposition $\mathcal{T}$.  

\vspace{-.35em}
\begin{example}\label{ex:running2}
    Recall our initial Example~\ref{ex:running} and the running formula~$\phi = c_1 \wedge c_2\wedge c_3$ with $c_1=\neg a \vee b \vee c$,
$c_2=a \vee \neg b\vee c$, and $c_3=\neg c$. 
Assume a labeled tree decomposition~$\mathcal{T}=(T,\chi,\delta)$ of $G_\varphi$ comprising the nodes~$t_0$, $t_1$, $t_2$, $t_3$ such that $\delta(t_1)=c_1$, $\delta(t_2)=c_2$, and $\delta(t_3)=c_3$.
Then, the reduction above constructs the following clauses, resulting in~$\varphi'$.
\vspace{-.75em}
\begin{flalign}
    &\text{\d{$c_1$}} \rightarrow a\qquad\;
    \text{\d{$c_2$}} \rightarrow b\qquad\;
    \text{\d{$c_3$}} \rightarrow c\tag{\ref{lab:wcnf}}&\\
    %d
    &b \rightarrow \overline{\text{\d{$c_1$}}} \qquad\;\,\;
    c \rightarrow \overline{\text{\d{$c_1$}}}\qquad\,\;a \rightarrow \overline{\text{\d{$c_2$}}}\qquad\,\;
    c \rightarrow \overline{\text{\d{$c_2$}}}\qquad\,\;\raisetag{1.15em}\tag{\ref{lab:wcnf3}}\\
    %
    %&\text{\d{$c_3$}} \rightarrow c\\
    %
    %
    &o1_{t_1} \rightarrow e_{t_0}\quad
   o1_{t_1} \rightarrow \text{\d{$c_1$}}\quad
   o1_{t_1} \rightarrow o_{t_1}\qquad\notag&\\
    &o1_{t_2} \rightarrow e_{t_1}\quad
   o1_{t_2} \rightarrow \text{\d{$c_2$}}\quad
   o1_{t_2} \rightarrow o_{t_2}\qquad \notag&\\
   &o1_{t_3} \rightarrow e_{t_2}\quad
   o1_{t_3} \rightarrow \text{\d{$c_3$}}\quad
   o1_{t_3} \rightarrow o_{t_3}\qquad \tag{\ref{eq:o1}}&\\[.5em]
   &o2_{t_1} \rightarrow o_{t_0}\quad
   o2_{t_1} \rightarrow \text{\d{$\overline{c_1}$}}\quad
   o2_{t_1} \rightarrow o_{t_1}\qquad\notag&\\
   &o2_{t_2} \rightarrow o_{t_1}\quad
   o2_{t_2} \rightarrow \text{\d{$\overline{c_2}$}}\quad
   o2_{t_2} \rightarrow o_{t_2}\qquad\notag&\\
   &o2_{t_3} \rightarrow o_{t_2}\quad
   o2_{t_3} \rightarrow \text{\d{$\overline{c_3}$}}\quad
   o2_{t_3} \rightarrow o_{t_3}\qquad  \tag{\ref{eq:o2}}&\\[.5em]
   &e1_{t_1} \rightarrow o_{t_0}\quad
   e1_{t_1} \rightarrow \text{\d{$c_1$}}\quad
   e1_{t_1} \rightarrow e_{t_1}\qquad \notag&\\
   &e1_{t_2} \rightarrow o_{t_1}\quad
   e1_{t_2} \rightarrow \text{\d{$c_2$}}\quad
   e1_{t_2} \rightarrow e_{t_2}\qquad \notag&\\
   &e1_{t_3} \rightarrow o_{t_2}\quad
   e1_{t_3} \rightarrow \text{\d{$c_3$}}\quad
   e1_{t_3} \rightarrow e_{t_3}\qquad \tag{\ref{eq:e1}}&\\[.5em]
   &e2_{t_1} \rightarrow e_{t_0}\quad
   e2_{t_1} \rightarrow \text{\d{$\overline{c_1}$}}\quad
   e2_{t_1} \rightarrow e_{t_1}\qquad \notag&\\
& e2_{t_2} \rightarrow e_{t_1}\quad 
   e2_{t_2} \rightarrow \text{\d{$\overline{c_2}$}}\quad
   e2_{t_2} \rightarrow e_{t_2}\qquad  \notag&\\
   &e2_{t_3} \rightarrow e_{t_2}\quad
   e2_{t_3} \rightarrow \text{\d{$\overline{c_3}$}}\quad
   e2_{t_3} \rightarrow e_{t_3}\qquad \tag{\ref{eq:e2}}&\\
   &x\rightarrow e_0\raisetag{1.25em}\tag{\ref{red:leaf}}
\end{flalign}
\vspace{-1.5em}

\noindent In order to count~$\#(\varphi)$, we compute $\#(\varphi' \cup \{x\rightarrow e_3\})$ $-\,\#(\varphi' \cup \{x\rightarrow o_3\})=$ $204{,}452-204{,}450=2$.
Note that it is not surprising that the constructed formulas admit a large number of models. Indeed, below, we will see that without the use of negation, there are even more satisfying assignments. Still, the reduction can be computed efficiently, and the key lies in the symmetrical construction and the use of subtraction.
\end{example}

\paragraph*{Extension to Tree Decompositions}
While the formula defined above already works for tree decompositions that are paths, for addressing tree decompositions\footnote{We assume a tree decomposition using a binary tree (largest degree~$3$) such that degree-$3$ (join) nodes~$t$ have an empty labeling~$\delta(t)$. Such a decomposition can be constructed in linear time in its size.} the following two cases are missing. %obtained by applying the above process to a formula $\phi$ and a corresponding tree decomposition $\mathcal{T}$.  

For tree decomposition nodes~$t$ with~$\delta(t)=\emptyset$ (and~$\children(t)=\{t'\}$) we do not even use variables~$o1_t, e1_t$ and only generate the following special case of Equations~(\ref{eq:o2}) and~(\ref{eq:e2}). 
\noindent\begin{flalign}
&o2_{t} \rightarrow o_{t'}\qquad\uselipics{\qquad} o2_t\rightarrow o_{t}\qquad\uselipics{\qquad\qquad\qquad}
e2_t \rightarrow e_{t'}\qquad\uselipics{\qquad} e2_t\rightarrow e_{t}&&\text{}\label{eq:simple}\useieee{\raisetag{1.15em}}
\end{flalign}

In fact, the reduction can also be updated to accommodate so-called join nodes.
For these tree decomposition nodes~$t$ with~$\delta(t)=\emptyset$ (and~$\children(t)=\{t', t''\}$) we generate the following clauses, which are similar to Equations~(\ref{eq:o1})--(\ref{eq:e2}).
\begin{flalign}
    & o1_t \rightarrow e_{t'} & o1_t \rightarrow o_{t''} && o1_t \rightarrow o_t &\text{\;\;Case 1: odd by even/}\notag\\[-.5em] &&&&&\text{\;\;odd child nodes}\hspace{1.55em}\raisetag{1.15em}\label{eq2:o1}\\
    & o2_t \rightarrow o_{t'} & o2_t \rightarrow e_{t''} && o2_t \rightarrow o_t & \text{\;\;Case 2: odd by odd/}\notag\\[-.5em] &&&&&\text{\;\;even child nodes}\hspace{1.55em}\raisetag{1.15em}\\[.75em]
    & e1_t \rightarrow o_{t'} & e1_t \rightarrow o_{t''} && e1_t \rightarrow e_t &\text{\;\;Case 1: even by odd/}\notag\\[-.5em] &&&&&\text{\;\;odd child nodes}\hspace{1.55em}\raisetag{1.15em}\\
    & e2_t \rightarrow e_{t'} & e2_t \rightarrow e_{t''} && e2_t \rightarrow e_t &\text{\;\;Case 2: even by even/}\notag\\[-.5em] &&&&&\text{\;\;even child nodes}\hspace{1em}\raisetag{1.15em}\label{eq2:e2}%\\[1.5em]
\end{flalign}

\noindent\textbf{Reduction~$R$.} Finally, let us denote with $R(\phi,\mathcal{T})$ the formulas obtained by applying the above process, comprising Equations~(\ref{lab:wcnf})--(\ref{eq2:e2}). %, to a formula $\phi$ and a corresponding labeled tree decomposition $\mathcal{T}$.  

\subsection{Solving \#SAT by Subtracting Two \#2SAT Calls}\label{sec:sharpsub}

%https://cocreate.csail.mit.edu/r/oSxApqeM8aHRFNrGS#pJt5ACE988XtqHxJf 

With the construction~$R$ from above, we can obtain the correct number of satisfying assignments
via two calls to a \Lang{\#2sat} oracle, one to
\(
\psi_1\coloneq R(\phi, \mathcal{T}) \cup \{x \rightarrow e_{\rootOf(T)} \}
\)
and one to
\(
\psi_2 \coloneq R(\phi, \mathcal{T}) \cup \{x \rightarrow o_{\rootOf(T)} \}
\). The goal in the following is to prove that
$\#(\phi)=\#(\psi_1)-\#(\psi_2)$, requiring 
the central definition of \emph{rogue~models.}

\begin{definition}[Rogue Model]\label{def:rogue}
 Let $t$ be a node in~$T$.
 A model~$M$ of a formula~$\phi'\supseteq R(\phi, \mathcal{T})$
 is referred to by \emph{rogue (at~$t$)} whenever
 \begin{enumerate}[label=(\roman*)]
 \item $x\notin M$,
 \item $|M\cap \{o1_t, o2_t, e1_t, e2_t\}|\neq 1$, 
 \item $|M\cap \{\text{\d{$c$}}, \overline{\text{\d{$c$}}}\}|\neq 1$ with $c\in\delta(t)$, or
 \item $|M\cap \{o_t, e_t\}|\neq 1$.
 \end{enumerate}
\end{definition}

\begin{figure}[t]
    \centering
    \begin{tikzpicture}[
        dot/.style = {
          draw = fg,
          semithick,
          circle,
          inner sep = 0pt,
          minimum width = 5pt,
        }
      ]
      \foreach \x in {0,1,...,6}{
        \node[dot] (d\x) at (0,-\x/2) {};
        \node[dot] (e\x) at (2,-\x/2) {};
      }
      \foreach \x in {2,3,5,6}{
        \node[dot, fill=fg!50] at (0,-\x/2) {};
        \node[dot, fill=fg!50] at (2,-\x/2) {};
      }

      \graph[use existing nodes, edges = {semithick}]{
        d0 -- d1 -- d2 -- d3 -- d4 -- d5 -- d6;
        e0 -- e1 -- e2 -- e3 -- e4 -- e5 -- e6
      };
      \node[above of = d0, node distance = 0.5cm] {\emph{odd}};
      \node[above of = e0, node distance = 0.5cm] {\emph{even}};
      \node[left of = d0, node distance = 0.25cm] {\small $o$};
      \node[left of = d1, node distance = 0.25cm] {\small $e$};
      \node[left of = d2, node distance = 0.25cm] {\small $o$};
      \node[right of = e0, node distance = 0.25cm] {\small $e$};
      \node[right of = e1, node distance = 0.25cm] {\small $o$};
      \node[right of = e2, node distance = 0.25cm] {\small $e$};

      \draw[semithick, color = blue!80] ($(d0)+(-0.5,0)$)
      -- ($(d2)+(-0.5,-0.25)$)
      -- ($(d2)+(+0.5,-0.25)$)
      -- ($(d0)+(+0.5,0)$);

      \draw[semithick, color = orange!80!black] ($(e0)+(-0.5,0)$)
      -- ($(e2)+(-0.5,-0.25)$)
      -- ($(e2)+(+0.5,-0.25)$)
      -- ($(e0)+(+0.5,0)$);

      \draw[semithick, ->, >={[round]Stealth}, color = orange!80!black] ($(e0)+(-0.75,0)$) -- ++(-0.5,0);
      \draw[semithick, ->, >={[round]Stealth}, color = blue!80] ($(d2)+(0.75,0)$) -- ++(0.5,0);           
    \end{tikzpicture}
    \qquad\qquad
    %
    %\vspace{-1em}
    %
    \begin{tikzpicture}[
        dot/.style = {
          draw = fg,
          semithick,
          circle,
          inner sep = 0pt,
          minimum width = 5pt,
        }
      ]
      \foreach \x in {0,1,2}{
        \node[dot] (d\x) at (0,-\x/2) {};
        \node[dot] (e\x) at (4,-\x/2) {};
      }
      \foreach \x in {3,4,5}{
        \node[dot] (d\x-l) at (-1,-\x/2) {};
        \node[dot] (d\x-r) at ( 1,-\x/2) {};

        \node[dot] (e\x-l) at (3,-\x/2) {};
        \node[dot] (e\x-r) at (5,-\x/2) {};
      }
      \node[dot] (d6-l) at (-1.5,-6/2) {};
      \node[dot] (d6-r) at (-0.5,-6/2) {};
      \node[dot] (e6-l) at (4-1.5,-6/2) {};
      \node[dot] (e6-r) at (4-0.5,-6/2) {};

      \graph[use existing nodes, edges = {semithick}]{
        d0 -- d1 -- d2 -- {
          (d3-l) -- (d4-l) -- (d5-l) -- {(d6-l),(d6-r)},
          (d3-r) -- (d4-r) -- (d5-r)
        };
        e0 -- e1 -- e2 -- {
          (e3-l) -- (e4-l) -- (e5-l) -- {(e6-l),(e6-r)},
          (e3-r) -- (e4-r) -- (e5-r)
        };
      };

      \node[dot, fill=fg!50] at (d4-l) {};
      \node[dot, fill=fg!50] at (d6-l) {};
      \node[dot, fill=fg!50] at (d3-r) {};
      \node[dot, fill=fg!50] at (d5-r) {};
            
      \node[dot, fill=fg!50] at (e4-l) {};
      \node[dot, fill=fg!50] at (e6-l) {};
      \node[dot, fill=fg!50] at (e3-r) {};
      \node[dot, fill=fg!50] at (e5-r) {};

      \node[above of = d0, node distance = 0.5cm] {\emph{odd}};
      \node[above of = e0, node distance = 0.5cm] {\emph{even}};

      \node[left of = d0, node distance = 0.25cm] {\small $o$};
      \node[left of = d1, node distance = 0.25cm] {\small $e$};
      \node[left of = d2, node distance = 0.25cm] {\small $o$};
      \node[left of = d3-l, node distance = 0.25cm] {\small $e$};
      \node[left of = d4-l, node distance = 0.25cm] {\small $o$};

      \node[right of = e0, node distance = 0.25cm] {\small $e$};
      \node[right of = e1, node distance = 0.25cm] {\small $o$};
      \node[right of = e2, node distance = 0.25cm] {\small $e$};
      \node[right of = e3-l, node distance = 0.25cm] {\small $o$};
      \node[right of = e4-l, node distance = 0.25cm] {\small $e$};

      \draw[semithick, color = blue!80] ($(d0)+(-0.5,0)$)
      -- ($(d2)+(-0.5,0)$)
      -- ($(d3-l)+(-0.5,0)$)
      -- ($(d4-l)+(-0.5,-0.25)$)
      -- ($(d4-l)+(+0.5,-0.25)$)
      -- ($(d3-l)+(+0.5,0)$)
      -- ($(d2)+(+0.5,0)$)
      -- ($(d0)+(+0.5,0)$);

      \draw[semithick, color = orange!80!black] ($(e0)+(-0.5,0)$)
      -- ($(e2)+(-0.5,0)$)
      -- ($(e3-l)+(-0.5,0)$)
      -- ($(e4-l)+(-0.5,-0.25)$)
      -- ($(e4-l)+(+0.5,-0.25)$)
      -- ($(e3-l)+(+0.5,0)$)
      -- ($(e2)+(+0.5,0)$)
      -- ($(e0)+(+0.5,0)$);

      \draw[semithick, ->, >={[round]Stealth}, color = orange!80!black] ($(e0)+(-1.75,0)$) -- ++(-0.5,0);
      \draw[semithick, ->, >={[round]Stealth}, color = blue!80] ($(d2)+(1.75,0)$) -- ++(0.5,0);           
    \end{tikzpicture}~\\[-.9em]
    \caption{Abstract visualization of the symmetric rogue model and its construction. Nodes of the tree $T$ are given in green. Nodes filled in green indicate that the model is rogue at this filled node. The intuition is that if $T$ is a path (Top), in the construction it suffices to change the parity of the rogue node closest to the root, which then also causes parity changes of remaining nodes upwards. These changes enable the transition from odd to even parity (and vice versa) and they are visualized in blue  (and orange). Intuitively, the (first) rogue node enables a free change of parity, which is then propagated towards the root. In case $T$ is not a path (Bottom), one can order root-to-leaf paths and just pick the lexicographic smallest path with a rogue node and continue similarly as in (Top).}
    \label{fig:rogue_transition}
\end{figure}
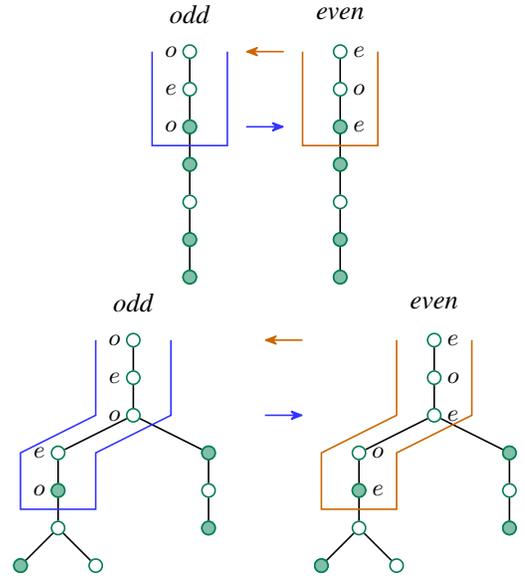

Intuitively, if there were zero rogue models, our reduction worked by the principle of inclusion-exclusion.
Now recall Figure~\ref{fig:bijection}, which demonstrates the intuition that we need a bijection between rogue models of the first formula and those of the second formula.
We rely on a construction that bijectively translates rogue models between $\psi_1$ and $\psi_2$. This aspect of symmetry for paths is visualized in Figure~\ref{fig:rogue_transition} (Top). 
The idea for constructing the symmetric model is to invert the parity of the rogue node closest to the root (and of all subsequent nodes, including the root). This immediately results in the corresponding symmetric rogue model, which preserves the rogue property of nodes. By construction, the symmetric rogue model of the symmetric rogue model is the rogue model itself (as desired). 

The construction can also be generalized to trees~$T$ as visualized in Figure~\ref{fig:rogue_transition} (Bottom), where we just need to uniquely pick a path containing rogue models. Here it is fine to order all \emph{root-to-leaf paths} of $T$ and then pick the lexicographic smallest path containing a rogue model (which is unique). For the sake of concreteness, we thereby assume in Equations (\ref{eq2:o1})--(\ref{eq2:e2}) that $t'$ is always the child node that is on this path. In turn, we are left with a unique path, so the remaining construction proceeds similarly to the path case.

Formally, we define the construction of the symmetric rogue model as follows. %covered next.

\begin{definition}[Symmetric Rogue Model]\label{def:roguesymm}
Let $M$ be a model of a formula~$\phi'$ with~$\phi'\supseteq R(\phi, {\mathcal{T}})$
that is rogue at~$t$. 
Assume that (1) there is no ancestor~$t'$ of~$t$ in~$T$ such that~$M$ is rogue at~$t'$,
and that (2) $t$ is on the lexicographic smallest root-to-leaf path in~$T$.
Then, the \emph{symmetric rogue model $M'$
(of~$M$)} is constructed as:
\begin{itemize}
    \item If $x\notin M$, we define $M'=M$
    \item Otherwise, if $x\in M$:
    \begin{itemize}
    \item Replace $o_t\in M$ by $e_t\in M'$ (and vice versa, i.e., $e_t\in M$ iff $o_t\in M'$).
    \item For every ancestor~$t'$ of~$t$ in~$T$, we replace $o_{t'}\in M$ by $e_{t'}\in M'$ and vice versa (i.e., $e_{t'}\in M$ iff $o_{t'}\in M'$), as well as $o1_{t'}\in M$ by $e1_{t'}\in M'$, $e1_{t'}\in M$ by $o1_{t'}\in M'$, $o2_{t'}\in M$ by $e2_{t'}\in M'$, $e2_{t'}\in M$ by $o2_{t'}\in M'$ (and vice versa)
    \item If (a) either\footnote{``Either $\ldots$ or'' refers to an exclusive disjunction.} $o_t\in M$ or $e_t\in M$, and (b) $|M\cap \{o1_t, o2_t, e1_t, e2_t\}|\geq 1$, 
    we additionally replace $o1_t\in M$ by $e2_t\in M'$, $o2_t\in M$ by $e1_t\in M'$ and vice versa (i.e., $e1_t\in M$ by $o2_t\in M'$, $e2_t\in M$ by $o1_t\in M'$)
    \end{itemize}
\noindent We say that $M$ is the \emph{symmetric rogue model} of~$M'$.
\end{itemize}
\end{definition}

With this definition at hand, we commence with proving correctness
of the reduction. To that end, we need to show that
symmetric rogue models are well-defined, i.e., that the construction
ensures that a symmetric rogue model~$M'$ of a rogue model~$M$ is
\emph{(a)} a model and \emph{(b)} rogue at a node~$t$ if and only $M$
is rogue at~$t$. This is established by Lemmas~\ref{lem:welldef} and \ref{lem:symmetry}, where full proof details are given in Appendix~\ref{sec:mainproof}. % the following two lemmas.
In Appendix~\ref{sec:structure} we show that \emph{structural parameters are linearly preserved}.

\subsection{Reducing to Monotone Formulas}\label{section:monotone}
We reuse the same construction as in
Equations~(\ref{lab:wcnf})--(\ref{eq:e2}), but in the following assume
\emph{fully labeled tree decompositions}, where also every variable is
a label of a tree decomposition node.  For a literal~$l$ over variable~$v$, we let~$inv(l)$ be the
variable~$\top_v$ if $l=\neg v$ and be~$\bot_v$ otherwise. Intuitively these auxiliary variables are used to refer to the truth value for $v$. We update
the inclusion-exclusion reduction such that
for every clause~$c=l_1 \vee l_2 \vee\cdots \vee l_k$ we construct \emph{positive clauses}:
\begin{flalign}
    \label{lab:wcnf2}& \overline{\text{\d{$c$}}} \vee inv(l_1)\qquad \overline{\text{\d{$c$}}} \vee inv(l_2)\qquad\ldots\uselipics{\qquad}\useieee{\quad} \overline{\text{\d{$c$}}} \vee inv(l_k) &&\text{}\useieee{\raisetag{1.15em}}
\end{flalign}

Additionally, for every node~$t$ with $t'\in\children(t)$ and $v\in\delta(t)\cap\vars(\phi)$, we add:
\begin{flalign}
    &\overline{v} \vee \top_v & \overline{v} \vee  \bot_v && \top_v\vee \bot_v&&\text{Choosing $v$ sets $v$ to true}\notag\\[-.5em] &&&&&&\text{and to false.\qquad\qquad\quad}\raisetag{1.15em} \label{eq:var}
\end{flalign}

We slightly adapt Equations~(\ref{eq:o1})--(\ref{eq:e2}) such that for every node~$t$ in~$T$ with $t'\in\children(t)$ and label $\alpha\in\delta(t)$, which can be either a clause or a variable, we construct:
\begin{flalign}
    & \overline{o1_t} \vee e_{t'} & \overline{o1_t} \vee \alpha && \overline{o1_t} \vee o_t &&& \text{Case 1: odd by}\notag\\[-.5em] &&&&&&&\text{choosing $\alpha$}\label{redpos:o1}\\
    & \overline{o2_t} \vee o_{t'} & \overline{o2_t} \vee \overline{\alpha} && \overline{o2_t} \vee o_t &&& \text{Case 2: odd by}\notag\\[-.5em] &&&&&&&\text{not choosing $\alpha$}\\[.75em]
    & \overline{e1_t} \vee o_{t'} & \overline{e1_t} \vee \alpha && \overline{e1_t} \vee e_t &&& \text{Case 1: even by}\notag\\[-.5em] &&&&&&&\text{choosing $\alpha$}\\
    & \overline{e2_t} \vee e_{t'} & \overline{e2_t} \vee \overline{\alpha} && \overline{e2_t} \vee e_t &&& \text{Case 2: even by}\notag\\[-.5em] &&&&&&&\text{not choosing $\alpha$}\hspace{.75em}\label{redpos:e2}%\\[1.5em]
\end{flalign}
By adapting Equation~\ref{red:leaf}, we obtain:
\begin{flalign}
    \label{redpos:leaf} & \overline{x} \vee e_t && \text{Initially, we choose $0$ clauses \& variables.}
\end{flalign}

\begin{example}\label{ex:running3}
    Recall the reduction given in Example~\ref{ex:running2}. Assume a labeled tree decomposition~$\mathcal{T}=(T,\chi,\delta)$ comprising the nodes~$t_0$, $t_1$, $\ldots$, $t_6$ such that $\delta(t_1)=c_1$, $\delta(t_2)=c_2$,  $\delta(t_3)=c_3$, $\delta(t_4)=a$, $\delta(t_5)=b$, and $\delta(t_6)=c$.
    Then, by constructing Equations~(\ref{lab:wcnf2})--(\ref{redpos:leaf}) similarly to Example~\ref{ex:running2}, we obtain a formula~$\varphi'$. 
In order to count~$\#(\varphi)$, we can compute~$\#(\varphi' \cup \{\overline{x} \vee e_6\})-\#(\varphi' \cup \{\overline{x} \vee o_6\})=$ $2{,}110{,}863{,}758-2{,}110{,}863{,}756=2$.
\end{example}

Note that, as above, one can easily adapt to the simpler types of tree decomposition nodes, see Equations~(\ref{eq:simple})--(\ref{eq2:e2}).
We refer to the adapted reduction comprising
Equations~(\ref{lab:wcnf2})--(\ref{eq:var}),
(\ref{redpos:o1})--(\ref{redpos:e2}), and (\ref{redpos:leaf})
by~${R}'(\phi,\mathcal{T})$. 
Roughly, the idea is to introduce an additional type of label (for variables) and ensure that Equations~(\ref{redpos:o1})--(\ref{redpos:e2}) work for \emph{both clause and variable labels}. By construction, a node of a labeled tree decomposition can only have one label (and therefore only one type). However, we do not care to manage these labels individually, but the idea is to keep track of the \emph{parity of the combined number} of corresponding variables being true.
Let us, as in the previous
section, also stipulate
\(
\psi_1'\coloneq R'(\phi, \mathcal{T}) \cup \{\overline{x} \vee e_{\rootOf(T)} \}
\)
and
\(
\psi_2'\coloneq
R'(\phi, \mathcal{T}) \cup \{\overline{x} \vee o_{\rootOf(T)} \}.
\)

Extensions of rogue models for \emph{monotone formulas} and proofs are given in Appendix~\ref{section:proofmonotone}.
Appendix~\ref{section:cubic-bipartite} generalizes the reduction to %monotone formulas, which is then further generalized to 
\emph{cubic and bipartite formulas}. % in Section~\ref{section:cubic-bipartite}.

\section{New Characterization of GapP}
\label{sec:gapp}

In this section, we show how Theorem~\ref{theorem:main} yields a more fine-grained characterization of $\Gap\Class{P}$. Below we show how one can still model a \emph{switch}, that enables us to change between satisfying assignments of one formula and to those of the other formula.

This switch construction has to be extended if % in the next section, 
%where 
we are only using monotone formulas (see Theorem~\ref{theorem:singlecalla}).

\gaptheorem*
\begin{proof}
    Proof details are given in Appendix~\ref{sec:proofgapP}.
\end{proof}
\section{A New Characterization of PH}\label{sec:toda}

Finally, we would like to give an outlook and some insights into many-one reductions that are enriched with additional postprocessing power on top of the resulting count. First, we observe the following.

\begin{restatable}{lemma}{actcinclusions}\label{cor:singlecall}%

%\vspace{-2em}
  \begin{align*}
    ~\\[-3.5em]\quad&&[\Lang{\#mon2sat}]^{\log}_{\Class{TC}^0}  &= [\Lang{\#mon2dnf}]^{\log}_{\Class{TC}^0}, \\
    &&[\Lang{\#impl2sat}]^{\log}_{\Class{AC}^0} &= [\Lang{\#0,1-2dnf}]^{\log}_{\Class{AC}^0}.
  \end{align*}
\end{restatable}
\begin{proof}
  %The first equality follows from Theorem~\ref{theorem:singlecalla}, the second from Theorem~\ref{theorem:singlecallb}. 
  In both cases we
  observe that the classes under consideration are closed under
  inversion, that is,
  \begin{align*}
      &[\Lang{\#mon2sat}]^{\log}_{\Class{TC}^0}
    = [2^n-\Lang{\#mon2dnf}]^{\log}_{\Class{TC}^0}\\
    =\,\, &\Lang{\#mon2dnf}]^{\log}_{\Class{TC}^0}
    =
    [2^n-\Lang{\#mon2sat}]^{\log}_{\Class{TC}^0}.
  \end{align*}
  Indeed, the subtraction for the inverse problem can be carried out in~$\Class{AC}^0$ and therefore the equations above hold.
Analogously,
  \begin{align*}      
    &[\Lang{\#impl2sat}]^{\log}_{\Class{AC}^0}
    = [2^n-\Lang{\#0,1-2dnf}]^{\log}_{\Class{AC}^0}\\
    =\,\,&[\Lang{\#0,1-2dnf}]^{\log}_{\Class{AC}^0} 
    = [2^n-\Lang{\#impl2sat}]^{\log}_{\Class{AC}^0}.\qquad\qedhere
    \end{align*}
  %\textcolor{red}{\bf not sure if that gets clear here}
\end{proof}%\medskip

With this lemma and the reduction techniques from above, we obtain the following (proven in Appendix~\ref{sec:proofscharact}).

\singlecalltheorema*

\singlecalltheoremb*

Now, we use both ingredients to establish a stronger characterization of~$\Lang{PH}$.

\todaimproved*

\section{Related Work}\label{sec:rel}
Recently, Laakkonen, Meichanetzidis, and Wetering~\cite{LaakkonenMW23}
also provided a new reduction from $\Lang{\#sat}$ to $\Lang{\#2sat}$
using the \Lang{zh}-calculus. Their work focuses on
producing a simple reduction that is representable in a pictorial
way. There, Laakkonen, Meichanetzidis, and
Wetering reduce to a \emph{single} $\Lang{\#2sat}$-instance, but
require \emph{modulo computations} in postprocessing; while we are parsimonious, requiring a \emph{single subtraction}.

Another downside of the \Lang{zh}-based reduction is a \emph{quadratic} blow-up: If the
original formula~$\phi$ has $n$ variables and $m$ clauses, the
produced formula will have size $O(n+mn)$. This blow-up
quickly accumulates if we perform further reductions. For
instance, the same set of authors also provided a reduction
from $\Lang{\#2sat}$ to $\Lang{\#mon2sat}$ that maps an instance with $n$
variables and $m$ clauses to a formula with $O(n+mn^2)$ variables and
$O(m+mn^2)$ clauses (Lemma~5 in~\cite{LaakkonenMW23}). Hence, the
whole reduction from $\Lang{\#sat}$ to $\Lang{\#mon2sat}$
constructs $O(n^3m^3)$ clauses. In contrast, our reductions produce
instances of \emph{linear} size, i.e., $O(n+m)$, for all the restricted fragments mentioned.

There is an extensive study on closure properties of~$\Class{\#P}$ and other counting complexity classes, see, e.g.,~\cite{FortnowReingold96,HoangThierauf05,OgiharaEtAl96,ThieraufEtAl94}.
There are also interesting findings regarding the closure of~$\Class{PP}$ under intersection~\cite{BeigelEtAl95}, which uses closure properties of $\#\Class{P}$. Closure properties of $\#\Class{AC}^0$ have been studied in \cite{Allender02} ($\#\Class{AC}^0$ is the set of functions computable by arithmetic circuits). They show negative results for functions such as max and division by 3. Connections are shown between $\#\Class{AC}^0$, $\Gap\Class{AC}^0$, and threshold circuits as well. The class $\#\Class{P}$ is also closed under so-called \emph{subtractive reductions} \cite{durand2005subtractive} along with other variants of counting classes in the polynomial hierarchy. 
These reductions use a different form of subtraction, % from the reductions we use here, 
as they are based on set difference, but not on the more general subtraction of counts (numbers). %preserve the function having only positive outputs. 
%
%\textcolor{red}{\textbf{TODO: what is so interesting about it? add more refs?}}

There have been many other classes \cite{beigel1990counting} defined similarly to our postprocessing strategy, such as $mod_k \Class{P}$ \cite{hertrampf1990relations}, i.e., counting the accepting paths of an $\Class{NP}$ machine and outputting whether the result is divisible by $k$. For $k=2$, this problem is known as $\oplus\Class{P}$ \cite{papadimitriou1982two}, which contains for instance graph isomorphism~\cite{arvind2006graph}.

\section{Discussion and Outlook}\label{sec:concl}

%\textcolor{red}{todo: since we showed: $gapp \subseteq spanl-spanl$. how powerful/hard is $fl-spanl$?!
%}

%\textcolor{red}{todo: spanL is in
%  FPRAS~\cite{ArenasEtAl21}. does this help us?}
  
%\textcolor{red}{todo: what about $\oplus P$ and $mod_k P$? mention other classes as well}

In this paper, we presented a new reduction from \Lang{\#sat} to
\Lang{\#2sat} and \Lang{\#2dnf}. Compared to the well-known reduction
from Valiant, our reduction is direct and only requires two calls to
a \Lang{\#2sat} (\Lang{\#2dnf}) oracle. This reduction is not only
conceptually simpler, but also computational: It can be carried out
either in logarithmic space or in linear time. In particular, it also
increases the size of the formula by at most a constant factor:
\[
\Lang{\#sat}\in[\Lang{\#2sat}-\Lang{\#2sat}]^{\log}=[\Lang{\#2dnf}-\Lang{\#2dnf}]^{\log}.
\]
As it turned out, the subtraction of two \Lang{\#2sat} or
\Lang{\#2dnf} calls is powerful enough to capture the larger class $\Gap\Class{P}\supsetneq\Class{\#P}$:
\[
\Gap\Class{P} = [\Lang{\#2sat}-\Lang{\#2sat}]^{\log}=[\Lang{\#2dnf}-\Lang{\#2dnf}]^{\log},
\]
which led to the title of the paper: Unless $\Class{NL}=\Class{NP}$, the class $\Class{\#P}$ is
\emph{strictly sandwiched} between one and two calls to a
$\Lang{\#2dnf}$ oracle:
\[
  [\Lang{\#2dnf}]^{\log}\subsetneq\Class{\# P}\subsetneq[\Lang{\#2dnf}-\Lang{\#2dnf}]^{\log}.
\]
We also observed that the \emph{subtraction} on the right side \emph{is stronger
than we thought} as it is enough to compensate for the absence of negation,
i.e., the \Lang{\#2dnf} formulas on the right side can be
monotone. This ``power'', however, can be simulated by a single
call to a \Lang{\#2dnf} oracle if we allow a mild postprocessing via
$\Class{TC}^0$ circuits:
\[
\Lang{\#sat}\in [\Lang{\#mon2dnf}]^{\log}_{\Class{TC}^0},
\]
which led to a new characterization of the polynomial hierarchy, i.e.,
a strengthening of Toda's Theorem:
\[
\Class{PH}\subseteq [\Lang{\#mon2dnf}]^{\log}_{\Class{TC}^0}.
\]
As a further byproduct of our reduction, we also obtain a new
algorithm that computes the count $\#(\phi)$ in time
$O\big(2^{\itw(\phi)}|\phi|\big)$, without the involved usage of zeta
and Möbius transforms. The strong parameter-preservation guarantees of this reduction allowed us to establish \emph{matching lower bounds} under $\Class{(S)ETH}$, confirming that we can't expect  significant~improvements. % this runtime. 

%% Our reduction and observations give a better, fine-grained picture of the complexity
%% of $\Lang{\#2sat}$, $\Lang{\#2dnf}$ and their monotone variants. We believe this is only an
%% initial exploration into a more detailed definition of
%% $\Class{\#P}$-hardness, as in Lemma \ref{lemma:2satvs3ssat} there
%% cannot exist a $c$-monious reduction from $\Lang{\#2sat}$ to
%% $\Lang{\#3sat}$. Proofs of $\Class{\#P}$-hardness thus must use
%% multiple calls such as in $[\Lang{\#2sat}-\Lang{\#2sat}]^{\log}[\Lang{\#mon2sat}-\Lang{\#mon2sat}]^{\log}=[\Lang{\#mon2dnf}-\Lang{\#mon2dnf}]^{\log}$,
%% however this function can output negative numbers so it isn't contained in $\Class{\#P}$. However this is captured
%% by $\Gap\Class{P}$ for which we give a stronger characterization in
%% Theorem \ref{theorem:gapp}.  

There is an other interesting consequence of our reduction. Existing work on sparsification~\cite{DellEtAl14} only translates $d$-\Lang{cnf} into sparse $3$-\Lang{cnf}. 
In contrast, our main construction in Lemma~\ref{lemma:main} enables \emph{sparsification} of any $d$-\Lang{cnf} \emph{into sparse $2$-\Lang{cnf} ($2$-\Lang{dnf})}, respectively.

\begin{corollary}[Sparsification into $2$-\Lang{cnf} ($2$-\Lang{dnf})]
    Let $d\geq 2$. For every $k\in\mathbb{N}$ and $d$-\Lang{cnf} formula $\gamma$ with $n$ variables, there exists $t\in \mathbb{N}$ such that in time $t\cdot \text{poly}(n)$ we obtain formulas $\gamma_1^{i}$ and $\gamma_{2}^i$ for every $i\in [t]$ and
\begin{enumerate}
    \item $\#(\gamma)=\sum_{i\in[t]}[\#(\gamma_1^i) - \#(\gamma_2^i)]$,
    \item $t\leq 2^{\frac{n}{k}}$,
    \item $\gamma_1^i$,  $\gamma_2^i$ are $2$-\Lang{cnf} ($2$-\Lang{dnf}) formulas in which each variable occurs at most $3$ times.
    \end{enumerate}
\end{corollary}
\begin{proof}
    By \cite[Lem.~A.1]{DellEtAl14}, we know there is a formula $\beta = \bigvee_{i\in [t]} \gamma^i$ such that $\#(\gamma)=\sum_{i\in [t]} \#(\gamma^i)$ and $\gamma^i$ in $3$-\Lang{cnf}. Claims 1), 3) are modifications of \cite[Lem.~A.1]{DellEtAl14} that follow from applying Lemma~\ref{lemma:main} on every $\gamma^i$.
\end{proof}

We would like to outline a few major directions for future work. The first path concerns
further improvements on $\Gap\Class{P}$.
%% It is known that $\Class{FP} - \Class{\#P} =
%% \Gap\Class{P}$~\cite{FennerFortnowKurtz94}, however the function in $\Class{FP}$ used is actually
%% very simple and $\Class{FP}$ can be replaced with
%% $\Class{FL}$.\footnote{The function actually reduced $\Class{\#P} -
%%   \Class{\#P}$ to $\Class{\#P} - 2^q$ for some $q$ smaller than instance size~$n$.} Can we
%% replace a $\Lang{\#2dnf}$ oracle call with $\Class{FL}$ in Theorem
%% \ref{theorem:gapp} to get even more fine-grained characterizations for $\Class{gapP}$? We know that the function must be more complex
%% than~$2^n$, but we do not expect that this is possible. 
%% How does this relate to Theorem \ref{theorem:todaimproved}?
%We know with AC$^0$ postprocessing we can use a single $\Lang{\#2dnf}$ oracle call but how much postprocessing power is actually required?
Since every function in $\Class{spanL}$ admits a \emph{fully
polynomial randomized approximation scheme}
($\Class{FPRAS}$)~\cite{ArenasEtAl21}, a natural question is in how
far our work relates to the difference of calls to approximation
algorithms?
We also wonder about the (indirect) relationship of our reduction to
the use of zeta and Möbius transforms in the recent algorithm by
Slivovsky and Szeider~\cite{SlivovskyS20}. Are there combinatorial or
algebraic properties that our reduction indirectly encodes, which allow
us to derive the result directly without the use of these techniques? Are
there deeper algebraic connections (e.g., to group theory) %that might
leading to further insights into complexity? 

The second direction is a better understanding of the exact power of a single $\Lang{\#2sat}$ ($\Lang{\#2dnf}$)
call. From Theorems~\ref{theorem:singlecalla}, \ref{theorem:singlecallb}, and \ref{theorem:todaimproved}, we know we only need one
oracle call if we allow $\Class{AC}^0$ or $\Class{TC}^0$
postprocessing to capture $\#\Class{P}$, $\Gap\Class{P}$, and
$\Class{PH}$. It would be interesting to precisely understand which
circuits (i.e., how much postprocessing) are needed for each of the
$\Lang{\#sat}$ variants to, say, still capture~$\Class{PH}$. From the
point of view of $\Lang{\#sat}$ variants, we know
$[\Lang{\#2dnf}]^{\log} \subseteq \Span\Class{L}$. On the other hand,
the precise complexity of $[\Lang{\#2sat}]^{\log}$ remains unclear as we are unaware of any $\Class{NL}$ machine that can output
a model of a $\Lang{2sat}$~formula. For completeness of $\Lang{\#2sat}$ we believe that we cannot achieve $\Tot\Class{P}$-hardness. What we can claim is that such a reduction would require super logarithmic space, since the decision version is $\Class{NL}$-complete. If we had a parsimonious reduction from any $\Tot\Class{P}$-complete problem, then its decision version can be solved in $\Class{NL}$ as well~--~implying $\Class{NL}=\Class{P}$~\cite{antonopoulos2022completeness}.

\begin{observation}
    $[\Lang{\#2sat}]^{\log} \subsetneq \Tot\Class{P}$ assuming $\Class{NL} \neq \Class{P}$. 
\end{observation}
    
Note that it is known that $\Tot\Class{P}$ - $\Tot\Class{P}$ and $\Class{FP} - \Tot\Class{P}$ are equivalent to $\Gap\Class{P}$~\cite{bakali2024power}, i.e.,  one call to a $\Tot\Class{P}$ oracle is enough to capture $\Gap\Class{P}$. However, our  formalisms is significantly weaker than $\Tot\Class{P}$, e.g., we believe that $\Lang{\#mon2dnf}$ 
is even weaker than $\Span\Class{L}$. While we showed that $[\Lang{\#mon2dnf} - \Lang{\#mon2dnf}]^{\log}$ and $[\Lang{\#mon2sat} - \Lang{\#mon2sat}]^{\log}$ are enough to capture the hardness of $\Gap\Class{P}$, we leave open whether one can replace one of the calls with an $\Class{FP}$ oracle. We believe this is not possible, as we expect the simulation of negation via subtraction to require symmetry.
Unsurprisingly, the answer to these questions probably lies in circuit complexity, as we showed that even slightly stronger postprocessing
is sufficient for only a single call (see Thm~\ref{theorem:singlecalla}).

\begin{conjecture*}
	$[\Class{FP}-\Lang{\#mon2sat}]^{\log}$ is strictly contained in $[\Lang{\#mon2sat}-\Lang{\#mon2sat}]^{\log}$.
\end{conjecture*}

%\textcolor{red}{Do we want to add a conj?}

To complete the picture and to generate the
examples, we implemented the reductions presented
within this paper via a first-order-like language (logic programming)~\cite{GebserEtAl12}. We parsimoniously translate these programs to
 $\Lang{\#sat}$ via existing translations~\cite{Janhunen06}. We tested the resulting $\Lang{\#sat}$ encodings using
sophisticated state-of-the-art model
counters~\cite{FichteHecherHamiti21}. On our tested examples, we can
confirm that these systems are reasonably fast, even if there are
billions of solutions. A
thorough \emph{practical} evaluation of our theoretical work, which covers all of the non-trivial implementation details of such a realization, is planned as a follow-up.

%% It might be an interesting challenge to define new normal forms for model counting. Given our linear reduction and the strong structural guarantees of this work, does it suffice to define future normal forms that are just (restricted variants) of~$\Lang{2cnf}$ or~$\Lang{2dnf}$?
%% This step probably depends on empirical results and practical
%% considerations during implementation. What about counters restricting
%% to $\Lang{2horn}$ formulas, e.g.,~\cite{DubraEtAl23}: Given our
%% results on~$\Lang{\#impl2sat}$, maybe these ``restrictions'' might
%% provide a good balance between simplification and practical
%% efficiency? Could we expect efficient~$\Lang{tqbf}$ solvers via a
%% single~$\Lang{\#2dnf}$ call?

%\clearpage
\bibliography{main}

\clearpage
%\appendix

%\setcounter{section}{0}
\renewcommand{\thesection}{\arabic{section}} 

\section*{Appendix}
\section{Proof of the Main Lemma}\label{sec:mainproof}

\begin{lemma}[Well-Definedness]
\label{lem:welldef}
    Let~$\phi$ be a \Lang{cnf}, $\mathcal{T}$ be a tree
    decomposition of it, and $M$ be a satisfying assignment of~$\psi_1$ that is rogue.
    The symmetric rogue model $M'$ of $M$ is a satisfying assignment of~$\psi_2$. Vice versa, the result holds if roles of $M$ and $M'$ are swapped. 
\end{lemma}
\begin{proof}
Let $M$ be rogue at a node $t$.  Assume that (1) there is no
indirect ancestor node of $t$ (e.g., parent node) such that $M$ is
rogue at this node. So $t$ is the node closest to the root of $T$ with
$M$ being rogue at this node. Further, (2) $t$ is on the lexicographic smallest root-to-leaf path in~$T$. Let $M'$ be the symmetric rogue model
of~$M$.  We proceed by case distinction on \emph{why} $M$ is rogue at $t$.

\begin{description}
  \item[Case \emph{(i)}.] We have $x\notin M$ and, therefore, $M'$ is
unique for $M$ and vice versa.
\end{description}

In the following cases we have $x\in M$ and, thus, $e_{\rootOf(T)}\in
M$. Furthermore,
since $t$ is also the first node such that $M$ is rogue at $t$
and $M$ is a model of $\psi_1$, we conclude that $o_t\in M$ or $e_t\in M$.
(Either $t=\rootOf(T)$ or $M$ is not rogue at the parent node of $t$, requiring $o_t\in M$ or $e_t\in M$).
Consequently, $M$ can only be rogue at $t$ due to property
\emph{(ii)}, \emph{(iii)}, or \emph{(iv)} in Definition~\ref{def:rogue} if both
$o_t\in M$ and $e_t\in M$.

\begin{description}
\item[Case \emph{(iv)}.]  Since both $o_t, e_t\in M$, the construction
  of $M'$ results in $M'$ being a model of $\psi_2$,
  i.e., the replacement of $o_t$ by $e_t$ (and vice versa) does not
  destroy model status of $\psi_2$.
\item[Case \emph{(iii)}.]  If $M \cap \{\text{\d{$c$}}, \overline{\text{\d{$c$}}}\}=\emptyset$,
  we have $M\cap \{o1_t, o2_t, e1_t, e2_t\}\allowbreak=\emptyset$.
  The replacement of $o_t\in M$ by $e_t\in M'$ (and vice versa)
  results in $M'$ being a model of $\psi_2$.  If $\text{\d{$c$}}, \overline{\text{\d{$c$}}}\in
  M$ and $|M\cap \{o1_t, o2_t, e1_t, e2_t\}|< 1$, again, it is easy to
  see that replacing $o_t\in M$ by $e_t\in M'$ (and vice versa)
  results in $M'$ being a model of $\psi_2$. Otherwise, if $\text{\d{$c$}},
  \overline{\text{\d{$c$}}}\in M$ and $|M\cap \{o1_t, o2_t, e1_t, e2_t\}|\geq 1$,
  note that $M'$ is obtained from~$M$ by \emph{additionally replacing}
  $o1_t\in M$ by $e2_t\in M'$ and $o2_t\in M$ by $e1_t\in M'$ (and
  vice versa, respectively). Since both $\text{\d{$c$}},\overline{\text{\d{$c$}}}\in M$, this
  ensures that therefore $M'$ is a model of $\psi_2$.
\item[Case \emph{(ii)}.]  If $M\cap \{o1_t, o2_t, e1_t,
  e2_t\}=\emptyset$, indeed, $M'$ is a model
  of~$\psi_2$ as well. Note that $|M\cap \{o1_t,
  o2_t, e1_t, e2_t\}|>1$ can never occur, assuming we have neither
  Case \emph{(iv)} nor Case \emph{(iii)}. The roles of~$M$ and~$M'$ can be easily
  switched, as the replacements of Definition~\ref{def:roguesymm} are
  completely symmetric. \qedhere
\end{description}
\end{proof}

\begin{lemma}[Symmetry]
\label{lem:symmetry}
    Let~$\phi$ be a \Lang{cnf}, $\mathcal{T}$ be a tree
    decomposition of it, and
    $M$ be a satisfying assignment of~$\psi_1$ that is rogue.
    Then, (I) a model $M$ of~$\psi_1$ is rogue at a node $t$ iff the symmetric rogue model $M'$ of~$\psi_2$ is rogue at $t$ (and vice versa with swapped $M$ and $M'$);
    and, (II) mapping the rogue model of~$\psi_1$ to its symmetric rogue model forms a bijection.
\end{lemma}
\begin{proof}
We define a function~$f$ mapping models~$M$ of~$\psi_1$ to corresponding symmetric rogue models~$M'$ of~$\psi_2$.
Suppose $M$ is rogue at a node~$t'$ of~$T$.
In order to show that \emph{(I)} $M'=f(M)$ is also rogue at $t'$, let
$t^\star$ be the node of $T$ such that $M$ is rogue at~$t^\star$ with
$M$ not being rogue at an ancestor of~$t^\star$.  We distinguish the following cases:
\begin{description}
\item[Case $t'$ is an ancestor of~$t^\star$.]%\useieee{~\\}
  By construction $M$ is not rogue at~$t^\star$ iff $M'$ is not rogue
  at~$t^\star$.
\item[Case $t'$ is a descendant of~$t^\star$.]%\useieee{~\\}
  Since $M$ and $M'$ are by construction identical regarding
  Definition~\ref{def:rogue} \emph{(i)}--\emph{(iv)}, we conclude that
  $M$ is rogue at~$t^\star$ iff $M'$ is not rogue at~$t^\star$.
\item[Case $t'=t^\star$.]%\useieee{~\\}
  Holds as the construction of Definition~\ref{def:roguesymm} does not change the rogue status of a model.
\end{description}

The proof works analogously if the roles of $M$ and $M'$ are swapped.
It remains to show~\emph{(II)}: $f$ is indeed a bijection. By
Lemma~\ref{lem:welldef}, $f$ is well-defined. Further, by the
construction given in Definition~\ref{def:roguesymm}, $f$ is also
injective. Suppose towards a contradiction that there was a rogue
model~$M_2$ of $\psi_2$ and two rogue models~$M_1, M_1'\in
f^{-1}(M_2)$. One can proceed by case distinction.  Since we start
replacing $o_{t'}$ by $e_{t'}$ (and vice versa) for nodes~$t'$ from
${t^\star}$ upwards in the direction towards the root of~$T$, $M_1$
and $M_1'$ coincide on the assignment of $o_{t'}$ and $e_{t'}$.  The
remaining interesting case is the last item of
Definition~\ref{def:roguesymm}. Observe that we only replace
$o1_{t^\star}$ by $e2_{t^\star}$, $o2_{t^\star}$ by $e1_{t^\star}$
(and vice versa) if without this replacement the result is not a
model.  Therefore, $M_1=M_1'$, which shows that $f$ is injective.

It remains to show that $f$ also is surjective. Suppose towards a contradiction that there is a rogue model~$M''$ of~$\psi_2$ such that $f^{-1}(M'')$ is not defined.
However, we can construct a rogue model~$M'''$ of~$M''$ according to
Definition~\ref{def:roguesymm}. Then, by Lemma~\ref{lem:welldef} and
(i) above, $M'''$ is indeed a rogue model
of~$\psi_1$. This contradicts the assumption that
$f^{-1}(M'')$ is not defined, since $M'''=f^{-1}(M'')$. 
\end{proof}

\begin{lemma}
\label{lem:nonmodels}
    %Let~$\phi$ be a \Lang{cnf} and $\mathcal{T}$ be a tree decomposition of it.
    %Then, 
    (1) For every model $M$ of $\psi_1$ or $\psi_2$ that is not rogue,
    $M\cap \vars(\phi)$ is an assignment over the variables of 
    $\phi$ that invalidates at
    least $nc=|\{\text{\d{$c$}} \mid c \in \clauses(\phi), \text{\d{$c$}}\in M\}|$ clauses. (2)
    $nc$ is odd iff $M$ is a model of $\psi_2$.
\end{lemma}
\begin{proof}
    By construction, $M$ invalidates at least $nc$ clauses.
    Further, since $M$ is not rogue,
    $nc$ is odd iff $M$ is a model of $\psi_2$.
\end{proof}

We now have all ingredients to prove Lemma~\ref{lemma:main}, which requires us to show that \emph{(a)}
$\#(\phi)=\#(\psi_1)-\#(\psi_2)$; \emph{(b)} that
$R(\phi,\mathcal{T})$ can be computed in linear time or logspace; and \emph{(c)}
$\max\{\tw(\psi_1),\allowbreak\tw(\psi_2)\}\leq\tw(\phi)$. The next three
results establish these statements.

\begin{proposition}[Correctness]\label{prop:corr}
  $\#(\phi)=\#(\psi_1)-\#(\psi_2)$
\end{proposition}
\begin{proof}
As discussed in the introduction, the construction simulates the principle of inclusion-exclusion.
We can count the number of models of $\phi$ as:
\begin{align*}
  \#(\phi) = 2^n -
  &\hspace{-3.5em}\phantom{+}\sum_{\substack{M\subseteq 2^{\vars(\phi)}, \exists c\in \clauses(\phi),\\ M\not\models \{c\}}}\hspace{-3.5em}1  %\\
  &+\sum_{\substack{M\subseteq 2^{\vars(\phi)}, \exists c,c'\in \clauses(\phi),\\ c\neq c',M\not\models c\vee c'}}\hspace{-3.5em}1 %\\
  &&+ \cdots \\
  &+\sum_{\mathclap{M\subseteq 2^{\vars(\phi)}, \not\exists c\in \clauses(\phi),M\models \{c\}}} (-1)^{|\phi|}, 
\end{align*}
which simplifies to
\vspace{-3.5em}\[
\qquad\qquad\qquad\qquad\qquad\qquad\qquad\quad\sum_{\mathclap{\substack{M\subseteq 2^{\vars(\phi)}, 0\leq nc\leq |\phi|,\\M\text{ does not satisfy $\geq nc$ clauses in }\phi}}}(-1)^{nc}.
\]

Therefore we can split this term into two parts, where we compute assignments that do not satisfy at least $nc$ clauses with $nc$ being even ($\psi_1$)
and then subtract those assignments with $nc$ being odd ($\psi_2$).
As a result, the goal is to encode the result of exclusive-or, i.e., whether an assignment does not satisfy at least $nc$ clauses with $nc$ being even (odd).
We refer to those assignments where $nc$ is even by
\[
E=\{M\subseteq 2^{\vars(\phi)} \mid 0\leq nc\leq |\phi|, M\text{ does not satisfy }\useieee{\]\vspace{-1.85em}\[}\geq nc\text{ clauses in }\phi, nc \equiv 0 (\bmod 2)\}.
\]
The assignments with $nc$ being odd are referred to by
\[
O=\{M\subseteq 2^{\vars(\phi)} \mid 0\leq nc\leq |\phi|, M\text{ does not satisfy }\useieee{\]\vspace{-1.85em}\[}\geq nc\text{ clauses in }\phi, nc \equiv 1 (\bmod 2)\}.
\]

By Lemma~\ref{lem:nonmodels}, every assignment 
in $E$ can be extended to a satisfying assignment of $\psi_1$, %i.e. $$ is considered in the number $\#(\psi_1)$,
i.e., $|E| \leq \#(\psi_1)$. Analogously, $|O| \leq \#(\psi_2)$.
However, by Lemma~\ref{lem:symmetry}, there is a bijective function between the set~$R$ of rogue models of $\psi_1$
and the rogue models of $\psi_2$.
Consequently, %$|E| \leq \#(\psi_1)$ and $|O| \geq \#(\psi_2)$. Let $R$ be the set of rogue assignments of $\psi_1$ and $\psi_2$. Then
\[
\#(\phi) = \big(|E| + {|R|}\big) - \big(|O| + {|R|}\big) =  \#(\psi_1) - \#(\psi_2).
\qedhere
\]
\end{proof}

\begin{proposition}
  The reduction $R(\phi,\mathcal{T})$ can be computed in linear time or logarithmic space for a given tree decomposition. %In
  %particular, 
  We have $|\psi_1| + |\psi_2|\in O\big(|\phi|+|\mathcal{T}|\big)$.
  %
  %If~$\mathcal{T}$ is not part of the input, we can 
\end{proposition}
\begin{proof}
Let~$\mathcal{T}=(T,\chi,\delta)$ be the given labeled TD.
Without loss of generality, we may assume that the size of~$\mathcal{T}$ is linear in $|\phi|$~\cite{BodlaenderKloks96}.
This still holds for labeled TD~$\mathcal{T}$, which only linearly increases the size in the worst case (compared to an unlabeled TD).
Then, Equation~(\ref{lab:wcnf}), (\ref{lab:wcnf3}) is computed for every clause~$c\in\clauses(\phi)$
and Equations~(\ref{eq:o1})--(\ref{eq2:e2}) are computed for every node in~$T$ and in the context of a single clause.
Consequently, it is easy to see that the size is linearly bounded for both~$\psi_1$ and~$\psi_2$.
The logspace bound follows, as we only need a constant number of pointers to the input.
\end{proof}

\subsection{Preserving Structural Parameters}\label{sec:structure}

Let~$\varphi$ be a~$\Lang{cnf}$ formula and recall~$R$, as well as~$\psi_1$ and~$\psi_2$ from above.
%Note that while~$R(\varphi,\mathcal{T})$ was presented only for (labeled) path decompositions, where every node~$t$, ensures $\delta(t)\neq \emptyset$, we can easily extend this to arbitrary tree decompositions.
%
%
%Indeed, tree decomposition nodes~$t$ with~$\delta(t)=\emptyset$ (and~$\children(t)=\{t'\}$) are a special case, where we only generate clauses~$o2_t \rightarrow o_{t}$, $o2_t\rightarrow o_{t'}$, $e2_t \rightarrow e_t$, and $e2_t\rightarrow e_{t'}$. So for this case, we do not even use variables~$o1_t, e1_t$.
%
%In fact, the reduction can also be updated to accommodate join nodes as follows.
%
%
%For tree decomposition nodes~$t$ with~$\delta(t)=\emptyset$ (and~$\children(t)=\{t', t''\}$) we generate the following clauses, which are similar to Equations~(\ref{eq:o1})--(\ref{eq:e2}).
%\begin{flalign}
%    & o1_t \rightarrow e_{t'} & o1_t \rightarrow o_{t''} && o1_t \rightarrow o_t && \text{Case 1: odd by even/odd child nodes}\hspace{1.55em}\label{eq2:o1}\\
%    & o2_t \rightarrow o_{t'} & o2_t \rightarrow e_{t''} && o2_t \rightarrow o_t && \text{Case 2: odd by odd/even child nodes}\hspace{1.55em}\\[1.5em]
    %
%    & e1_t \rightarrow o_{t'} & e1_t \rightarrow o_{t''} && e1_t \rightarrow e_t && \text{Case 1: even by odd/odd child nodes}\hspace{1.55em}\\
%    & e2_t \rightarrow e_{t'} & e2_t \rightarrow e_{t''} && e2_t \rightarrow e_t && \text{Case 2: even by even/even child nodes}\hspace{1em}\label{eq2:e2}%\\[1.5em]
    %
%\end{flalign}
%
%
We are ready to show our strong guarantees for structural parameters. 

\begin{lemma}\label{lem:param}
  $\max\{\tw(\psi_1),\tw(\psi_2)\}\leq\tw(\phi)+13$ and $\max\{\itw(\psi_1),\itw(\psi_2)\}\leq\itw(\phi)+14$.
\end{lemma}
\begin{proof}
Take any tree decomposition~$\mathcal{T}=(T,\chi)$ of~$G_\varphi$ (of width~$\tw(\varphi)+1$). Without loss of generality, we may assume that every node in~$T$ has at most~$2$ child nodes. Indeed, this comes with a factor $\tw(\varphi)$  overhead in the number of nodes by just adding intermediate copies of nodes.
From this, we can easily obtain a labeled TD~$\mathcal{T}'=(T',\chi',\delta)$ of width~$\tw(\phi)+1$.

%
%\textcolor{red}{Todo: prove that treewidth is preserved}
Then, we obtain a tree decomposition~$\mathcal{T}''=(T',\chi'')$ of both~$G_{\psi_1}$ and~$G_{\psi_2}$, where we define~$\chi''$ as follows. 
For every node~$t$ of~$T''$,
let~$\chi''(t)=\chi'(t) \cup \{x, o1_t, o2_t, e1_t, e2_t, e_t, o_t\}\cup \{e_{t'}, o_{t'} \mid t'\in \children(t)\} \cup \{\text{\d{$c$}}, \overline{\text{\d{$c$}}} \mid c\in\delta(t)\}$. It is therefore easy to see that~$|\chi''(t)| \leq |\chi'(t)| + (9+2\cdot|\children(t)|) = |\chi'(t)| + 13$, since~$|\children(t)| \leq 2$.

For the second claim, suppose that~$\mathcal{T}'$ was a tree decomposition of the incidence graph~$I_\varphi$. Then, we can reuse the same construction as above, %of~$\mathcal{T}''$ above, 
to convert the resulting tree decomposition~$\mathcal{T}''$ into a tree decomposition~$\mathcal{T}'''$ of both~$I_{\psi_1}$ and~$I_{\psi_2}$.
Indeed, we additionally need to add vertices for the clauses in~$I_{\psi_1}$ ($I_{\psi_2}$) we generated in Equations (\ref{lab:wcnf})--(\ref{red:leaf}). 
However, for each bag it suffices to add one of these vertices at a time (and just duplicate the bag several times). This results in a tree decomposition~$\mathcal{T}'''$ of~$I_{\psi_1}$ and~$I_{\psi_2}$, where the bag sizes are just increased by~$1$, establishing the desired claim. Note that due to the resulting chain of (copied) bags, the resulting tree decomposition might just be of linear size in the instance size (as each variable occurrence in a clause is treated at most once).
\end{proof}

Observe that the same argument immediately applies to pathwidth, where every TD node has at most one child node, so we obtain even~$+12$ instead of~$+14$.
For the bandwidth parameter, we obtain similar results, which yield tight lower bounds (see Theorem~\ref{sec:sethlb}).

\begin{lemma}\label{lem:bw}
We can slightly modify~$\psi_1$ and~$\psi_2$ such that the following claim holds:
  \[\max\{\bw(\psi_1),\bw(\psi_2)\}\leq\bw(\phi){+}11 \text{ and }\useieee{\]\vspace{-1.8em}\[}\max\{\ibw(\psi_1),\ibw(\psi_2)\}\leq\ibw(\phi){+}12.\]
\end{lemma}
\begin{proof}
Take a bijective mapping~$f$ of~$G_\varphi$ with~$n=|\vars(\varphi)|$.
In this proof we will process~$f$ in batches of size~$k$, where the~$i$th batch ($0\leq i < b$ with~$b=\floor{\frac{n}{k}}$) ranges from index~$ik$ to index~$i(k+1)-1$.
The variables (vertices) of the~$i$th batch are addressed by~$B_i=\{v_{ik}, \ldots, v_{i(k+1)-1}\}$.
Consequently, we can simulate a tree decomposition~$\mathcal{T}_f$ of~$G_\varphi$, which we can pass to~$R$. This decomposition is a path where the~$i$th bag comprises elements of the~$i$th batch as well as the~$(i-1)$st batch (if it exists). Hence, the width of~$\mathcal{T}_f$ is~$2k$.

However, for bandwidth, the construction of~$\psi_1$ and~$\psi_2$ needs to be slightly adapted as follows. Instead of variable~$x$, we use copies~$x_0, \ldots, x_{b-1}$ and replace~$x$ in Equation~(\ref{red:leaf}) by~$x_{b-1}$ (and every other occurrence of~$x$ is replaced by~$x_0$). Further, we add implications~$x_0\rightarrow x_1$, $x_0\leftarrow x_1$, \dots, $x_{b-2}\rightarrow x_{b-1}$, $x_{b-2}\leftarrow x_{b-1}$.

Then, similarly to Lemma~\ref{lem:param} above, we can look at clauses in~$\varphi$ one by one.
If, say, there is more than one clause over variables contained in the~$i$th batch, we could just have created an intermediate batch between the~$i$th and the $(i+1)$st batch. 
This batch is over copy variables~$v'_{ik}, \ldots, v'_{i(k+1)-1}$ and we would have constructed clauses (implications) for every index~$j$ in this batch: $v_{j} \rightarrow v'_j$ and $v_j \rightarrow v'_j$ to~$\varphi$ (and renamed occurrences of~$v_j$ in~$\varphi$ involving variables in later batches following~$i$).

Hence, we can construct a modified bijective mapping~$f'$ of both~$G_{\varphi_1}$ and~$G_{\varphi_2}$, using similar constructions as in Lemma~\ref{lem:param}.
Thereby, analogously to above, every batch~$i$ gets extended by $\{x_i, o1_i, o2_i, e1_i, e2_i, e_i, o_i\}\cup \{e_{i-1}, o_{i-1} \mid i-1 > 0\} \cup \{\text{\d{$c$}}, \overline{\text{\d{$c$}}} \mid c\in\clauses(\varphi)\text{ considered in}\xspace\allowbreak\text{batch }i\}$.
Observe that this establishes the first claim, as~$f'$ is a bijective mapping of dilation at most~$\bw(\phi)+11$.
\medskip

The argument for incidence bandwidth works similarly to Lemma~\ref{lem:param} above, thereby increasing from~$+11$ to~$+12$. However, in addition, we need to add intermediate copies of batches as demonstrated above.
\end{proof}

\subsection{Proof for Monotone Formulas}\label{section:proofmonotone}

We are ready to extend our definition of a the rogue model (Definition~\ref{def:rogue}) for monotony below. 
Note that Equation~(\ref{eq:var}) introduces another reason, \emph{why a model can be rogue} at a node. This is formalized in Definition~\ref{def:extrogue} by the added item (iiib). %Again, by construction of labeled tree decompositions, (iii) and (iiib) can not be true simultaneously. However, %we do not care to distinguish between (iii) and (iiib), but just need to correctly keep track of the \emph{parity of the combined number} of both. 
As in Section~\ref{sec:sharpsub}, by establishing a bijection we then utilize the \emph{power of subtraction} to eliminate rogue~models.

\begin{definition}[Rogue Model for Monotony]\label{def:extrogue}
 Let~$\phi$ be a \Lang{cnf}, $\mathcal{T}=(T,\bag,\delta)$ be a fully labeled TD of~$\phi$, and $t$
 be a node in~$T$. % with~$\alpha\in\delta(t)$.
 Then, a model~$M$ of a formula~$\phi'\supseteq R'(\phi, \mathcal{T})$
 is referred to by \emph{rogue (at~$t$)} whenever 
  \begin{enumerate}[label=(\roman*)]
 \item $\overline{x}\in M$,
 \item $|M\cap \{\overline{o1_t}, \overline{o2_t}, \overline{e1_t}, \overline{e2_t}\}|\neq %3 
 |\{\overline{o1_t}, \overline{o2_t}, \overline{e1_t}, \overline{e2_t}\} \cap \vars(\phi')|-1$,
 \item $|M\cap \{\alpha, \overline{\alpha}\}|\neq 1$ with $\alpha\in\delta(t)$,
 \item [\textcolor{\uselipics{lipicsGray}\useieee{gray}}{(iiib)}] $\{\top_v, {\bot_v}\}\subseteq M$ with $v\in\delta(t)\cap\vars(\varphi)$, or 
 \item  $|M\cap \{o_t, e_t\}|\neq 1$.
 \end{enumerate}
\end{definition}

\noindent Condition (ii) usually means $|M\cap \{\overline{o1_t}, \overline{o2_t}, \overline{e1_t}, \overline{e2_t}\}|\neq 3$.
The construction of the symmetric rogue model works analogously as in Definition~\ref{def:roguesymm},
where instead of $x$ we use $\overline{x}$ and instead of $o1_t$, $o2_t$, $e1_t$, $e2_t$ 
we put $\overline{o1_t}$, $\overline{o2_t}$, $\overline{e1_t}$, $\overline{e2_t}$, respectively.

\begin{definition}[Symmetric Rogue Model for Monotony]\label{def:extroguesymm}
Let $M$ be a model of a formula~$\phi'$ with $\phi'\supseteq R(\phi, {\mathcal{T}})$
that is rogue at~$t$ with~$\alpha\in\delta(t)$.
Assume that (1) there is no ancestor~$t'$ of~$t$ in~$T$ such that~$M$ is rogue at~$t'$ and that (2) $t$ is on the lexicographic smallest root-to-leaf path of~$T$.
The \emph{symmetric rogue model $M'$
(of~$M$)} is constructed~by:
\begin{itemize}
    \item If $\overline{x}\in M$, we define $M'=M$
    \item Otherwise, if $\overline{x}\notin M$:
    \begin{itemize}
    \item Replace $o_t\in M$ by $e_t\in M'$ (and vice versa, i.e.\ $e_t\in M$ iff $o_t\in M'$).
    \item For every ancestor~$t'$ of~$t$ in~$T$, we replace $o_{t'}\in M$ by $e_{t'}\in M'$ and vice versa (i.e.\ $e_{t'}\in M$ iff $o_{t'}\in M'$), as well as $\overline{o1_{t'}}\notin M$ by $\overline{e1_{t'}}\notin M'$,  $\overline{e1_{t'}}\notin M$ by $\overline{o1_{t'}}\notin M'$,  $\overline{o2_{t'}}\notin M$ by $\overline{e2_{t'}}\notin M'$,  $\overline{e2_{t'}}\notin M$ by $\overline{o2_{t'}}\notin M'$ (and vice versa) %In other words for $M'$, we have $o_{t'}\in M$ iff $e_{t'}\in M'$, and $e_{t'}\in M$ iff $o_{t'}\in M'$
    \item If (a) either $o_t\in M$ or $e_t\in M$, and (b) $|M\cap \{\overline{o1_t}, \overline{o2_t}, \overline{e1_t}, \overline{e2_t}\}|\leq 3$, %and (c) $\{\alpha,\overline{\alpha}\}\subseteq M$, 
    we additionally replace $\overline{o1_t}\notin M$ by $\overline{e2_t}\notin M'$, $\overline{o2_t}\notin M$ by $\overline{e1_t}\notin M'$ and vice versa (i.e.\ $\overline{e1_t}\notin M$ by $\overline{o2_t}\notin M'$, $\overline{e2_t}\notin M$ by $\overline{o1_t}\notin M'$) 
    \item If (a)%either $o_t\in M$ or $e_t\in M$
    , (b)%$|M\cap \{\overline{o1_t}, \overline{o2_t}, \overline{e1_t}, \overline{e2_t}\}|= 3$
    , % (c) $|\{\alpha,\overline{\alpha}\}\cap M|=1$,
    and (c) $\{\top_\alpha, \bot_\alpha\} \subseteq M$% %and $\overline{\alpha}\in M\cap\vars(\phi)$
, we additionally replace $\alpha\in M$ by $\overline{\alpha}\in M'$ and vice versa %, $\overline{o1_t}\notin M$ by $\overline{o2_t}\notin M'$, $\overline{o2_t}\notin M$ by $\overline{e1_t}\notin M'$ and vice versa 
(i.e.\ $\overline{\alpha}\in M$ by ${\alpha}\in M'$%, $\overline{e1_t}\notin M$ by $\overline{o2_t}\notin M'$, $\overline{e2_t}\notin M$ by $\overline{o1_t}\notin M'$
)
    \end{itemize}
   \noindent We say $M$ is the \emph{symmetric rogue model} of~$M'$.
\end{itemize}
\end{definition}

\begin{lemma}
\label{lem:extwelldef}
    Let ~$\phi$ be a \Lang{cnf}, $\mathcal{T}$ be a tree
    decomposition of it, and
    $M$ be a satisfying assignment of~$\psi_1'$ that is rogue.
    Then, the symmetric rogue model $M'$ of $M$ is a satisfying
    assignment of~$\psi_2'$. Vice versa, the result
    holds if roles of $M$ and $M'$ are swapped. 
\end{lemma}
\begin{proof}
Let $M$ be rogue at a node $t$ and assume that
(1) there is no indirect ancestor node of $t$ (e.g., parent node) such that $M$ is rogue at this node and that (2) $t$ is on the lexicographic smallest root-to-leaf path of~$T$. So $t$
is the node closest to the root of $T$ with $M$ being rogue at this node.
Let $M'$ be the symmetric rogue model of~$M$.

The proof works analogously to Lemma~\ref{lem:welldef}.
The only missing case is where~$M$ is rogue at~$t$ only due (iiib) of Definition~\ref{def:extrogue}.
Then we simply perform the same replacements, as in Case (iii). Since
these replacements require that if~$\alpha\in M$ then
$\overline{\alpha}\in M'$ (and vice versa), the last item of
Definition~\ref{def:extroguesymm} ensures that~$M'$ is a model
of~$\phi_2'$. As for Lemma~\ref{lem:welldef}, the roles of~$M$
and~$M'$ can be switched by symmetry. 
\end{proof}

\begin{lemma}
\label{lem:possymmetry}
    Let~$\phi$ be a \Lang{cnf}, $\mathcal{T}$ be a fully labeled
    tree decomposition of it, and
    $M$ be a satisfying assignment of~$\psi_1'$ that is rogue.
    Then, (I) a model $M$ of~$\psi_1'$ is rogue at a node $t$ iff the
    symmetric rogue model $M'$ of~$\psi_2'$ is rogue at $t$ (and vice
    versa with swapped $M$ and $M'$); and
    (II) mapping the rogue model of~$\psi_2'$ to its symmetric rogue model forms a bijection.
\end{lemma}
\begin{proof}
We define a function~$f$ by mapping a model~$M$ of~$\psi_1'$ to its corresponding symmetric rogue model~$M'$ of~$\psi_2'$.
Suppose $M$ is rogue at a node~$t'$ of~$T$.
In order to show that \emph{(I)} $M'=f(M)$ is also rogue at $t'$, let
$t^\star$ be the node of $T$ such that $M$ is rogue at~$t^\star$ with
$M$ not being rogue at an ancestor of~$t^\star$. 
We distinguish the following cases.

\begin{description}
\item[Case $t'$ is an ancestor of~$t^\star$.]%\useieee{~\\}
  By construction $M$ is not rogue at~$t^\star$ iff $M'$ is not rogue
  at~$t^\star$.
\item[Case $t'$ is a descendant of~$t^\star$.]\useieee{~\\}
  Since $M$ and $M'$ are by construction identical regarding
  Definition~\ref{def:extrogue} \emph{(i)}--\emph{(iv)}, we follow that $M$ is rogue
  at~$t^\star$ iff $M'$ is not rogue at~$t^\star$.
\item[Case $t'=t^\star$.]\useieee{~\\}
  Holds since the construction of Definition~\ref{def:extroguesymm}
  does not change the rogue status of a model. 
\end{description}

The proof works analogously if the roles of $M$ and $M'$ are swapped.
It remains to show that \emph{(II)}~$f$ is indeed a bijection. By
Lemma~\ref{lem:extwelldef}, $f$ is well-defined. Further, by the
construction given in Definition~\ref{def:extroguesymm}, $f$ is also
injective. Indeed, suppose towards a contradiction that there was a
rogue model~$M_2$ of $\psi_2'$ and two rogue models~$M_1, M_1'\in
f^{-1}(M_2)$. One can proceed by case distinction.  Since we start
replacing $o_{t'}$ by $e_{t'}$ (and vice versa) for nodes~$t'$ from
${t^\star}$ upwards in the direction towards the root of~$T$, $M_1$
and $M_1'$ coincide on the assignment of $o_{t'}$ and $e_{t'}$.  The
remaining interesting case is the last item of
Definition~\ref{def:extroguesymm}.  Observe that we only replace
$\alpha$ by $\overline{\alpha}$ (and vice versa) if without this
replacement the result is not a model. Indeed, if
$\{\alpha,\overline{\alpha}\}\subseteq M$, the replacement does not
achieve anything.  Otherwise, \emph{(a)} and \emph{(b)} of
Definition~\ref{def:extroguesymm} implies that $|M\cap
\{\overline{o1_t}, \overline{o2_t}, \overline{e1_t},
\overline{e2_t}\}|= 3$. Consequently, $M$ is rogue at $t^\star$, only
due to (iiib) of Definition~\ref{def:extrogue}, i.e.\ the replacement
of~$\alpha$ ($\overline{\alpha}$) in~$M$ is required for the resulting
assignment to be a model of Equation~(\ref{eq:var}).  Therefore,
$M_1=M_1'$, which shows that $f$ is injective.

Analogously to Lemma~\ref{lem:symmetry}, $f$ is
surjective.
\end{proof}

Clearly, $\psi_1'$ and $\psi_2'$ are \emph{monotone} formulas in
\Lang{2cnf}. It is easy to see that the modifications from
$R'(\phi,\mathcal{T})$ compared to $R(\phi,\mathcal{T})$ do not
increase the treewidth and increase the size of the produced
formulas only by a constant factor.
By utilizing Lemma~\ref{lem:extwelldef} and
Lemma~\ref{lem:possymmetry} as in the previous section, we can
conclude:

\begin{corollary}\label{corollary:monotone}
    There is a linear-time algorithm that maps formulas $\phi$
    to formulas $\psi_1$ and $\psi_2$ \textbf{without negation} and with at most two variables per clause such that
    \begin{align*}
      \#(\phi) = \#(\psi_1) - \#(\psi_2)
      \quad\text{and}\quad\useieee{\\}
      \max\{\tw(\psi_1),\tw(\psi_2)\}\leq\tw(\phi)
      + 13, \\
      \max\{\itw(\psi_1),\itw(\psi_2)\}\leq\itw(\phi)
      + 14.
    \end{align*}
\end{corollary}

\medskip
\subsection{Reducing to Cubic and Bipartite Formulas}\label{section:cubic-bipartite}

Before we discuss stronger results by restricting~$\Lang{\#impl2sat}$ to formulas of degree at most~$3$ and bipartite primal graphs, we briefly mention the following.

\begin{proposition}\label{prop:3sat}
There is a linear-time conversion from a formula~$\varphi$ in~\Lang{cnf} to a formula~$\varphi'$ in~\Lang{3cnf}, such that the $\tw(\varphi')\leq \tw(\varphi)+2$. If additionally, every variable in~$\varphi'$ occurs at most~$3$ times (not of the same sign), we still obtain $\tw(\varphi')\leq 3\tw(\varphi){+}2$, $\itw(\varphi')\leq 3\itw(\varphi){+}3$. %, as well as $\bw(\varphi')\leq 2\bw(\varphi)$. %. ($\itw(\varphi')\leq \itw(\varphi)+2$).
%{\color{red}\textbf{TODO: cnf->3cnf constant treewidth/bandwidth overhead. This even holds if a variable occurs at most~$3$ times and those occurrences are not of the same sign}}
\end{proposition}
\begin{proof}
The first claim can be easily established by taking any labeled tree decomposition~$\mathcal{T}=(T,\chi, \delta)$ of~$G_{\varphi}$ (of width~$\tw(\varphi)+1$). Then, for every node~$t$ in $T$, we split up long clauses~$c=l_1 \vee l_2 \vee \cdots \vee l_k$ such that~$c\in\delta(t)$, via auxiliary variables~$a_1, \ldots, a_{k-1}$ and by constructing auxiliary clauses of the form~$l_1 \vee l_2 \vee a_1$, $\neg a_1 \vee l_3 \vee a_2$, $\neg a_2 \vee l_4 \vee a_3$, $\ldots$, $\neg a_{k-1} \vee l_k$. We refer to the resulting formula containing every auxiliary clause by~$\varphi'$.  
It is easy to see how we obtain a tree decomposition~$\mathcal{T}'$ of~$\varphi'$ from~$\mathcal{T}$. We take~$\mathcal{T}$ and basically duplicate nodes (i.e., we replace a node in~$T$ by a path, similarly to Lemma~\ref{lem:param}) and add to each duplicate bag at most two auxiliary variables~$a_i, a_{i+1}$.
Then, as $|\delta(t)|\leq 1$, the width of the resulting tree decomposition~$\mathcal{T}'$ is bounded by~$\tw(\varphi)+1+2$.
Indeed, this can be done such that all variables of every constructed auxiliary clause are covered by~$\mathcal{T}'$, i.e., $\mathcal{T}'$ is a tree decomposition of~$G_{\varphi'}$.

If instead~$\mathcal{T}$ were a tree decomposition of~$I_{\varphi}$ of width~$\itw(\varphi)+1$, we could still apply the same idea as above, but the labeling~$\delta$ is insufficient since clauses span over several bags. Consequently, when we guide the construction of auxiliary variables along~$\mathcal{T}$, 
we require to add for \emph{each} clause vertex $c$ in a bag the corresponding auxiliary variable~$a_i$ used in the previous bag(s).
After~$a_i$ is not used anymore, we could duplicate the bag and add~$a_{i+1}$ to the fresh bag. 
It is therefore easy to see that this, unfortunately, causes a factor~$2$: $\itw(\varphi') \leq 2\itw(\varphi)$, as there can be up to~$\itw(\varphi)$ clause vertices in a bag (each of these might need to keep an auxiliary vertex~$a_i$).

\medskip

    For the second claim, we can ensure that a variable occurs at most~$3$ times in a clause, using established techniques~\cite{Tovey84}. Thereby, we create a copy~$v_i$ for every variable appearance~$v$ and chain these, e.g., $v \rightarrow v_2$, $\ldots$, $v_u \rightarrow v$. However, while the creation of such a (chain of) implications can be guided along a tree decomposition~$\mathcal{T}$ (similar to above), in the worst case this requires that for every element in the bag, we also need to add the directly preceding copy variable form the previous bag, as well as the first copy variable to the bag (which we need for ``closing'' the cycle).  
    Unfortunately, this already causes a factor~$3$ (worst-case) overhead: $\tw(\varphi') \leq 3\tw(\varphi)$, but it is easy to see that this can be combined with splitting clauses from above.
    Further, we may assume that not all of occurrences of each variable are of the same sign.
    If they were, we could combine the previous step of copying variables to remove those: Suppose a variable~$x$ occurs~$3$ times in the form of the same literal~$l$. Then we replace the three occurrences of~$l$ by literals~$l, l_2, l_3$ of the same sign as~$l$, but over variables~$x, x_2, x_3$. Then we add clauses~$l \rightarrow l_2$, $l_2\rightarrow l_3$, $l_3\rightarrow l$, which ensures equivalence.
    Observe that this can be carried out with an overhead of~$+2$, as we can do this for each variable~$x$ independently by copying bags. This results in $\tw(\varphi') \leq 3\tw(\varphi)+2$.
    It is easy to see that %the generated clauses can 
    then we obtain  $\itw(\varphi') \leq 3\itw(\varphi)+3$ since we might need to add vertices for constructed clauses one-by-one.
\end{proof}

% \begin{proposition}
% {\color{red}\textbf{TODO: cnf->3cnf constant treewidth/bandwidth overhead.}}
% %This even holds if a variable occurs at most~$3$ times and those occurrences are not of the same sign}
% \end{proposition}
% \begin{proof}
%     % Further, assume that every variable occurs at most~$3$ times~\cite{Tovey84}. Then, for every variable not all of its occurrences are of the same sign. If they were, we could change that. Suppose a variable~$x$ occurs~$3$ times in form of the same literal~$l$. Then, we replace the three occurrences of~$l$ by literals~$l_1, l_2, l_3$ of the same sign as~$l$, but over fresh variables~$x_1, x_2, x_3$. Then we add clauses~$l_1 \rightarrow l_2$, $l_2\rightarrow l_3$, $l_3\rightarrow l_1$, which ensures equivalence.
% %
% \end{proof}

By Proposition~\ref{prop:3sat}, %without loss of generality, 
we may assume a formula~$\varphi$ in $\Lang{3cnf}$, where every variable occurs at most~$3$ times, but not with a single sign. %, see also~\cite{Tovey84}. %Then, for every variable not all of its occurrences are of the same sign. If they were, we could change that. Suppose a variable~$x$ occurs~$3$ times in form of the same literal~$l$. Then, we replace the three occurrences of~$l$ by literals~$l_1, l_2, l_3$ of the same sign as~$l$, but over fresh variables~$x_1, x_2, x_3$. Then we add clauses~$l_1 \rightarrow l_2$, $l_2\rightarrow l_3$, $l_3\rightarrow l_1$, which ensures equivalence.
%
%Consequently, variables of the form~$\bot_v, \top_v$ have already degree at most three.
Observe that the formula~$\psi$ constructed by Equations~(\ref{lab:wcnf})--(\ref{eq:e2}) on~$\varphi$ and a labeled tree decomposition~$\mathcal{T}=(T,\chi,\delta)$ is already bipartite.
Indeed, edges only occur between sets~$U=\{v, e1_t, e2_t, o1_t, o2_t, x \mid v\in\vars(\varphi), t\text{ in }T\}$ and~$V=\{\text{\d{$c$}}, e_t, o_t\mid  c\in\clauses(\varphi), t\text{ in }T\}$.

To preserve this bipartite property and ensure maximum degree~$3$, we need to update~$R(\varphi,\mathcal{T})$ by adding additional clauses.
For each clause~$c\in\clauses(\varphi)$, we add additional auxiliary variables~$c'$, $c''$, $\overline{c}'$, and $\overline{c}''$ and construct the following clauses.
\begin{flalign}
    \label{lab:wcnf4}&\text{\d{$c$}}\rightarrow c'\qquad c'\rightarrow c''\qquad\qquad \overline{\text{\d{$c$}}}\rightarrow \overline{c}'\qquad \overline{c}'\rightarrow \overline{c}''&%\\ 
\end{flalign}
Further, we replace every occurrence of~$\text{\d{$c$}}$ and~$\overline{\text{\d{$c$}}}$ in Equations~(\ref{lab:wcnf}) and~(\ref{lab:wcnf3}) by~$c''$ and~$\overline{c}''$, respectively. Observe that this requires to add an additional condition to Definition~\ref{def:rogue}, which is fulfilled if one of the copies of~$\text{\d{$c$}}$ ($\overline{\text{\d{$c$}}}$) are assigned differently.
Indeed, it could be that in a satisfying assignment, e.g., $c''$ is assigned to~$1$, but~$c'$ is not.
To accommodate this, we update Definition~\ref{def:roguesymm} on symmetric rogue models, as outlined below.

Analogously, for every non-root node~$t^*$ in~$T$, we add auxiliary variables~$e_{t^*}'$, $e_{t^*}''$, $o_{t^*}'$, $o_{t^*}''$ and construct:
\begin{flalign}
\label{lab:auxeo}&e_{t^*}\rightarrow e_{t^*}'\useieee{\quad}\uselipics{\qquad\qquad} e_{t^*}'\rightarrow e_{t^*}''\useieee{\qquad}\uselipics{\qquad\qquad} o_{t^*}\rightarrow o_{t^*}'\useieee{\quad}\uselipics{\qquad\qquad} o_{t^*}'\rightarrow o_{t^*}''&%\\ 
\end{flalign}
Then, it remains to replace in Equations~(\ref{eq:o1})--(\ref{eq:e2})
%Idea: Edges between $e_t/e_{t'}$, $o_t/o_{t'}$, $c$'s, $nc$'s, $v$'s, $nv$'s are sufficient.
those occurrences of~$e_{t'}$ and~$o_{t'}$ where $t'=t^*$, by $e_{t^*}''$ and~$o_{t^*}''$, respectively. We refer to the adapted Reduction by~$R(\varphi, \mathcal{T})$.

In turn, these additional clauses not only preserve the bipartite property, but they also ensure maximum degree~$3$. We refer by~$R^{\text{C+B}}(\varphi,\mathcal{T})$ to the reduction obtained from modifying~$R(\varphi,\mathcal{T})$ as outlined above.
In order to show correctness for the reduction similar to Lemmas~\ref{lem:welldef}--\ref{lem:nonmodels}, we require an updated definition of the (symmetric) rogue model below. As above, %we define 
\(
\psi^{\text{C+B}}_1\coloneq R^{\text{C+B}}(\phi, \mathcal{T}) \cup \{x \rightarrow e_{\rootOf(T)} \}
\)
and 
\(
\psi^{\text{C+B}}_2 \coloneq R^{\text{C+B}}(\phi, \mathcal{T}) \cup \{x \rightarrow o_{\rootOf(T)} \}
\). %The goal in the following is to prove that $\#(\phi)=\#(\psi_1)-\#(\psi_2)$.

\begin{definition}[Rogue Model For Cubic and Bipartite]\label{def:rogue_3}
 Let $t$ be a node in~$T$.
 A model~$M$ of a formula~$\phi'\supseteq R^{\text{C+B}}(\phi, \mathcal{T})$
 is referred to by \emph{rogue (at~$t$)} whenever
 \begin{enumerate}[label=(\roman*)]
 \item $x\notin M$,
 \item $|M\cap \{o1_t, o2_t, e1_t, e2_t\}|\neq 1$, 
 \item $|M\cap \{\text{\d{$c$}}, \overline{\text{\d{$c$}}}\}|\neq 1$, $|M\cap \{\text{\d{$c$}}, c', c''\}| \notin \{0,3\}$, or $|M\cap \{\overline{\text{\d{$c$}}}, \overline{c}', \overline{c}''\}| \notin \{0,3\}$ with $c{\in}\delta(t)$,~or
 %
 %
 %\item  with $c\in\delta(t)$,
 %
 %
 %\item  with $c\in\delta(t)$, or
 %
 %
 \item $|M\cap \{o_t, e_t\}|\neq 1$, $|M\cap \{o_t, o_t', o_t''\}| \notin \{0,3\}$, or $|M\cap \{e_t, e_t', e_t''\}| \notin \{0,3\}$.
 \end{enumerate}
\end{definition}

Then, we can still bijectively translate rogue models between $\psi^{\text{C+B}}_1$ and $\psi^{\text{C+B}}_2$.

\begin{definition}[Symmetric Rogue Model For Cubic and Bipartite]\label{def:roguesymm_3}
Let $M$ be a rogue model at~$t$ of a formula~$\phi'$ with~$\phi'\supseteq R^{\text{C+B}}(\phi, {\mathcal{T}})$. 
Assume that (1) there is no ancestor~$t'$ of~$t$ in~$T$ such that~$M$ is rogue at~$t'$ and that (2) $t$ is on the lexicographic smallest root-to-leaf path of~$T$.
The \emph{symmetric rogue model $M'$
(of~$M$)} is constructed as follows.
\begin{itemize}
    \item If $x\notin M$, we define $M'=M$
    \item Otherwise, if $x\in M$:
    \begin{itemize}
    \item Replace $o_t\in M$ by $e_t\in M'$, $o_t'\in M$ by $e_t'\in M'$, and $o_t''\in M$ by $e_t''\in M'$ (as well as vice versa, i.e., $e_t\in M$ iff $o_t\in M'$, $e_t'\in M$ iff $o_t'\in M'$, and $e_t''\in M$ iff $o_t''\in M'$).
    \item For every ancestor~$t'$ of~$t$ in~$T$, we replace $o_{t'}\in M$ by $e_{t'}\in M'$, $o_{t'}'\in M$ by $e_{t'}'\in M'$, $o_{t'}''\in M$ by $e_{t'}''\in M'$, and vice versa (i.e., $e_{t'}\in M$ iff $o_{t'}\in M'$, $e_{t'}'\in M$ iff $o_{t'}'\in M'$, $e_{t'}''\in M$ iff $o_{t'}''\in M'$), as well as $o1_{t'}\in M$ by $e1_{t'}\in M'$, $e1_{t'}\in M$ by $o1_{t'}\in M'$, $o2_{t'}\in M$ by $e2_{t'}\in M'$, $e2_{t'}\in M$ by $o2_{t'}\in M'$ (and vice versa)
    \item If (a) either $o_t\in M$ or $e_t\in M$, and (b) $|M\cap \{o1_t, o2_t, e1_t, e2_t\}|\geq 1$, 
    we additionally replace $o1_t\in M$ by $e2_t\in M'$, $o2_t\in M$ by $e1_t\in M'$ and vice versa (i.e., $e1_t\in M$ by $o2_t\in M'$, $e2_t\in M$ by $o1_t\in M'$).
    \item If (a), (b), and (c) $|M\cap \{\text{\d{$c$}}, \overline{\text{\d{$c$}}}\}|=1$ with $c\in\delta(t)$, 
    we additionally replace $\text{\d{$c$}}\in M$ by $\overline{\text{\d{$c$}}}\in M'$, $c'\in M$ by $\overline{c}'\in M'$, $c''\in M$ by $\overline{c}''\in M'$ and vice versa (i.e., $\overline{\text{\d{$c$}}}\in M$ by $\text{\d{$c$}}\in M'$, $\overline{c}'\in M$ by $c'\in M'$, $\overline{c}''\in M$ by $c''\in M'$).
    \end{itemize}
   \noindent We say that $M$ is the \emph{symmetric rogue model} of~$M'$.
\end{itemize}
\end{definition}

With these key definitions, we can establish correctness similarly to Lemmas~\ref{lem:welldef}--\ref{lem:nonmodels} and Proposition~\ref{prop:corr}.
There, the crucial observation is that we can always perform the translations required by the symmetric rogue model of Definition~\ref{def:roguesymm_3}. Indeed, even if there is only one of~$\text{\d{$c$}}$ or $\overline{\text{\d{$c$}}}$ in a model~$M$, in~$M'$ we still need to precisely flip between copy variables for~$\text{\d{$c$}}$ and those for~$\overline{\text{\d{$c$}}}$ (see the added last case in Definition~\ref{def:roguesymm_3}).

\begin{proof}[Proof of Proposition~\ref{proposition:planar}]
	Suppose towards a contradiction that such a reduction exists.
	Then, we can decide $\Lang{3sat}$ with $n$ variables via linear many $\Lang{\#planar3sat}$ calls.
	However, each of these calls can be decided in time~$2^{\mathcal{O}(\sqrt{n})}\cdot n^{\mathcal{O}(1)}$,
	as planar graphs can be grid-embedded; in the worst case both dimensions are roughly equal ($\mathcal{O}(\sqrt{n})\times\mathcal{O}(\sqrt{n})$)
	since in a grid the treewidth is the smaller of both.
	%of size $\sqrt(n)\cdot\sqrt(n)$ has treewidth planar 
	This contradicts $\Class{ETH}$, deciding $\Lang{3sat}$ in $2^{o(n)}\cdot n^{\mathcal{O}(1)}$.
\end{proof}

\clearpage
\section{Proofs for New Characterization of GapP}
%Proof of Theorem~\ref{theorem:gapp}}
\label{sec:proofgapP}

\gaptheorem*
\begin{proof}
Class $\Gap\Class{P}$ is equivalent to the subtraction of two $\Class{\#P}$ calls~\cite[Proposition 3.5]{FennerFortnowKurtz94}.
We show the inclusions from left to right (and then close the cycle).

Case ``$\Gap\Class{P}\subseteq[\Lang{\#2sat}-\Lang{\#2sat}]^{\log}$'':
    Since $\Gap\Class{P}$ is equivalent to  $\Class{\#P}-\Class{\#P}$, %calls~\cite[Proposition 3.5]{FennerFortnowKurtz94}, 
    it is equivalent to $[\Lang{\#3sat}-\Lang{\#3sat}]^{\log}$.
    Indeed, each of these $\Class{\#P}$ calls can be parsimoniously translated into $\Lang{\#(3)sat}$~\cite[Lemma 3.2]{Valiant79} and it is easy to see that these translations can be carried out using a constant number of pointers to the input. 
    This results in two formulas~$\varphi$ and~$\varphi'$.
    Then, we apply our reduction as in Theorem~\ref{theorem:main} %{cor:sharpmon} 
    on $\varphi$, resulting in \Lang{impl2sat} formulas~$\psi_1, \psi_2$ such that~$\#(\varphi)=\#(\psi_1)-\#(\psi_2)$. Similarly, we obtain $\psi'_1$ and $\psi'_2$ from~$\varphi'$. 
    We compute $\#(\varphi)-\#(\varphi')$ by~$(\#\psi_1-\#\psi_2)-(\#\psi'_1-\#\psi'_2)=(\#\psi_1+\#\psi'_2)-(\#\psi_2+\#\psi'_1)$.
    From this, we construct formula~$\alpha=\psi_1\cup \psi''_2$, where~$\psi''_2$ is obtained from~$\psi'_2$ by replacing every variable with a fresh variable.
    Analogously, we construct $\beta=\psi_2\cup \psi''_1$, where~$\psi''_1$ is obtained from~$\psi'_1$ by replacing variables with fresh variables.
    Observe that $\#(\alpha)=\#(\psi_1)\cdot\#(\psi''_2)$. To go from~``$\cdot$'' to~``$+$'', we need to switch between both formulas alternatively.
    Such a switch between any two sets~$V_1, V_2$ of variables of \Lang{impl2sat} formulas~$\gamma_1,\gamma_2$, is modeled by an~\Lang{impl2sat} formula $switch(V_1, V_2)=\{(s\rightarrow v), (v' \rightarrow s)\mid v \in V_1, v'\in V_2\}$, where~$s$ is a fresh variable. Observe that $v'\rightarrow s$ is equivalent to $\neg s\rightarrow \neg v'$ by contraposition, since~$(\neg \neg s \vee \neg v') = (\neg v' \vee s) = (v'\rightarrow s)$.
    Since both~$\gamma$ and~$\delta$ are in \Lang{impl2sat}, depending on~$s$, we set the variables of~$\gamma_1$ to~$1$ (if~$s=1$) or we set variables of~$\gamma_2$ to~$0$ (if~$s=0$), as in both cases the formulas are satisfied by the default value.
    Indeed, with $\alpha'=\alpha \cup switch(\vars(\psi_1), \vars(\psi''_2))$ and $\beta'=\beta \cup switch(\vars(\psi_2), \vars(\psi''_1))$ we establish the claim, as $\#(\alpha')-\#(\beta')=\#(\varphi)-\#(\varphi')$.

Case ``$[\Lang{\#2sat}-\Lang{\#2sat}]^{\log}\subseteq [\Lang{\#impl2sat}-\Lang{\#impl2sat}]^{\log}$'': Without loss of generality, we may assume two formulas $\psi$ and~$\psi'$ in \Lang{impl2sat}, where the goal is to compute~$\#(\psi)-\#(\psi')$.
Indeed, if either~$\psi$ or~$\psi'$ were not in \Lang{impl2sat}, it can be translated as shown above.

Case ``$[\Lang{\#impl2sat}-\Lang{\#impl2sat}]^{\log}\subseteq[\Lang{\#0,1-2dnf}-\Lang{\#0,1-2dnf}]^{\log}$'':
%Without loss of generality, we may 
Assume two formulas $\psi$ and~$\psi'$ in \Lang{impl2sat}, where the goal is to compute~$\#(\psi)-\#(\psi')$.
%
%Indeed, if either~$\psi$ or~$\psi'$ were not in \Lang{impl2sat}, it can be translated as shown above.
%
The goal is to obtain this number by computing $2^{|\vars(\psi)|}-\#(\neg\psi)-(2^{|\vars(\psi')|}-\#(\neg\psi'))$. However, in general, the number of variables of~$\psi$ might differ from those in~$\psi'$.
To compensate for this difference~$n=|\vars(\psi)-\vars(\psi')|$, we need to add additional variables to the smaller formula.
Let~$\alpha,\beta$ be the formulas~$\psi,\psi'$ such that~$|\vars(\alpha)|<|\vars(\beta)|$.
We will reuse the $switch$ construction from above, where we let $V=\{v_1, \ldots, v_n\}$ be fresh variables, $\alpha'=\alpha\cup switch(\vars(\alpha)\cup V, V)$ and~$\beta'=\beta \cup switch(\vars(\beta), \emptyset)$.
Observe that due to $switch$, both~$\alpha'$ and~$\beta'$ have one additional satisfying assignment.
Indeed, the construction ensures that $2^{|\vars(\alpha')|}-\#(\neg\alpha')-(2^{|\vars(\beta')|}-\#(\neg\beta')) = \#(\neg\beta')-\#(\neg\alpha')=\#(\neg\psi')-1-(\#(\neg\psi)-1)=\#(\neg\psi')-\#(\neg\psi)=\#(\psi)-\#(\psi')$.

Case ``$[\Lang{\#0,1-2dnf}-\Lang{\#0,1-2dnf}]^{\log}\subseteq[\Lang{\#2dnf}-\Lang{\#2dnf}]^{\log}$'': By definition.

Case ``$[\Lang{\#2dnf}-\Lang{\#2dnf}]^{\log}\subseteq [\Lang{\#3dnf}-\Lang{\#3dnf}]^{\log}$'': By definition. 

Case ``$ [\Lang{\#3dnf}-\Lang{\#3dnf}]^{\log} \subseteq [\Lang{\#mon2sat}-\Lang{\#mon2sat}]^{\log}$'':
%First, we show that $\Gap\Class{P} \subseteq [\Lang{\#mon2sat}-\Lang{\#mon2sat}]^{\log}$.
    %We know that~$\#\Class{P}-\#\Class{P}=\Gap\Class{P}$. Further, 
    As mentioned above, there is a parsimonious many-one reduction from any problem in $\Class{\#P}$ to $\Lang{\#sat}$, as the Cook-Levin construction
is solution preserving~\cite[Lemma 3.2]{Valiant79}, which works in logspace.
    Consequently, for~$\Gap\Class{P}$ we obtain two propositional formulas~$\varphi, \varphi'$ such that the goal is to compute~$c=\#(\varphi)-\#(\varphi')$. 
Then, we use our reduction~$R$ on~$\varphi$, resulting in two~$\Lang{\#mon2cnf}$ formulas~$\psi_1$, $\psi_2$. Similarly, applying~$R$ on~$\varphi'$ yields the~$\Lang{\#mon2cnf}$ formulas~$\psi'_1$, $\psi'_2$. 
Then, we have that~$c=\#(\psi_1)-\#(\psi_2)-(\#(\psi'_1)-\#(\psi'_2))=(\#(\psi_1)+\#(\psi'_2))-(\#(\psi_2)+\#(\psi'_1))$.
Now, it remains to construct formulas~$\alpha, \alpha'$ such that~$\#(\alpha)-\#(\alpha')=c=(\#(\psi_1)+\#(\psi'_2))-(\#(\psi_2)+\#(\psi'_1))$.
To this end, we build a formula~$monswitch(\iota, \tau, \kappa)$ over three formulas~$\iota,\tau,\kappa$ (such that~$\iota\subseteq\kappa$, $\tau\subseteq\kappa$), which uses fresh variables~$s_\iota, s_\tau$ and constructs %~$s_\iota \vee s_\tau$, 
$s_\iota \vee v$ for every~$v\in \vars(\kappa\setminus \iota)$, as well as~$s_\tau \vee v'$ for every~$v'\in \vars(\kappa\setminus \tau)$.
This ensures that if~$s_\iota$ is set to false, we obtain~$\#(\iota)$ models, whereas setting~$s_\tau$ to false, we receive~$\#(\tau)$ models.
If both $s_\iota$ and $s_\tau$ are set to true, the goal is to obtain~$\#(\kappa)$ many models, and if both $s_\iota$ and $s_\tau$ are set to false, we get~$1$ model.

Without loss of generality, we assume that~$\psi_1, \psi_2, \psi'_1, \psi'_2$ do not share variables, which can be easily achieved by renaming. Let~$\beta = \psi_1 \cup \psi_2 \cup \psi'_1 \cup \psi'_2$.
Then, we build $\alpha = \beta \cup monswitch(\psi_1, \psi'_2, \beta)$ as well as~$\alpha'=\beta\cup monswitch(\psi_2, \psi'_1, \beta)$.
Consequently, we have that~$\#(\alpha)=\#(\psi_1)+\#(\psi'_2)+\#(\beta)+1$ and~$\#(\alpha')=\#(\psi_2)+\#(\psi'_1)+\#(\beta)+1$, resulting in~$c=\#(\alpha)-\#(\alpha')$.

%\medskip

Case ``$[\Lang{\#mon2sat}-\Lang{\#mon2sat}]^{\log}\subseteq[\Lang{\#mon2dnf}-\Lang{\#mon2dnf}]^{\log}$'': Similar to above we assume two formulas $\psi$ and~$\psi'$ in \Lang{mon2sat}, with the goal of computing~$\#(\psi)-\#(\psi')$.
%
%Indeed, if either~$\psi$ or~$\psi'$ were not in \Lang{impl2sat}, it can be translated as shown above.
%
Then, this equals to $2^{|\vars(\psi)|}-\#(\neg\psi)-(2^{|\vars(\psi')|}-\#(\neg\psi'))$. However, in general, the number of variables of~$\psi$ might differ from those in~$\psi'$.
To compensate, adding additional variables to the smaller formula seems challenging (without negation).
%
%Let~$\alpha,\beta$ be the formulas~$\psi,\psi'$ such that~$|\vars(\alpha)|<|\vars(\beta)|$.
%
However, we can do the following.
Without loss of generality, assume that $\psi$ and $\psi'$ do not share variables and let~$\beta=\psi\cup\psi'$.
We will reuse the $monswitch$ construction from above, where we construct $\alpha=\psi\cup monswitch(\psi, \emptyset, \beta)$ and~$\alpha'=\psi' \cup monswitch(\psi', \emptyset, \beta)$.
Indeed, the construction ensures that $\#(\neg\alpha')-\#(\neg\alpha)=2^{|\vars(\alpha')|}-\#(\alpha')-(2^{|\vars(\alpha)|}-\#(\alpha)) = \#(\alpha)-\#(\alpha')=\#(\psi)+1+1+\#(\psi)\#(\psi')-(\#(\psi')+1+1+\#(\psi)\#(\psi'))=\#(\psi)-\#(\psi')$.

Case ``$[\Lang{\#mon2dnf}-\Lang{\#mon2dnf}]^{\log}\subseteq  \Span\Class{L}-\Span\Class{L} \subseteq \Gap\Class{P}$'': Trivial, since $\Lang{\#dnf}\in\Span\Class{L}$~\cite[Proof of Theorem 4.8]{AlvarezJenner93}. Further, we have $\Span\Class{L}-\Span\Class{L}\subseteq\Class{\#P}-\Class{\#P}=\Gap\Class{P}$ since $\Span\Class{L}\subseteq\Class{\#P}$ %as $\Lang{\#2dnf}$ is in $\Class{\#P}$ and $\Gap\Class{P}$ is equivalent to the subtraction of two $\Class{\#P}$ calls~\cite[Proposition 3.5]{FennerFortnowKurtz94}.
and $\Gap\Class{P}$ is equivalent to the subtraction of two $\Class{\#P}$ calls~\cite[Proposition 3.5]{FennerFortnowKurtz94}.

 \medskip

% \noindent Now we show the inclusions from right to left.

% Case ``$\Gap\Class{P}\supseteq[\Lang{\#impl2sat}-\Lang{\#impl2sat}]^{\log}$'': Trivial, as $[\Lang{\#impl2sat}]^{\log}\subseteq \Class{\#P}$ and $\Gap\Class{P}=[\Class{\#P}-\Class{\#P}]^{\Class{P}}\supseteq[\Class{\#P}-\Class{\#P}]^{\log}$~\cite[Proposition 3.5]{FennerFortnowKurtz94}.

% Case ``$[\Lang{\#impl2sat}-\Lang{\#impl2sat}]^{\log}\supseteq[\Lang{\#2sat}-\Lang{\#2sat}]^{\log}$'': We showed the stronger statement~$[\Lang{\#3Sat}-\Lang{\#3sat}]^{\log}\subseteq[\Lang{\#impl2sat}-\Lang{\#impl2sat}]^{\log}$.

% Case ``$[\Lang{\#2sat}-\Lang{\#2sat}]^{\log}\supseteq[\Lang{\#2dnf}-\Lang{\#2dnf}]^{\log}$'': 
% This direction is easy to see, as a formula in DNF can be encoded in 3CNF using the translation by Tseitin~\cite{Tseitin1983}, thereby preserving the number of satisfying assignments. 
% %
% Above we showed~$[\Lang{\#3Sat}-\Lang{\#3sat}]^{\log}\subseteq[\Lang{\#impl2sat}-\Lang{\#impl2sat}]^{\log}$.
% %

% Case ``$[\Lang{\#2sat}-\Lang{\#2sat}]^{\log}\supseteq[\Lang{\#0,1-2dnf}-\Lang{\#0,1-2dnf}]^{\log}$'': Subsumed by  previous~case. 

% Case ``$[\Lang{\#0,1-2dnf}-\Lang{\#0,1-2dnf}]^{\log}\supseteq[\Lang{\#2dnf}-\Lang{\#2dnf}]^{\log}$'': We showed $[\Lang{\#2sat}-\Lang{\#2sat}]^{\log}\supseteq[\Lang{\#2dnf}-\Lang{\#2dnf}]^{\log}$ and $[\Lang{\#2sat}-\Lang{\#2sat}]^{\log}\subseteq[\Lang{\#0,1-2dnf}-\Lang{\#0,1-2dnf}]^{\log}$.

%
Then, we can follow the chain of inclusions above to %land in $[\Lang{\#2dnf}-\Lang{\#2dnf}]^{\log}$, thereby 
finally establish the claim. %$[\Lang{\#2dnf}-\Lang{\#2dnf}]^{\log}$.
%$spanL \subsetneq_{NL\neq NP} \#P$

\bigskip
In order to establish stronger claims involving properties bipartiteness and max-degree~$3$ for $\Lang{\#impl2sat}$ and $\Lang{\#0,1-2dnf}$, we replace the $switch$ construction above.
We provide $cycswitch$ over formulas~$\varphi_1$ and $\varphi_2$ that preserves the required properties (at most degree~$3$ and bipartiteness: the edges of primal graphs $G_{\varphi_1}$ and $G_{\varphi_2}$ are in~$V_1^e\times V_1^o$ and~$V_2^e\times V_2^o$, respectively). The idea is precisely the same as above, but the actual construction, referred to by~$cycswitch(V_1,V_2)$ with~$V_1=\vars(\varphi_1)$ and $V_2=\vars(\varphi_2)$, is more involved. 
Indeed, to preserve both properties of degree at most~$3$ and bipartiteness, we will construct chains of implications, using $4m$ fresh switch variables of the form~$S=\{s_i^e, s_i^o \mid 1\leq i\leq 2m\}$ (where $m=\max(|V_1|, |V_2|)$), as well as~$5$ copy variables~$v^1, \ldots, v^5$ for every variable~$v$ in~$V_1\cup V_2$ that is of degree~$3$. 

The switch consists of cyclic implications of the form~$(s_1^e \rightarrow s_1^o), (s_1^o \rightarrow s_2^e), \ldots, (s_{2m}^e \rightarrow s_{2m}^o), (s_{2m}^o \rightarrow s_1^e)$. This ensures that either each of these bits is set to~$1$ or all are set to~$0$. 
Variables~$u_i\in V_1$ of degree~$\leq 2$ can be easily connected to the switch cycle, using implications~$(s_i^e\rightarrow u_i)$ if~$u_i\in V_1^e$, and~$(s_i^o\rightarrow u_i)$ if~$u_i\in V_1^o$.
Analogously, variables~$u_i\in V_2$ of degree~$\leq 2$ can be connected using implications~$(u_i \rightarrow s_{m+i}^e)$ if~$u_i\in V_2^e$, and~$(u_i \rightarrow s_{m+i}^o)$ if~$u_i\in V_2^o$.
For variables~$v_j\in V_1$ of degree~$3$, we need to rewrite such that we can reduce to the case of degree~$2$ above. % those can be connected to any of the switch variables above. 
Let therefore~$w_1$ and~$w_2$ be two neighbors of~$v_j$ in~$G_{\varphi_1}$ such that both~$w_1,w_2$ form outgoing
implications (to~$v_j$) or incoming implications (to~$w_1$/$w_2$). 
For the sake of concreteness, we assume both~$(w_1\rightarrow v_j)$ and~$(w_2\rightarrow v_j)$ are in $\varphi_1$,
as the other case $(v_j\rightarrow w_1)$ and~$(v_j\rightarrow w_2)$ works analogously (which covers all cases we need to consider).
We additionally construct~$(v_j \rightarrow v_j^1)$, $(v_j^1 \rightarrow v_j^2)$, $(v_j^2 \rightarrow v_j^3)$, $(v_j^3 \rightarrow v_j^4)$, $(v_j^4 \rightarrow v_j^5)$, and $(v_j^5 \rightarrow v_j)$.
Then, we can construct implications~$(w_1\rightarrow v_j^2)$ and $(w_2\rightarrow v_j^4)$. As above, observe that~$v_j^1$ is of degree~$2$ and can be connected to the switch cycle.

In the end, for every such variable~$v_j$ we also need to remove implications $(w_1\rightarrow v_j)$ and~$(w_2\rightarrow v_j)$ from $\varphi_1$, resulting in~$\varphi'_1$.
Analogously, we proceed for variables~$v_j\in V_2$, construct implications in~$cycswitch(V_1,V_2)$, and remove implications from $\varphi_2$ as above, resulting in~$\varphi'_2$.
%, we replace~$v_j$ let
%
The overall construction preserves max.\ degree~$3$ and bipartiteness.
\end{proof}

\clearpage
\section{Proofs for A New Characterization of PH}\label{sec:proofscharact}

\singlecalltheoremb*
\begin{proof}
  
``$\Gap\Class{P}\subseteq [\Lang{\#impl2sat}]^{\log}_{\Class{AC}^0}$'': In contrast to above, we instead apply Theorems~\ref{theorem:main} (B) and~\ref{theorem:gapp}, thereby obtaining two $\Lang{impl2cnf}$ formulas~$\varphi_1$ and~$\varphi_2$, which have the desired properties.
We assume that~$\vars(\varphi_1)\cap\vars(\varphi_2)=\emptyset$, which can be obtained by renaming all variables.
Let~$n=|\vars(\varphi_1)|$ and~$n'=|\vars(\varphi_2)|$.
We only need to find a basis that is provably 
larger than~$\max(\#(\varphi_1), \#(\varphi_2))$, namely $>2^{\max(n,n')}$. Therefore, $m=\max(n,n')$ bits are sufficient.

Similarly to the switch construction in Theorem~\ref{theorem:singlecalla} above, 
we need to provide an updated switch version. We extend $cycswitch$, as defined in the proof of Theorem~\ref{theorem:gapp}, that preserves the required properties (at most degree~$3$ and bipartiteness: the edges of primal graphs $G_{\varphi_1}$ and $G_{\varphi_2}$ are in~$V_1^e\times V_1^o$ and~$V_2^e\times V_2^o$, respectively). The idea is precisely the same as above, but the actual construction, referred to by~$extcycswitch(B,V_1,V_2)$ with~$V_1=\vars(\varphi_1)$ and $V_2=\vars(\varphi_2)$, is more involved. 
As above, we will construct chains of implications, using $4m$ fresh switch variables of the form~$S=\{s_i^e, s_i^o \mid 1\leq i\leq 2m\}$, $m$ bit variables~$B=\{b_1, \ldots, b_m\}$ as above, as well as~$5$ copy variables~$v^1, \ldots, v^5$ for every variable~$v$ in~$V_1\cup V_2$ of degree~$3$. 

The switch $extcycswitch$ consists of  implications of the form~$(s_1^e \rightarrow s_1^o), (s_1^o \rightarrow s_2^e), \ldots,\allowbreak (s_{2m}^e \rightarrow s_{2m}^o), (s_{2m}^o \rightarrow s_1^e)$. %This ensures that either every single one of these bits is set to~$1$ or all are set to~$0$. 
Variables~$u_i\in V_1$ of degree~$\leq 2$ can be easily connected to the switch cycle, using implications~$(s_i^e\rightarrow u_i)$ if~$u_i\in V_1^e$, and~$(s_i^o\rightarrow u_i)$ if~$u_i\in V_1^o$.
Analogously, variables~$u_i\in V_2$ of degree~$\leq 2$ can be connected using implications~$(u_i \rightarrow s_{m+i}^e)$ if~$u_i\in V_2^e$, and~$(u_i \rightarrow s_{m+i}^o)$ if~$u_i\in V_2^o$.
Again, variables~$v_j\in V_1$ of degree~$3$ need to be rewritten such that we can reduce to the case of degree~$2$ above. % those can be connected to any of the switch variables above. 
Let $w_1$ and~$w_2$ be two neighbors of~$v_j$ in~$G_{\varphi_1}$ such that~$w_1,w_2$ form outgoing
implications (to~$v_j$) or incoming implications (to~$w_1$/$w_2$). 
For concreteness, assume both~$(w_1\rightarrow v_j)$ and~$(w_2\rightarrow v_j)$ are in $\varphi_1$; the other case $(v_j\rightarrow w_1)$ and~$(v_j\rightarrow w_2)$ works analogously.
We additionally construct~$(v_j \rightarrow v_j^1)$, $(v_j^1 \rightarrow v_j^2)$, $(v_j^2 \rightarrow v_j^3)$, $(v_j^3 \rightarrow v_j^4)$, $(v_j^4 \rightarrow v_j^5)$, and $(v_j^5 \rightarrow v_j)$.
We construct implications~$(w_1\rightarrow v_j^2)$ and $(w_2\rightarrow v_j^4)$; now~$v_j^1$ is of degree~$2$ and can be connected to the switch cycle.

For every variable~$v_j$ we need to remove implications $(w_1\rightarrow v_j)$ and~$(w_2\rightarrow v_j)$ from $\varphi_1$, resulting in~$\varphi'_1$.
Analogously, we proceed for variables~$v_j\in V_2$, construct implications in~$extcycswitch(B,V_1,V_2)$, and remove implications from $\varphi_2$ as above, resulting in~$\varphi'_2$.
%, we replace~$v_j$ let
%
Finally, we need to connect~$B$ to the switch. Since there are at least~$3m$ possibilities in total (and connecting the variables in~$V_1\cup V_2$ already used up at most~$2m$ of them), this can be achieved by connecting any~$b_i$ in~$B$ to a variable~$s_j^*$ in~$S$ that is not yet connected to a variable in~$V_1\cup V_2$. %The overall construction preserves max.\ degree~$3$ and bipartiteness.

Consequently, we construct the formula~$\alpha=\varphi'_1\cup\varphi'_2\cup extcycswitch(B, \vars(\varphi_1), \vars(\varphi_2))$. 
Then, if every variable in~$S$ is set to~$1$, $\#(\alpha)$ corresponds to the number of models of~$\varphi_2$. Otherwise, every variable in~$S$ is set to~$0$ by construction, yielding~$2^m\#(\varphi_1)$ many assignments. This results in~$\#(\alpha)=\#(\varphi_2)+ 2^m\cdot\#(\varphi_1)$. 
It is easy to see that this reduction works in logspace, using a constant number of pointers to the input.

After counting, we can integer-divide the result by~$2^m$, and obtain the result~$\#(\varphi_1)$ as well as the remainder~$\#(\varphi_2)$ of the division. As above, this works in~$\Class{AC^0}$ using bit-wise AND (see also Theorem~\ref{theorem:todaimproved}). %Then, we obtain the integer part~$\#(\varphi_1)\cdot\#(\varphi_2)$ of the division (also in~$\Class{AC^0}$ using bit-wise AND). %also be obtained by a bit-wise AND gate (in~$\Class{AC^0}$). 
%
%Finally, by dividing the integer part by the remainder~$\#(\varphi_2)$ we can reconstruct~$\#(\varphi_1)=\frac{\#(\varphi_1)\cdot\#(\varphi_2)}{\#(\varphi_2)}$, which requires~$\Class{TC^0}$~\cite{HesseEtAl02}.
%
Finally, the result is established by a single subtraction, computing~$\#(\varphi_1)-\#(\varphi_2)$ in~$\Class{AC^0}$, which establishes the required result.

\medskip

``$[\Lang{\#impl2sat}]^{\log}_{\Class{AC}^0}= [\Lang{\#0,1-2dnf}]^{\log}_{\Class{AC}^0}$'': Follows from Lemma~\ref{cor:singlecall}.
\end{proof}

\singlecalltheorema*
\begin{proof}
%First, we parsimoniously reduce from $\Class{\#P}$ to $\Lang{\#(3)sat}$~\cite[Lemma 3.2]{Valiant79}.
%
%\medskip
%
``$\Gap\Class{P}\subseteq [\Lang{\#mon2sat}]^{\log}_{\Class{TC}^0}$'': We use Theorems~\ref{theorem:main} (A) and~\ref{theorem:gapp} to obtain two $\Lang{mon2cnf}$ formulas~$\varphi_1$ and~$\varphi_2$.
Then, we rename all the variables in~$\varphi_2$, obtaining~$\varphi'_2$, which does not share a variable with~$\varphi_1$.
Let~$n=|\vars(\varphi_1)|$ and~$n'=|\vars(\varphi_2)|$.
Now, in order to represent both formulas in a single call, we need to find a basis that is provably 
larger than the product~$\#(\varphi_1)\cdot\#(\varphi_2)$, namely $>2^{n+n'}$. Therefore, $m=n+n'+1>n+n'$ bits are sufficient. 
We will construct a (relaxed) version of the switch construction, as used in the proof of Theorem~\ref{theorem:gapp}. This construction adds clauses of the form~$s \vee b_1, \ldots, s \vee b_m$ for fresh variables~$s, b_1, \ldots, b_m$, thereby ensuring that if~$s$ is assigned to~$0$, all the bits of the basis must be fixed (set to~$1$). 
Further, we add $s \vee v_1, \ldots, s \vee v_n$ for the variables~$\{v_1, \ldots, v_n\}=\vars(\varphi_1)$ of the first formula~$\varphi_1$. This ensures that if~$s$ is~$0$, we only obtain satisfying assignments of~$\varphi_2$. More precisely, for any two sets~$B, V$ of variables, we construct~$relswitch(B,V)=\{(s \vee b), (s \vee v) \mid b\in B, v\in V\}$.

Consequently, we construct the formula~$\alpha=\varphi_1\cup\varphi'_2\cup relswitch(\{b_1, \ldots, b_m\}, \vars(\varphi_1))$. 
Then, if~$s=1$, $\#(\alpha)$ corresponds to the number of models of~$\varphi_1$ multiplied by those of~$\varphi_2$ multiplied by~$2^m$. This results in~$\#(\alpha)=\#(\varphi_2)+ \#(\varphi_1)\cdot\#(\varphi_2)\cdot 2^m$. 
It is easy to see that this reduction works in logspace, as we only need a constant number of pointers to the input.
After counting, we can integer-divide the result by~$2^m$, and obtain the remainder~$\#(\varphi_2)$ of the division, which works in~$\Class{AC^0}$ using bit-wise AND on $\#(\alpha)$ (see also Theorem~\ref{theorem:todaimproved}). Then, we obtain the integer part~$\#(\varphi_1)\cdot\#(\varphi_2)$ of the division (also in~$\Class{AC^0}$ using bit-wise AND). %also be obtained by a bit-wise AND gate (in~$\Class{AC^0}$). 
Finally, by dividing the integer part by the remainder~$\#(\varphi_2)$ we can reconstruct~$\#(\varphi_1)=\frac{\#(\varphi_1)\cdot\#(\varphi_2)}{\#(\varphi_2)}$, which requires~$\Class{TC^0}$~\cite{HesseEtAl02}.
Finally, the result is obtained by a single subtraction, computing~$\#(\varphi_1)-\#(\varphi_2)$ in~$\Class{AC^0}$, which establishes the required result.
%
%\medskip
%

``$[\Lang{\#mon2sat}]^{\log}_{\Class{TC}^0}= [\Lang{\#mon2dnf}]^{\log}_{\Class{TC}^0}$'': Follows from Lemma~\ref{cor:singlecall}.
%
%``$[\Lang{\#mon2sat}]^{\log}_{\Class{TC}^0}\subseteq [\Lang{\#mon2dnf}]^{\log}_{\Class{TC}^0}$'': This immediately follows from the observation that $[\Lang{\#mon2sat}]^{\log}=[2^n-\Lang{\#mon2dnf}]^{\log}\subseteq[\Lang{\#mon2dnf}]^{\log}_{\Class{TC}^0}$.
%
%
%``$[\Lang{\#mon2dnf}]^{\log}_{\Class{TC}^0}\subseteq [\Lang{\#mon2sat}]^{\log}_{\Class{TC}^0}$'': As above, $[\Lang{\#mon2dnf}]^{\log}=[2^n-\Lang{\#mon2sat}]^{\log}\allowbreak\subseteq[\Lang{\#mon2sat}]^{\log}_{\Class{TC}^0}$.
\end{proof}

\todaimproved*
\begin{proof}
We perform the known reduction~\cite{Toda91} from $\Lang{ph}$
to~$\Class{P}^{\Class{\#P}[1]}$ that uses a single $\Lang{\#sat}$
call, followed by computing the remainder of a division by~$2^m$, where~$m$ is polynomial in the size of the input.  
The whole reduction can be computed in logspace. The key ingredient~\cite{Toda91} is actually Lemma~2.1 of the Valiant-Vazirani theorem~\cite{ValiantVazirani86}. However, each step does not only work in linear time (as claimed), but there is no need to keep more than a constant number of pointers to the input to output the formula since~$w$ is picked randomly for each round.

Then, we apply Theorem~\ref{theorem:singlecalla}, obtaining a
single~$\Lang{\#mon2sat}$ formula~$\varphi$ and encode the $2$
shifting operations, $1$ division, as well as $1$ subtraction into
the~$\Class{TC^0}$ circuit~\cite{HesseEtAl02} for postprocessing.  Finally, we also encode the final division by~$2^m$ into the~$\Class{TC^0}$ circuit%~\cite{HesseEtAl02}
, such that the result equals~$1$ iff the result does not divide by~$2^m$, i.e., $\not\equiv 0 (\bmod 2^m)$. Observe that this operation actually works in~$\Class{AC^0}$. Indeed, one can encode this via a binary AND operation with a bit-mask where every bit is set to~$1$, except the $m$ least significant bits, which are set to~$0$. The inclusion $[\Lang{\#mon2sat}]^{\log}_{\Class{TC}^0}\subseteq  \Class{P}^{\Span\Class{L}}$ is easy to see, as $\Lang{\#dnf}$ is contained in~$\Span\Lang{L}$ and $[\Lang{\#mon2sat}]^{\log}_{\Class{TC}^0}=[\Lang{\#mon2dnf}]^{\log}_{\Class{TC}^0}$, see Lemma~\ref{cor:singlecall}.

Alternatively, we may apply Theorem~\ref{theorem:singlecallb} to obtain
a single~$\Lang{\#impl2sat}$ formula~$\varphi'$ that has the claimed
properties. Then, we can separate both counts using binary AND with a
bitmask similar to above (and one on the negated bitmask), which works
in~$\Class{AC^0}$. The resulting subtraction between both parts can
also be carried out in~$\Class{AC^0}$.  Finally, we check whether the
resulting number is $\not\equiv 0 (\bmod 2^m)$, which also works
in~$\Class{AC^0}$ as mentioned above.
For the closure under negation see Lemma~\ref{cor:singlecall}.
\end{proof}

Our reduction and Lemma~\ref{lemma:main} almost immediately allows us to derive the following strong lower bound.
\sethlb*
\begin{proof}
$\Class{SETH}$~\cite{ImpagliazzoPaturi01} implies that we cannot decide~$s$-$\Class{SAT}$ in time~$o(2^{\rho})\cdot|\phi|^{O(1)}$.
    (A) follows from strong parameter guarantees of Lemma~\ref{lemma:main} and a slightly modified reduction. % for (A)).
 Thereby, in Equations~(\ref{eq:var}),
 we replace clauses of the form $\top_v \vee \bot_v$
 (preventing bipartiteness) by $\top_v \vee a_1$, $\bot_v \vee b_1$, $\ldots$,  $\top_v \vee a_c$, $\bot_v \vee b_c$ for fresh variables~$a_i,b_i$, constant~$c{>}s$.
 This enables integer-dividing the resulting count by~$2^{c|\vars(\phi)|}$, to recover number of solutions.
 %
%\textcolor{red}{TODO: replace $t_v \vee f_v$ by $t_v \vee x_v$, $f_v \vee y_v$ for fresh $x_v, y_v$, enabling bipartite, degree 4+12 for mon2dnf$_{tc0}$}
 %
\end{proof}

\ethlb*
\begin{proof}
    For (A) and (B), we normalize to at most $3$ occurrences (degree) per variable, resulting in a linear increase of $\tw(\phi)$. For (A), the proof of Theorem~\ref{sec:sethlb} causes indeed only a constant-degree blowup, (B) holds by Lemma~\ref{lemma:main}. By Cor.~\ref{corollary:sharp-itw}, the result holds for $\itw(\phi)$.
\end{proof}

\end{document}